\documentclass[aps,prd,notitlepage,amsfonts,amssymb,amsmath,nofootinbib,superscriptaddress,twocolumn,longbibliography]{revtex4-2}
\usepackage[utf8]{inputenc}
\usepackage[english]{babel}
\usepackage[titletoc,toc,title]{appendix}
\usepackage{amsmath}
\usepackage{amsfonts}
\usepackage{amssymb}

\usepackage[colorlinks=true,linkcolor=blue,citecolor=blue,urlcolor=blue]{hyperref}
\usepackage{graphicx}
\graphicspath{{figs/}} 

\usepackage{comment} 

\usepackage{soul} 

\usepackage{enumerate}
\usepackage{csquotes}
\usepackage[caption=false]{subfig}
\usepackage{dsfont}
\usepackage{color}
\usepackage{bbold}
\usepackage{mathtools}
\usepackage{gensymb}
\usepackage{ulem}

\begin{document}

\title{Phonon decay in 1D atomic Bose quasicondensates via Beliaev-Landau damping}
\date{\today}

\author{Amaury Micheli}
\affiliation{Universit\'e Paris-Saclay, CNRS/IN2P3, IJCLab, 91405 Orsay, France}
\affiliation{Sorbonne Universit\'e, CNRS, UMR 7095, Institut d’Astrophysique de Paris, 98 bis bd Arago, 75014 Paris, France}
\author{Scott Robertson}
\affiliation{Universit\'e Paris-Saclay, Institut d'Optique Graduate School, CNRS, Laboratoire Charles Fabry, 91127 Palaiseau, France}
\affiliation{Universit\'e Paris-Saclay, CNRS/IN2P3, IJCLab, 91405 Orsay, France}

\begin{abstract}

In a 1D Bose gas, there is no non-trivial scattering channel involving three Bogoliubov quasiparticles that conserves both energy and momentum. Nevertheless, we show that such 3-wave mixing processes (Beliaev and Landau damping) account for their decay via interactions with thermal fluctuations. Within an appropriate time window where the Fermi Golden Rule is expected to apply, the occupation number of the initially occupied mode decays exponentially and the rate takes a simple analytic form. The result is shown to compare favorably with simulations based on the Truncated Wigner Approximation. It is also shown that the same processes slow down the exponential growth of phonons induced by a parametric oscillation.

\end{abstract}

\maketitle


\section{Introduction}

Ultracold gases have proved to be a fruitful arena for both theoretical and experimental research.
In particular, the propagation of elementary excitations on top of a macroscopic condensed background provides an accessible realisation of a quantum field in an effective curved spacetime~\cite{Garay2000,LivingReview}.
This can be exploited to, {\it e.g.}, mimic a black-hole horizon so as to induce the analogue of Hawking radiation (as recently achieved in~\cite{deNova2019}).
It can also be used as a platform for realising an analogue of preheating, or the Dynamical Casimir Effect, which is a topic of current interest~\cite{Jaskula2012, silkePrimaryThermalisation2022,carusottoNumericalStudies2022}. The degree of experimental control, combined with the intrinsically quantum nature of ultracold gases, makes them well-suited for such experiments.

In studies of this kind, an important issue concerns the effect of dissipation on the expected signal.  
Dissipation arises in closed systems as an effective phenomenon due to quasiparticle interactions~\cite{Pylak-Zin-2018,Robertson_2018}.  This entails the existence of an intrinsic quasiparticle decay.
In 3D Bose gases, the principle mechanisms behind this decay 
are the 3-wave mixing processes of Beliaev and Landau damping \cite{Pitaevskii_1997,Fedichev_1998}.
However, in 1D Bose gases with only two-body contact interactions ({\it i.e.}, of the Lieb-Liniger model \cite{Lieb1963}), two objections have been raised against the possibility of such processes.
The first is that the 
integrability of the model prevents quasiparticle decay in principle, and that integrability-breaking perturbations must therefore be included before damping can occur~\cite{Tan2010,Ristivojevic2016,imambekov_one-dimensional_2012}.
The second is that quasiparticle decay requires the existence of non-trivial scattering channels conserving both momentum and energy, a criterion that has been routinely applied in many systems of all dimensionalities~\cite{Maris-1977,Tsuchiya-Griffin-2004,Natu-DasSarma-2013,Van_Regemortel_2017,Kurkjian-Ristivojevic-2020}.
As there are no such channels involving three collinear Bogoliubov excitations due to their gapless and convex spectrum, it has been concluded that 3-wave mixing in 1D cannot induce decay~\cite{Tan2010,Ristivojevic_2014, Van_Regemortel_2017}.

In this paper, we wish to push back a little against these conclusions. 
First, integrability does not seem to prevent relaxation within a physical model, but only thermalization: the system tends towards an equilibrium state with rather more structure than a thermal state~\cite{Kinoshita2006}, described by a generalized Gibbs ensemble~\cite{Rigol2007}, see~\cite{bouchouleRelaxation2022} for a recent application to the Lieb-Liniger gas.  In addition, in the context of 1D Bose gases, this relaxation could even be necessary in order to comply with the Mermin-Wagner-Hohenberg theorem~\cite{Hohenberg1967, Pitaevskii2016}, as it provides a mechanism by which the long-range order induced by a sufficiently narrowband excitation spectrum is washed away.

Second, we wish to show and emphasise that the apparent absence of an elastic scattering channel 
does not necessarily preclude any quasiparticle decay. 
In the Fermi Golden Rule (FGR), the Dirac delta enforcing energy conservation is typically interpreted in a binary way: either an energy-conserving channel exists and quasiparticle decay occurs, or there is no such channel and the quasiparticle is stable.
However, the Dirac delta is an idealization of a narrow distribution with a small but finite width, so we must consider those final states that are {\it in the vicinity} of the exactly energy-conserving one.
In particular, in 1D quasicondensates, the trivial 3-wave mixing channel involving the zero-energy mode is not physical itself. Yet the divergent thermal population of nearby infrared modes yields a well-defined contribution to the FGR decay rate.  
As anticipated, this mechanism leads to a broadening of the excitation spectrum and, within an appropriate time interval, the occupation of the initially occupied mode decays exponentially at a constant rate.
The calculation of this rate is our main result.

The paper is organized as follows.
In Sec.~\ref{sec:Settings}, we recall the Lieb-Liniger model of a 1D Bose gas, including its hydrodynamical description in the weakly-interacting regime, and define the quasiparticles whose decay we are interested in.
In Sec.~\ref{sec:Theory}, we present a derivation of the intrinsic quasiparticle decay rate, as well as the outline of a more precise formulation that includes the main deviations from exponential decay.
In Sec.~\ref{sec:numerical_confirmation}, we show that the decay rate extracted from numerical simulations is in good agreement with the prediction of Sec.~\ref{sec:Theory}, both in the relaxation of an initial injection of phonons and in the slowing of exponential growth induced by a parametric oscillation (the latter example being inspired by experiments~\cite{Jaskula2012}).
We conclude and make links with other works in Sec.~\ref{sec:Conclusion}, while some further details are relegated to Appendices.

\section{Settings}
\label{sec:Settings}

Here we recall the necessary preliminaries for describing a 1D Bose gas.  The quasicondensate nature, and the appropriateness of the hydrodynamical description, are introduced.  The quasiparticles are defined in the hydrodynamical framework.  A brief description of the numerical procedure adopted for simulating the dynamics of the gas is also given, and some preliminary numerical observations of the quasiparticle decay are presented.

\subsection{Quasicondensate description
\label{subsec:quasicondensate}}

Consider a 1D gas of identical bosons of mass $m$ with only two-body contact interactions and no atom losses, described by the Hamiltonian
\begin{equation}
\hat{H} = \int_{0}^{L} dx \, \left\{ \frac{\hbar^{2}}{2m} \partial_{x}\hat{\Psi}^{\dagger} \, \partial_{x}\hat{\Psi} + \frac{g}{2} \hat{\Psi}^{\dagger 2} \hat{\Psi}^{2} \right\} \, ,
\label{eq:Hamiltonian_full}
\end{equation}
where $\hat{\Psi}$ is the atomic field and $g$ is the 1D interaction constant.
We impose periodic boundary conditions, so that the gas effectively lives on a torus of length $L$.  
The gas is ``trapped'' by the finiteness of the torus, but its dynamics and statistics are completely homogeneous. This allows an equivalent description using Fourier modes, each characterized by a wave number $k = 2\pi n/L$ for $n \in \mathbb{Z}$.
Each Fourier amplitude $\hat{\Psi}_{k}$ is the annihilation operator for atoms of momentum $\hbar k$.  The state of the system is fully characterized by the expectation values of the $\hat{\Psi}_{k}^{(\dagger)}$ and all their products.

A condensate occurs when one of the states of the system (typically the lowest-energy state $k=0$) contains a macroscopic fraction of the total atom number $N_{\rm at}$ or, equivalently, when the gas demonstrates long-range order: the one-particle density matrix $g_{1}\left(x-x^{\prime}\right) = \left\langle \hat{\Psi}^{\dagger}\left(x\right) \hat{\Psi}\left(x^{\prime}\right) \right\rangle$ has a finite limit when $\left|x-x^{\prime}\right| \to \infty$~\cite{Penrose-Onsager-1956}.  However, the Mermin-Wagner-Hohenberg theorem~\cite{Hohenberg1967, Pitaevskii2016} precludes the apparition of such long-range order in a 1D system, essentially because the excitations of the system induce large fluctuations in the relative phase of $\hat{\Psi}$ between widely separated points.  The one-particle density matrix then decays exponentially: $g_{1}\left(x-x^{\prime}\right) \approx {\rm exp}\left(-\left|x-x^{\prime}\right|/r_{0}\right)$, where the coherence length $r_{0}$ is given by~\cite{Pitaevskii2016}
\begin{equation}
\label{def:coherence_length}
\frac{r_0}{\xi}  =  \left( \frac{k_{B}T}{mc^{2}} \right)^{-1} \, 2 \rho_0 \xi \, .
\end{equation}

A {\it quasi}-condensed state can however be achieved in a 1D Bose gas at sufficiently low temperature~\cite{Petrov2004}.  As indicated above, the quasicondensate is characterized by large relative phase fluctuations over large distances, while the density fluctuations remain small.
It is therefore appropriate to adopt these as the field variables. They are related to the atomic field via the Madelung transformation: $\hat{\Psi} = e^{i \hat{\theta}} \sqrt{\hat{\rho}}$.
The representation in terms of the density and phase fields is known as the {\it hydrodynamical} description, since $\hat{\theta}$ acts like a potential for the flow velocity: $\hat{v} = \frac{\hbar}{m} \partial_{x}\hat{\theta}$. At the classical level where these are all $c$-numbers, this is an exact canonical transformation, with $\rho$ and $\theta$ being conjugate variables.  At the quantum level, this is approximately true as long as the discreteness of the atoms can be neglected, which requires a sufficiently weak interaction and coarse-graining over sites containing many atoms~\cite{Mora_2003}. We may then write the hydrodynamical version of Eq.~(\ref{eq:Hamiltonian_full}), up to some irrelevant term coming from normal ordering:
\begin{equation}
    \hat{H} = \int_{0}^{L} dx \left\{ \frac{\hbar^2}{2m} \frac{\partial \hat{\theta}}{\partial x} \hat{\rho} \frac{\partial \hat{\theta}}{\partial x} +  \frac{\hbar^2}{8m \hat{\rho}} \left( \frac{\partial \hat{\rho}}{\partial x} \right)^2  + \frac{g}{2} \hat{\rho}^2 \right\} \,,
\end{equation}
while imposing the canonical commutation relation
\begin{equation}
    \left[ \hat{\rho}\left(x\right) \,,\, \hat{\theta}\left(x^{\prime}\right) \right] = i \, \delta\left(x-x^{\prime}\right) \,.
    \label{eq:rho-theta-CCR}
\end{equation}

\subsection{Perturbative expansion of Hamiltonian}

We wish to study elementary excitations, which requires a well-defined splitting of the total field into a background plus perturbations. The background is defined as the homogeneous solution of
the classical equation associated to the Hamiltonian (\ref{eq:Hamiltonian_full}) ({\it i.e.}, the Gross-Pitaevskii equation~\cite{Pitaevskii2016}), working in the rest frame of the gas. 
The density $\rho_0$ is then constant (the total number of atoms is $N_{\rm at} = \rho_{0} L$) and 
the phase  $\theta_0 = - g \rho_0 t / \hbar$. The density fluctuations $ \delta \hat{\rho}$ around this background are assumed small while only the spatial variation of the phase fluctuations $\partial_x  \delta \hat{\theta} $ is assumed small. We then expand the Hamiltonian in $(\delta \hat{\rho} , \partial_x \delta \hat{\theta} )$ 
\begin{equation}
	\hat{H} = E_{0} \hat{\mathds{1}} + \hat{H}_{2} + \hat{H}_{3} + \sum_{i \geq 0} \hat{H}_{4+i} \, ,
\end{equation}
where the zeroth-order term $E_{0} = g\rho_{0}^{2}L/2$ is the energy of the homogeneous background, before any fluctuations are included.  Since the background is an exact solution of the classical equation of motion, the first-order term vanishes identically. The higher orders are given by
\begin{align}
	\begin{split}
		\label{def:expressions_pert_hamiltonian}
		\hat{H}_2 & = \int_{0}^{L} \left[ \frac{\hbar^2 \rho_0}{2m}\left(  \frac{\partial \hat{ \delta \theta} }{\partial x} \right)^2  + \frac{\hbar^2}{8 m \rho_0 } \left( \frac{\partial \delta \hat{\rho} }{\partial x} \right)^2  +  \frac{g}{2} \delta\hat{\rho}^{2} \right]  \mathrm{d}x  \, , \\
		\hat{H}_{3} & = \int_{0}^{L} \left[ \frac{\hbar^2}{2m} \frac{\partial \delta \hat{\theta} }{\partial x} \delta\hat{\rho} \frac{\partial \delta \hat{\theta} }{\partial x}   - \frac{\hbar^2}{8 m \rho_0} \left( \frac{\partial \delta \hat{\rho} }{\partial x} \right)^2  \frac{\delta \hat{\rho}}{\rho_0} \right]  \mathrm{d}x \, , \\
		\hat{H}_{4 + i} & = (-1)^i \int_{0}^{L} \frac{\hbar^2}{8 m \rho_0} \left( \frac{\partial \delta \hat{\rho}}{\partial x}\right)^2  \left( \frac{\delta \hat{\rho}}{\rho_0} \right) ^{(2+i)}  \mathrm{d}x \,, 
	\end{split}
\end{align}
for $i \geq 0$.

A couple of remarks are in order here. Notice first that the quasicondensate perturbative scheme clearly differs from the standard Bogoliubov treatment since the standard non-linear term $\hat{\Psi}^{\dagger \,  2} \hat{\Psi}^2$ is fully included in the quadratic Hamiltonian, while the infinite series of perturbative corrections comes entirely from the kinetic term.  
Second, Eq.~(\ref{def:expressions_pert_hamiltonian}) indicates that each order is suppressed by an additional factor of $\delta\hat{\rho}/\rho_{0}$ which suggests that $\left\langle \delta\rho^{2} \right\rangle/\rho_{0}^{2}$ can be used as a measure of the importance of taking these higher orders into account. For the typical values of parameters used in this work we have $\left\langle \delta\hat{\rho}^{2}\right\rangle / \rho_{0}^{2} \sim 10^{-3}$ so that we will consider only the second-order term $\hat{H}_{2}$ and the first perturbation Hamiltonian $\hat{H}_{3}$.

\subsection{Quasiparticle definition}

Working in the canonical ensemble, 
the atom number $N_{\rm at}$ is a fixed parameter, and therefore so is the background density $\rho_{0}$. The zero mode of the density fluctuations 
thus vanishes identically: $\delta \hat{\rho}_{k=0} = 0$. Consequently, the conjugate variable 
$\delta \hat{\theta}_{k=0}$ is non-dynamical and can be ignored. The fluctuations $\delta \hat{\rho}$ and $\delta \hat{\theta}$ on top of this background are then composed of the non-zero Fourier modes $\delta \hat{\rho}_{k}$ and $\delta \hat{\theta}_{k}$:
\begin{equation}
		\delta\hat{\rho}(x) = \sqrt{\frac{\rho_{0}}{L}} \sum_{k \neq 0} e^{i k x} \delta\hat{\rho}_{k} \,, \quad \delta\hat{\theta}(x) = \frac{1}{\sqrt{\rho_{0} L}} \sum_{k \neq 0} e^{i k x} \delta\hat{\theta}_{k} \,.
\end{equation}
With this writing  $\delta\hat{\rho}_{k}$ and $\delta\hat{\theta}_{k}$ are dimensionless, and they satisfy $\left[\delta\hat{\rho}_{k} \,,\, \delta\hat{\theta}_{k^{\prime}} \right] = i \, \delta_{k,-k^{\prime}}$.  Since $\delta\hat{\rho}(x)$ and $\delta\hat{\theta}(x)$ are Hermitian operators, the Fourier components satisfy $\delta\hat{\rho}_{-k} = \delta\hat{\rho}_{k}^{\dagger}$ and $\delta\hat{\theta}_{-k} = \delta\hat{\theta}_{k}^{\dagger}$. $\hat{H}_{2}$ can be diagonalized into normal modes, called {\it phonons}~\footnote{In this paper, we use the term ``phonon'' to refer to a quasiparticle of any wavelength, and \textit{not} just to those well within the linear part of the dispersion relation~\eqref{eq:dispersion}.}, represented by operators $\hat{\varphi}_{k}^{(\dagger)}$ such that
\begin{equation}
	\hat{H}_{2} = \sum_{k \neq 0} \hbar \omega_{k} \left( \hat{\varphi}^{\dagger}_{k}\hat{\varphi}_{k} + \frac{1}{2} \right) \,, 
	\label{eq:Hamiltonian_quadratic}
\end{equation}
where the phonon frequency
\begin{equation}
	\omega_{k} = c\left|k\right| \sqrt{1 + \frac{1}{4} k^{2} \xi^{2}} \,,
	\label{eq:dispersion}
\end{equation}
with $c = \sqrt{g \rho_{0} / m}$ the speed of sound and $\xi = \hbar/mc$ the healing length. In the limit $k \xi \xrightarrow{} 0$, we have an exactly linear dispersion relation like that of the Luttinger liquid. The phononic operators are related to the density and phase fluctuations by
\begin{equation}
	\hat{\varphi}_{k} = \frac{1}{\sqrt{2}} \left( C_{k}^{-1} \, \delta\hat{\rho}_{k} + i \, C_{k} \, \delta\hat{\theta}_{k} \right) \,,
\end{equation}
where $C_{k}^{2} = \hbar k^{2}/\left(2m \omega_{k}\right)$. The use of inverse coefficients $C_{k}$ and $C_{k}^{-1}$ ensures that the transformation is canonical and hence that the phonon operators satisfy the bosonic commutation relation $\left[ \hat{\varphi}_{k} \,,\, \hat{\varphi}_{k^{\prime}}^{\dagger} \right] = \delta_{k,k^{\prime}}$.  

$\xi$ and the associated healing time, $t_{\xi} = \xi/c$, provide natural units in which to express quantities adimensionally.
There are three dimensionless parameters describing the system: the 1D density $\rho_{0} \xi$, the length $L/\xi$, and the temperature $k_{B}T/mc^{2}$ (where $mc^{2} = \hbar/t_{\xi}$ is the chemical potential $\partial E_{0}/\partial N_{\rm at}$).
The interaction strength is characterised by $\gamma_{\rm LL} = 1/\left(\rho_{0} \xi\right)^{2}$, the dimensionless Lieb-Liniger constant~\cite{Lieb1963,Petrov2004}. Numerical simulations presented in this work typically have $\gamma_{\rm LL} \sim 10^{-5} - 10^{-3}$, placing us firmly in the weakly-interacting regime. We also choose a grid spacing $\Delta x$ such that $\rho_0 \Delta x \sim 20$ atoms per site, justifying our use of the hydrodynamical description. 

Just as for the atom operators $\hat{\Psi}_{k}$, the phonon operators $\hat{\varphi}_{k}$ provide a complete description of the system, whose state is fully characterized by the expectation values of the $\hat{\varphi}_{k}^{(\dagger)}$ and all their products.  Since the $\hat{\varphi}_{k}$ come close to diagonalizing the full Hamiltonian $\hat{H}$ , the phonons are close to the exact normal modes of the system and their mutual interactions are relatively weak.  The phonons therefore provide the most natural basis in which to examine the state of the system and interpret its dynamical behavior.  However, the simplicity of the full Hamiltonian~(\ref{eq:Hamiltonian_full}) makes the atom basis more convenient for numerical treatments of the evolution.

\subsection{Quasiparticle interactions}

Turning now to $\hat{H}_{3}$, and neglecting terms of the form $\varphi^{\dagger}\hat{\varphi}^{\dagger}\hat{\varphi}^{\dagger}$ ($\hat{\varphi}\hat{\varphi}\hat{\varphi}$) which cause the unbalanced appearance (disappearance) of three phonons typically associated with strong violation of energy conservation, the relevant part of the interaction Hamiltonian takes the form
\begin{equation}
    \hat{V}_{3} = \frac{1}{\sqrt{N_{\rm at}}} \sum_{\substack{p,q \neq 0 \\ p+q \neq 0}} \hbar V_3\left(p,q\right) \left\{ \hat{\varphi}^{\dagger}_{p} \hat{\varphi}^{\dagger}_{q} \hat{\varphi}_{p+q} + \hat{\varphi}_{p+q}^{\dagger} \hat{\varphi}_{p}\hat{\varphi}_{q} \right\} \, ,
\label{eq:Hamiltonian_fluctuations}
\end{equation}
where
\begin{align}
\begin{split}
\label{def:V3_on_shell}
V_3 (p,q) = \sqrt{ \frac{\hbar}{32 m  } }  & \sqrt{\frac{ \left| p q (p+q) \right| }{v^{\rm ph}_p v^{\rm ph}_q v^{\rm ph}_{p+q}}} 
\Biggl\{ -\left(\frac{\hbar}{2m}\right)^{2} \left[ p^{2} + p q  + q^{2} \right] \\
& + v^{\rm ph}_p v^{\rm ph}_q + v^{\rm ph}_q v^{\rm ph}_{p+q} + v^{\rm ph}_p v^{\rm ph}_{p+q}  \Biggr\} \, ,
\end{split}
\end{align}
and $v^{\rm ph}_k = \omega_{k}/k$ is the phase velocity at $k$.

$\hat{V}_{3}$ describes the decomposition of a single phonon into two phonons, as well as the inverse process where two phonons combine into one. Note that it is momentum-conserving, reflecting the homogeneity of the system, and that if $p$ is held fixed and $q \to 0$, $V_{3}\left(p,q\right)$ vanishes as $\sqrt{q}$.
A similar writing of  $\hat{V}_{3}$ can be found in~\cite{Ristivojevic2016}, in a form that is equivalent to the one given here.

\subsection{Numerical simulations}
\label{subsec:Numerical_simulations}

The system is modeled numerically using the truncated Wigner approximation (TWA)~\cite{steelDynamicalQuantumNoise1998} (see Appendix~\ref{app:TWA} for more details). The operators $\hat{\Psi}$ are replaced by classical variables $\Psi$, and products of these variables are identified with the corresponding fully symmetrized quantum operators.  A series of {\it ab initio} Monte Carlo simulations are performed, with quantum indeterminacy appearing through the statistical ensemble describing the initial state.  The field is then evolved according to the dynamics of Hamiltonian~(\ref{eq:Hamiltonian_full}). This is repeated for a large number of independent initial realisations, so as to get good statistics when computing averages.

\begin{figure}
    \centering
         \begin{minipage}{0.49\textwidth}
        \centering
        \includegraphics[width=\textwidth]{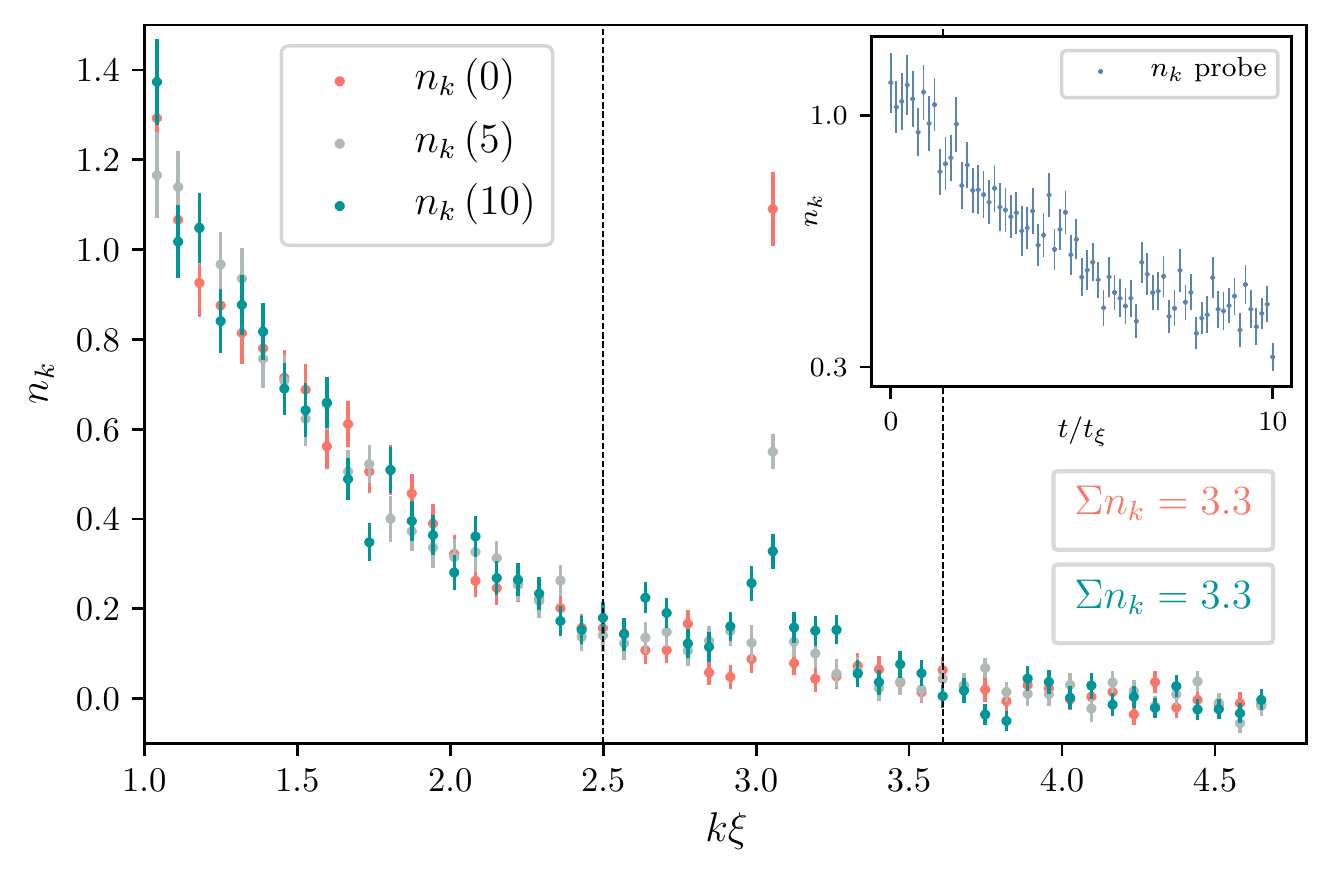}
    \end{minipage} \hfill
          \begin{minipage}{0.49\textwidth}
        \centering
         \includegraphics[width=\textwidth]{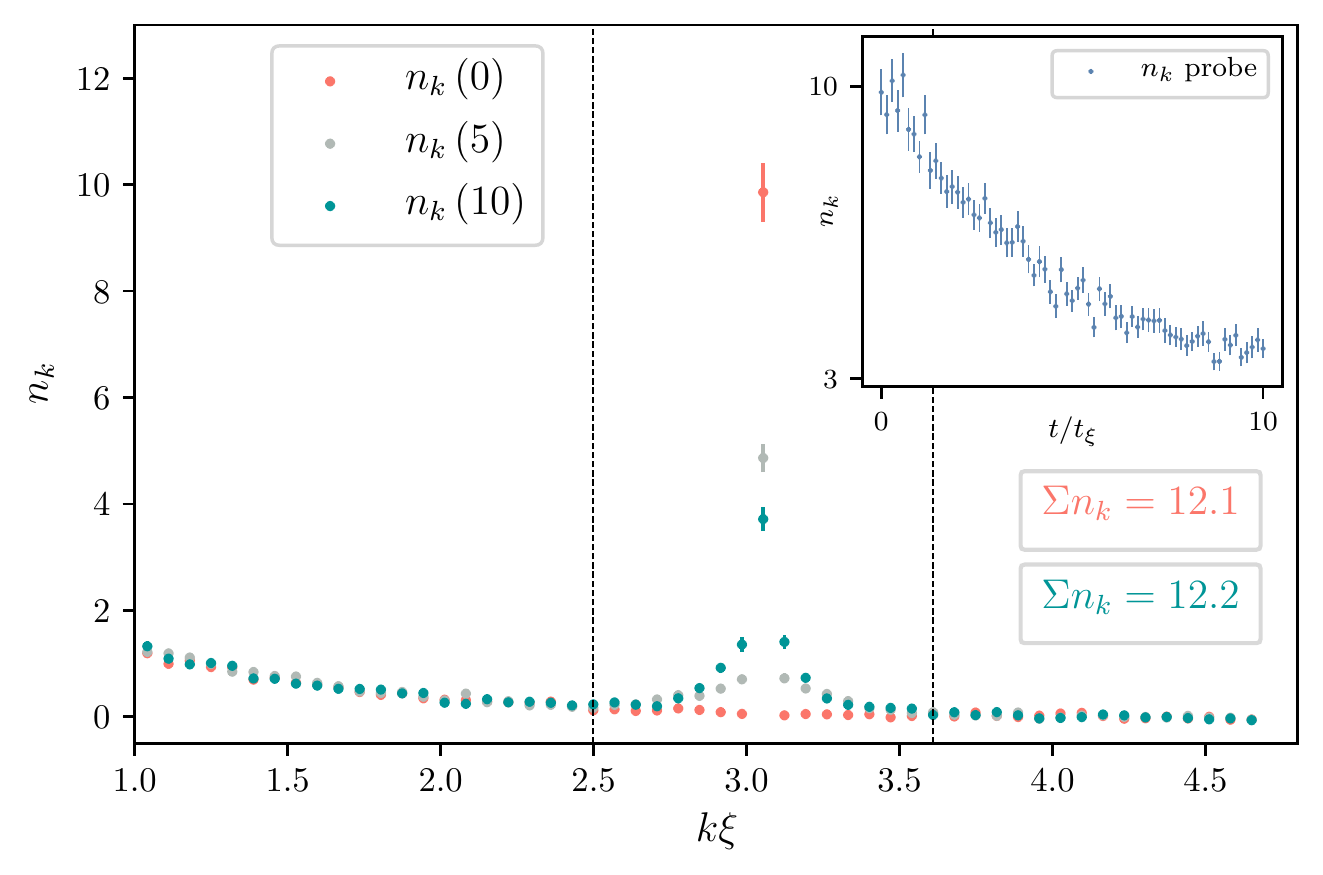}
  
    \end{minipage}
    
    \caption{Snapshots of the phonon number spectrum at  times $t/t_{\xi}=0,5,10$ assuming an initial
    thermal state at temperature $k_{B} T/m c^2 = 2$ on top of which an additional $\delta n = 1$ (top) or $\delta n = 10$ (bottom) phonons are added in the mode $k \xi = 3.1$. The size of the system is $L/\xi = 90.5$ and its atomic density $\rho_0 \xi =49.9$. The averages are calculated from an ensemble of $400$ independent realisations, while the error bars represent the standard deviation. There are $256$ points on the grid. The inset shows a more time-resolved evolution of the population in the probe mode. The total number of phonons within the vertical dotted lines is given for the initial and final times, and is seen to be conserved.
    \label{fig:phonon_spectrum_different_probes}}

\end{figure}

The phenomenon of interest is illustrated in Fig.~\ref{fig:phonon_spectrum_different_probes}, which shows the typical evolution observed in numerical simulations.  Starting from a thermal state, the occupation number of a mode is increased by  $\delta n$, to be considered as the initial number of phonons in the probe.  Indeed, throughout this paper we shall adopt the following decomposition:
\begin{equation}
    n_{k} = n_{k}^{\rm th} + \delta n_{k} \,,
    \label{eq:nk_decomposition}
\end{equation}
where $n_{k} = \left\langle \hat{\varphi}_{k}^{\dagger}\hat{\varphi}_{k} \right\rangle$ is the full phonon spectrum and $n_{k}^{\rm th}$ is the (thermal) spectrum in the absence of any probe.
Further details of the simulations are given in Sec.~\ref{sec:numerical_confirmation}. The figure shows $n_{k}$ at a series of different times.  
We see that $n_{k}^{\rm probe}$, the population of the initially occupied mode, decays. The observed behaviour is essentially linear, in the sense that the relative change $\delta n_{k}(t) / \delta n$ is independent of $\delta n$, the number of phonons injected in the probe. Moreover,
it is clear that the spectrum of the probe has broadened.  Summing over nearby modes (those within the vertical dotted lines), the total $n$ is found to be constant in time. The phonons seem not to have been lost, but rather to have been kicked into neighboring modes.  
Note that the broadening is essentially symmetric: the phonons are just as likely to be kicked towards a higher momentum as a lower one.
The small shifts in momentum suggest that the evolution is primarily driven by interactions with infrared phonons from the thermal bath.

\section{Theoretical derivation}
\label{sec:Theory}

In this section, we provide an analytical description of the dissipative process at play.  In Sec.~\ref{subsec:FGR}, we derive the decay rate using the FGR; as this applies to a discrete eigenstate in interaction with a continuum, we expect it to describe the decay of the population of a singularly occupied mode ({\it i.e.}, when the spectrum is sufficiently narrow).  In Sec.~\ref{subsec:Response_functions}, we sketch a formalism which takes us beyond such an approximation, using response functions to model the dynamics of the spectrum more accurately and to describe the deviations from purely exponential behavior.  A fuller treatment of this formalism is given in Appendix~\ref{app:phonon_decay}, while a comprehensive analysis of the main deviations observed numerically is given in Appendix~\ref{app:deviations}.

\subsection{Using the Fermi Golden Rule}
\label{subsec:FGR}

Given an initial eigenstate $\left|i\right\rangle$ that couples to a continuum of final eigenstates $\left|\nu_f\right\rangle$ via a time-independent weak perturbation Hamiltonian $\hat{H}^{\prime}$, the FGR gives the transition rate into the continuum
\cite{CohenTannoudji2020}:
\begin{equation}
\label{eq:Fermi_Golden_Rule}
   dW_{i \to f} = \frac{2\pi}{\hbar} \left| \left\langle \nu_f \right| \hat{H}^{\prime} \left| i \right\rangle \right|^{2} \delta\left(E_{f}-E_{i}\right) \, d\nu_{f} \,,
\end{equation}
where $E_{i}$ and $E_{f}$ are the energies of the initial and final states, the latter being labeled by the dimensionless number $\nu_{f}$.  The Dirac delta imposes energy conservation, {\it i.e.} the rate is evaluated as a matrix element between states with the same (unperturbed) energy. The total rate is found by integrating over $\nu_{f}$.

Consider a singularly occupied phonon mode $k$, which decays due to interactions with the thermal population of phonons.  Each available momentum is a multiple of $2\pi/L$. For sufficiently large $k L$ the thermal distribution of phonons, as well as the states available to decay to, can be approximated as a continuum from the point of view of the mode $k$. We thus expect the FGR~(\ref{eq:Fermi_Golden_Rule}) to be applicable.  We need only determine the relevant perturbation and corresponding initial and final states.

In the phonon basis, the role of the perturbation entering the FGR is played by $\hat{V}_{3}$ of Eq.~(\ref{eq:Hamiltonian_fluctuations}).
It describes two distinct processes involving the annihilation of a probe phonon at wave number $k$ :
\begin{itemize}
\item Phonons of wave-numbers $k$ and $q$ combine to produce a single phonon with wave number $k+q$. The relevant term is $\hat{\varphi}_{k+q}^{\dagger}\hat{\varphi}_{q}\hat{\varphi}_{k}$, taking $\left|i\right\rangle = \left|n_{q}, n_{k}, n_{k+q}\right\rangle$ to $\left|f\right\rangle = \left|n_{q}-1,n_{k}-1,n_{k+q}+1\right\rangle$, and the corresponding squared matrix element is $\frac{1}{N_{\rm at}} \hbar^{2} \left|2\, V_3\left(k,q\right)\right|^{2} \times n_{k} n_{q} \left(n_{k+q}+1\right)$. The factor $2$ comes from the $p \xleftrightarrow[]{} q$ symmetry of the sum in Eq.~(\ref{eq:Hamiltonian_fluctuations}). The factors of $n$ come from the action of the phonon operators on the state $\left|i\right\rangle$, with the `$n$' and `$+1$' terms encoding stimulated and spontaneous processes, respectively.  The energy difference is $E_{f}-E_{i} = \hbar \, \delta\omega_{L} = \hbar \left( \omega_{k+q}-\omega_{q}-\omega_{k} \right)$;
 at small $q$, this gives $\delta \omega_{L} \approx v^{\rm gr}_k q - c \left|q\right| = q \left(v^{\rm gr}_k \mp c\right)$, where $v^{\rm gr}_k = d\omega_{k}/dk$ is the group velocity at $k$.

\item $k$ decays to two phonons, with wave numbers $k-q$ and $q$.  The relevant term is $\hat{\varphi}_{k-q}^{\dagger}\hat{\varphi}_{q}^{\dagger}\hat{\varphi}_{k}$, taking $\left|i\right\rangle = \left|n_{q}, n_{k-q}, n_{k}\right\rangle$ to $\left|f\right\rangle = \left|n_{q}+1,n_{k-q}+1,n_{k}-1\right\rangle$, and the corresponding squared matrix element is $\frac{1}{N_{\rm at}} \hbar^{2} \left|2 \, V_3\left(k-q,q\right)\right|^{2} \times n_{k} \left(n_{q}+1\right) \left(n_{k-q}+1\right)$.  The energy difference is $E_{f}-E_{i} = \hbar \, \delta\omega_{B} = \hbar \left( \omega_{k-q}+\omega_{q}-\omega_{k} \right)$; at small $q$, this gives $\delta \omega_{B} \approx -q \left(v^{\rm gr}_k \mp c\right) = -\delta \omega_{L}$.
\end{itemize}
The two scattering processes described above lead to the well-known Landau and Beliaev damping of phonons in BECs: the former is associated with the absorption of a thermal phonon, the latter with a splitting into two phonons.~\footnote{While Beliaev damping often occurs spontaneously (as it is a $1 \to 2$-phonon process), we are concerned here with a regime of sufficiently high temperature in which it is stimulated by the presence of a large number of thermal infrared phonons.}
The associated frequency differences are thus labeled $\delta\omega_{L}$ and $\delta\omega_{B}$.

Imposing also energy conservation $\delta\omega_{L , B} \left( q;k \right) = 0$, as demanded by the Dirac delta in Eq.~(\ref{eq:Fermi_Golden_Rule}), we encounter the problem anticipated above: the only exact solution respecting both momentum {\it and} energy conservation is the trivial one, $q=0$.~\footnote{Notice that for a Luttinger liquid where $\xi =0$ the problem is exactly the opposite as there is an infinite number of elastic scattering channels leading to a divergence of the FGR decay rate \cite{andreev_hydrodynamics_1980,samokhin_lifetime_1998}.}
In the vicinity of this channel, the interaction vanishes as $\left|V_{3}(k,q)\right|^{2} \propto q$. However, the frequency $\omega_{q}$ also vanishes as $\omega_{q} \to c \left|q\right|$, and therefore the thermal population $n_{q} \approx k_{B}T/\hbar \omega_{q}$ simultaneously diverges. It therefore makes sense to refer to the product $\left| V_{3}(k,q) \right|^2 n_{q} $ as an effective interaction strength for the corresponding channel, as it has a finite limit as $q \to 0$, and it is this finite limit that is picked up by the Dirac delta in Eq.~(\ref{eq:Fermi_Golden_Rule}). Although the delta is centered at $q=0$, it does not actually matter that the trivial elastic channel is unphysical and thus removed from the dynamics: the Dirac delta is a placeholder for a steadily narrowing distribution of final states in the vicinity of this channel (see Appendix~\ref{app:phonon_decay}). As long as there are sufficiently many modes within this distribution, the single removed mode at $q=0$ has a relative measure of zero.  This allows the application of the FGR, as if the limiting processes at $q \to 0^{\pm}$ were physically allowed.~\footnote{One important modification with respect to the standard case arises because the effective interaction strength $\left|V_{3}(k,q)\right|^{2} n_{q}$ is discontinuous across the resonant channel $q=0$. This obliges us to consider separately the coupling with positive and negative $q$, leading to two applications of the FGR where each channel picks up ``half'' of the Dirac delta around $q=0$.
This factor is cancelled by the co-occurrence of the Landau and Beliaev channels, which contribute equally to the decay rate.}

The Dirac delta in Eq.~(\ref{eq:Fermi_Golden_Rule}) also serves to multiply by the density of available states with respect to the energy.
Since the available states are evenly spaced in momentum, in a window of size $\Delta q$ the number of states is $\Delta N = L/2\pi \times \Delta q \approx L/2\pi\hbar \times \left|dq/d\left(\delta \omega_{L/B}\right)\right| \times \Delta E$.  Recalling from above that, for small $q$, we have $\delta \omega_{L} \approx -\delta \omega_{B} \approx q \left(v^{\rm gr}_k \mp c\right)$, this gives the following density of states:
\begin{equation}
    \rho_{E}\left(k\right) \equiv \frac{dN}{dE} = \frac{1}{2\pi\hbar} \, \frac{L}{v^{\rm gr}_k \mp c} \,.
\end{equation}
For ease of notation we restrict to $k > 0$, so that $v^{\rm gr}_k \mp c$ is always positive.  Due to isotropy, $-k$ will behave in exactly the same way as $k$.  We recall that the sign in the denominator is related to the sign of $q$, the momentum of the infrared phonons with which the interaction takes place.  The abrupt change in the velocity as $q$ crosses zero means that there are two distinct limits to be taken: $q \to 0$ from above and from below.  Putting everything together, we thus derive two decay rates, corresponding to the coupling with positive and negative $q$:
\begin{equation}
    \Gamma_{\pm}\left(k\right) = \frac{{\rm lim}_{q \to 0^{\pm}} \left[ \frac{L}{N_{\rm at}} \left|2 \, V_3\left(k,q\right)\right|^{2} n_{q} \right]}{\left(v^{\rm gr}_k \mp c\right)} \,.
\end{equation}
Inserting the explicit form of $V_3\left(k,q\right)$ yields:
\begin{equation}
\label{def:decay_rate}
    \Gamma_{\pm} t_{\xi} = \frac{k_{B}T}{m c^2} \frac{1}{\rho_{0}\xi} f_{\pm}\left(k\xi\right)
\end{equation}
where
\begin{equation}
    f_{\pm}\left(k\xi\right) = \frac{1}{2} \, \frac{\left(k\xi\right)^{2}}{\left(v^{\rm ph}_k/c\right)^{2}} \, \frac{\left(v^{\rm ph}_k/c \pm 1/2\right)^{2}}{v^{\rm gr}_k/c \mp 1} \, .
\end{equation}
The total decay rate presented in the figures is $\Gamma_{k} = \Gamma_{+} + \Gamma_{-} $.

Equation~(\ref{def:decay_rate}) constitutes our main result: an explicit expression for the rate at which the population of a singularly populated mode will decay, due to interactions with thermal phonons.

Since at small $q$ we have $\delta \omega_{L} \approx -\delta \omega_{B} \approx v^{\rm gr}_k q - c \left|q\right|$, at fixed $\left|q\right|$ the magnitude of the frequency difference is smaller when $q$ has the same sign as $k$. We therefore expect that such modes provide the dominant contribution to the decay, while the coupling to modes of the opposite sign induces a subdominant correction. So the Landau channel tends to kick a phonon at $k$ into a higher-energy mode, while the Beliaev channel acts in the opposite direction. Since the two channels contribute equally, there is no preference for a phonon at $k$ to be kicked towards either a higher or lower momentum.  This explains the approximate symmetry of the broadening  observed in Fig.~\ref{fig:phonon_spectrum_different_probes}.


\subsection{Using response functions}
\label{subsec:Response_functions}

Here, we give the outline of a more precise description of how a perturbation to the phonon number spectrum $\delta n_k$ evolves.  For brevity, we shall avoid details here, though they can be found in Appendix~\ref{app:phonon_decay}.  However, the writing will allow us to point out significant deviations from the exponential decay rate of Eq.~(\ref{def:decay_rate}), which are studied in more detail in Appendix~\ref{app:deviations}.

The time-derivative of $\delta n_{k}$ depends both on $\delta n_{k}$ and on phonon correlations, the most important of which are the 3-point correlations induced by the Beliaev and Landau processes described above: $C^{(3)}_{p,q} = \left\langle \hat{\varphi}_{p}^{\dagger} \hat{\varphi}_{q}^{\dagger} \hat{\varphi}_{p+q} \right\rangle$. Neglecting other connected correlation functions, the equations of motion can be written entirely in terms of $n_{k}$, though they become integro-differential equations that include response functions, which encode how the system ``remembers'' and responds to its past behaviour. Neglecting any self-interaction of the perturbation, we linearize the equations in $\delta n_{k+q}$:
\begin{align}
\begin{split}
\partial_{t}\left(\delta n_{k}\right) & = - \int_{0}^{t} dt^{\prime} \, D_{k}\left(t-t^{\prime}\right) \, \delta n_{k}\left(t^{\prime}\right) \\
& + \int_{0}^{t} dt^{\prime} \, \sum_{q \neq - k } M_{k,k+q}\left(t-t^{\prime}\right) \, \delta n_{k+q}\left(t^{\prime}\right) \,.
\label{eq:nk_eom_response_functions}
\end{split}
\end{align}

The first term on the right-hand side  of~(\ref{eq:nk_eom_response_functions}), governed by $D_k$, describes the essential behavior of $\delta n_{k}$, when back-reaction from its near-neighbors can be neglected and finite-size effects can also be ignored (see below).
$D_{k}\left(t-t^{\prime}\right)$ decays to zero within a time scale $t_{\rm crit}$, and its integral approaches $\Gamma_{k}$ of Eq.~(\ref{def:decay_rate}).  As long as $t \gg t_{\rm crit}$ and $\delta n_{k}$ does not vary much on time scales of order $t_{\rm crit}$, its value at $t$ can be taken outside the integral, and the first term becomes $-\Gamma_{k} \, \delta n_{k}$.  This is the regime in which the analysis of the previous subsection applies.  We can thus expect deviations when $\delta n_{k}$ varies significantly over times of order $t_{\rm crit}$.  This can happen either when $\partial_{t}\left(\delta n_{k}\right)$ is particularly large, or for a time $t_{\rm crit}$ after a sudden injection of phonons.  The latter case will be relevant for some of our numerical simulations.

The second term of Eq.~(\ref{eq:nk_eom_response_functions}), governed by $M_{k,k+q}$, describes the influence of other modes on the evolution of $n_{k}$.  These typically act to slow down the decay of $n_{k}$, because some of the phonons in neighboring modes will be kicked into the mode of interest.  This is particularly relevant in situations where $k$ is at the center of a peak with a finite width, and in such a scenario we expect the net decay rate to be smaller than that predicted by Eq.~(\ref{def:decay_rate}).
Note that this is not in contradiction with the FGR, which applies when a single discrete mode loses energy to a continuum of modes; this picture becomes less applicable when the mode losing energy is itself part of a continuum.
Generally, the effect of this back-reaction term is difficult to take fully into account, but in certain situations (particularly in the case of parametric resonance) we can make approximations and derive the expected qualitative behavior for the decay rate as a function of the peak width.

\subsection{Suppression due to finite size of system
\label{subsec:Finite-size}}

 The second term of Eq.~(\ref{eq:nk_eom_response_functions}) includes a contribution from $q=0$, which will be present even when neighboring modes are not significantly populated.  
This $q=0$ term slows down the decay, but it is proportional to $1/L$ and vanishes entirely in the limit $L \to \infty$ where we have a continuum in $k$-space.
The primary function of this term is to account for the finite resolution in $k$ by keeping track of those phonons which, in the continuum limit, would be lost to nearby modes within $1/L$ of the main decaying mode.  With the finite resolution induced by the finiteness of $L$, these phonons remain in the same bin as the decaying mode. Therefore, their contribution to the variation of $n_k$ is removed, and the net decay is effectively slowed down.

In Appendix \ref{app:deviations} we derive the slowing down of the decay rate induced by this contribution. On sufficiently short time scales, it enters as a quadratic correction to the exponential decay:
\begin{equation}
	\delta n_{k} \approx \delta n_{k}(0) \, e^{-\Gamma_{k}t + \frac{1}{2} \gamma_{k} t^{2}} \, ,
	\label{eq:nk_finite_size_correction}
\end{equation}
where 
\begin{equation}
	\label{def:prediction_small_gamma}
	\gamma_{k} t_{\xi}^2  = \frac{k_{B}T}{mc^{2}} \, \frac{1}{\rho_{0}L} \, g \left(k \xi\right)\,
\end{equation}
and where we have defined
\begin{equation}
	g\left(k\xi\right) = \frac{\left(k\xi\right)^{2}}{v_{\rm ph}^{2}(k)/c^{2}}  \left\{ \left(\frac{v_{\rm ph}(k) }{c} - \frac{1}{2}\right)^{2} + \left(\frac{v_{\rm ph}(k) }{c} + \frac{1}{2}\right)^{2} \right\}  \,.
\end{equation}
The finite-size effect grows in importance as time progresses, 
on a time scale $t_{\rm fs}$ such that $\gamma_{k} t_{\rm fs}^{2} = \Gamma_{k} t_{\rm fs}$ (where the decay in Eq.~(\ref{eq:nk_finite_size_correction}) slows to zero).
Its relevance therefore depends on how $t_{\rm fs}$ compares to the typical time-scale of decay $\Gamma_k^{-1}$, {\it i.e.}, the relevant quantity is
\begin{eqnarray}
\label{eq:relevant_finite_size}
\Gamma_{k} t_{\rm fs} &=& \frac{\Gamma_{k}^{2}}{\gamma_{k}} = \frac{2 L}{r_{0}}  \frac{f^2\left(k\xi\right)}{g\left(k\xi\right)} \nonumber \\
&\approx& \frac{L}{r_{0}} \quad {\rm when} \quad k\xi \gg 1 \,.
\end{eqnarray}
In the second line we have taken the large wavenumber limit,
where our exponential prediction for the decay is most accurate (see Sec.~\ref{sec:numerical_confirmation} below).  $r_0$ 
is the (one-body) coherence length of the quasicondensate defined in Eq.~(\ref{def:coherence_length}).  So, when $L/r_{0} \ll 1$, the decay proceeds hardly at all before it is stopped by the finite-size effect.  On the other hand, when $L/r_{0} \gg 1$, the decay is significant before the slowing-down kicks in, and the finite-size effect becomes irrelevant.

As mentioned in Sec.~\ref{subsec:quasicondensate}, when considering a quasicondensate over distances larger than $r_0$ the long-range order characterising condensation is lost due to large thermal fluctuations in the phase. 
Conversely, if we restrict attention to distances much shorter than $r_{0}$, the one-body correlation is preserved and the gas looks like a ``true'' condensate. Equation~\eqref{eq:relevant_finite_size} then shows that if we consider a small enough system 
that appears as a true condensate, the decay will be strongly suppressed by finite-size effects. 
In effect, the decay
processes we have identified manifest in position space
as an $x$-dependent drift in the phase of the excited phonon, whose variance becomes significant only over distances larger than $r_{0}$.
Therefore, the decay is only effective when we reach the quasicondensate regime $L \gg r_0$. 

This correspondence is further explored in Appendix~\ref{app:deviations}, where it is also shown that binning modes in momentum space yields an evolution that is very similar to that on a shorter torus.  We thus conclude that the system size $L$ appearing in Eq.~(\ref{eq:relevant_finite_size}) can be generalized to the size of any subsection of a larger system.  If this size is small compared to $r_{0}$, local measurements will be insensitive to the decay.

\section{Numerical confirmation}
\label{sec:numerical_confirmation}

We have run several numerical simulations in order to test our prediction for the phonon damping rate.  The essentials of the numerical method have already been described in Sec.~\ref{subsec:Numerical_simulations}, and more details can be found in Appendix~\ref{app:TWA}.

The simulations fall into two types.  First, we perform a series of simulations like that of Fig.~\ref{fig:phonon_spectrum_different_probes}, where a number of phonons is injected into a single mode $k$, and the evolution of the phonon number spectrum is followed in time.  This setup allows us to study the response of a thermal system to a single perturbation, the addition of phonons in the mode $k$. We use it to demonstrate that the thermal processes discussed in the previous section are indeed the main culprit involved in the scattering of injected phonons, and to check that the scaling properties of the decay rate match those predicted by Eq.~(\ref{def:decay_rate}). Moreover, this controlled scenario enables us to track precisely how phonons scatter to other modes (as shown in Fig.~\ref{fig:phonon_spectrum_different_probes}), informing our study of the decay process and its generalization to a peak of finite width (see Sec.~\ref{subsec:slowing_of_exponential_growth} below, and Appendix~\ref{app:deviations}).

The second series of simulations is inspired by the parametric resonance experiments of~\cite{Jaskula2012} and theoretically studied in~\cite{Busch-2014,Robertson_2018}: starting from a thermal state, a sinusoidal modulation of the 1D atomic interaction strength is applied, inducing exponential growth of the phonon occupation number within a resonance window. (This can be achieved experimentally by modulating the transverse stiffness of the trap~\cite{Robertson_2018,Pylak-Zin-2018,Jaskula2012}.)
The phonon damping is then observed as a reduction in the rate of exponential growth.
As suggested above, the finite width of the resonant peak induces a deviation from the decay rate~(\ref{def:decay_rate}), which strictly speaking is only applicable in the limit of a singularly occupied mode.  However, this deviation occurs in a controlled fashion, and is found to be consistent with the back-reaction term appearing in Eq.~(\ref{eq:nk_eom_response_functions}).

\subsection{Initial injection of phonons}
\label{subsec:initial_injection}

In the first run of simulations, the initial state is taken to be a thermal state for the quadratic Hamiltonian $\hat{H}_{2}$, with the possible addition of a probe in a single phonon mode of wave vector $k$.  When the probe is present, it is given an initial occupation number $\delta n$ on top of the thermal distribution by simply multiplying the amplitude of the relevant mode by a constant factor. This ensures that the phonon modes are initially independent of each other, and separately exhibit Gaussian statistics.  
Each realisation is then evolved twice, with and without the probe, to take account of the slight degree of time-dependence that occurs even when the probe is absent. The spectra are calculated independently at different times in the interval $t/t_{\xi} \in \left[0 \,,\, 10\right]$, and the probe spectrum $\delta n_{k}(t)$ is defined as the difference between the two, in accordance with Eq.~\eqref{eq:nk_decomposition}.
$\delta n_{k}(t)$ is then fitted to the template 
$A\, {\rm exp}\left(- \Gamma t + \gamma t^{2}/2\right)$
of Eq.~(\ref{eq:nk_finite_size_correction}), 
the fitted value of $\Gamma$ being the extracted decay rate.  
As mentioned in Sec.~\ref{subsec:Finite-size} above,
the $t^{2}$ correction is related to the limited resolution in $k$-space induced by the finite length of the condensate, coming from the $q=0$ term in the second line of Eq.~(\ref{eq:nk_eom_response_functions}). It is included here as it makes a small but noticeable difference to the fit.
More details on the extraction of the $\gamma$ term are given in Appendix~\ref{app:deviations}.

\begin{figure}
    \centering
    \includegraphics[width=0.49\textwidth]{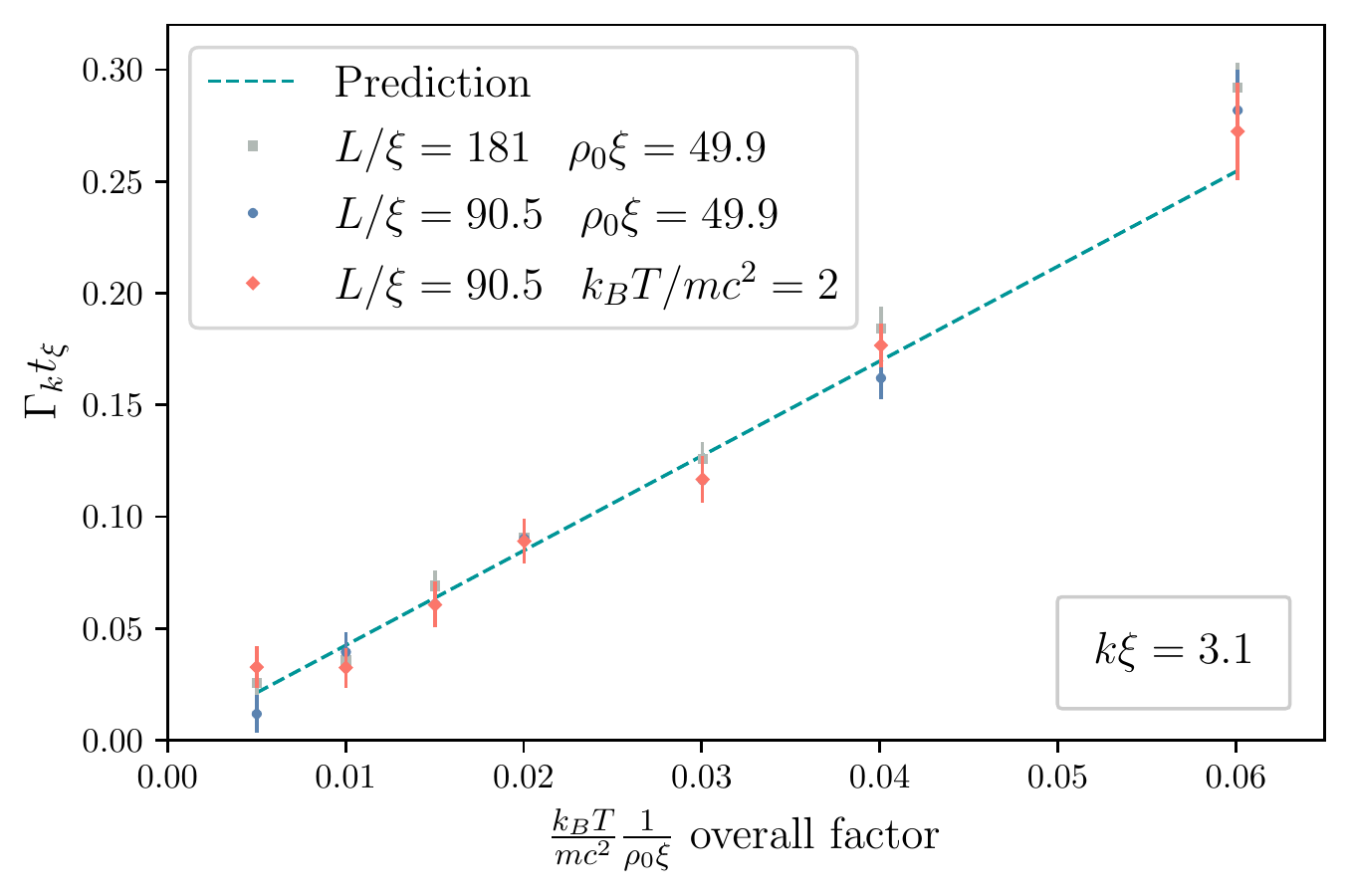}
    \caption{Best fit values for $\hbar \Gamma_k$ extracted from the TWA simulations, as a function of $k_{B} T/ \rho_{0}\xi$.
    All points are extracted from simulations performed using an addition of $\delta n = 3$ phonons on average.
    We used $400$ realisations, a grid spacing $\Delta x/\xi = 0.35$,
    and 
    a time window $t / t_{\xi} \in \left[ 0, 10\right]$ with $140$ time steps. 
    For the grey and blue points the temperature varies $k_{B} T/ m c^2 \in \left[ 0.25 , 3 \right]$, while for the red points the density varies in $\rho_0 \xi \in \left[ 33 , 399 \right]$. The parameters that are fixed in each run are listed in the legend.
    \label{fig:scaling_Gamma_asap_T}}
\end{figure}

\begin{figure}
    \centering
    \includegraphics[width=0.49\textwidth]{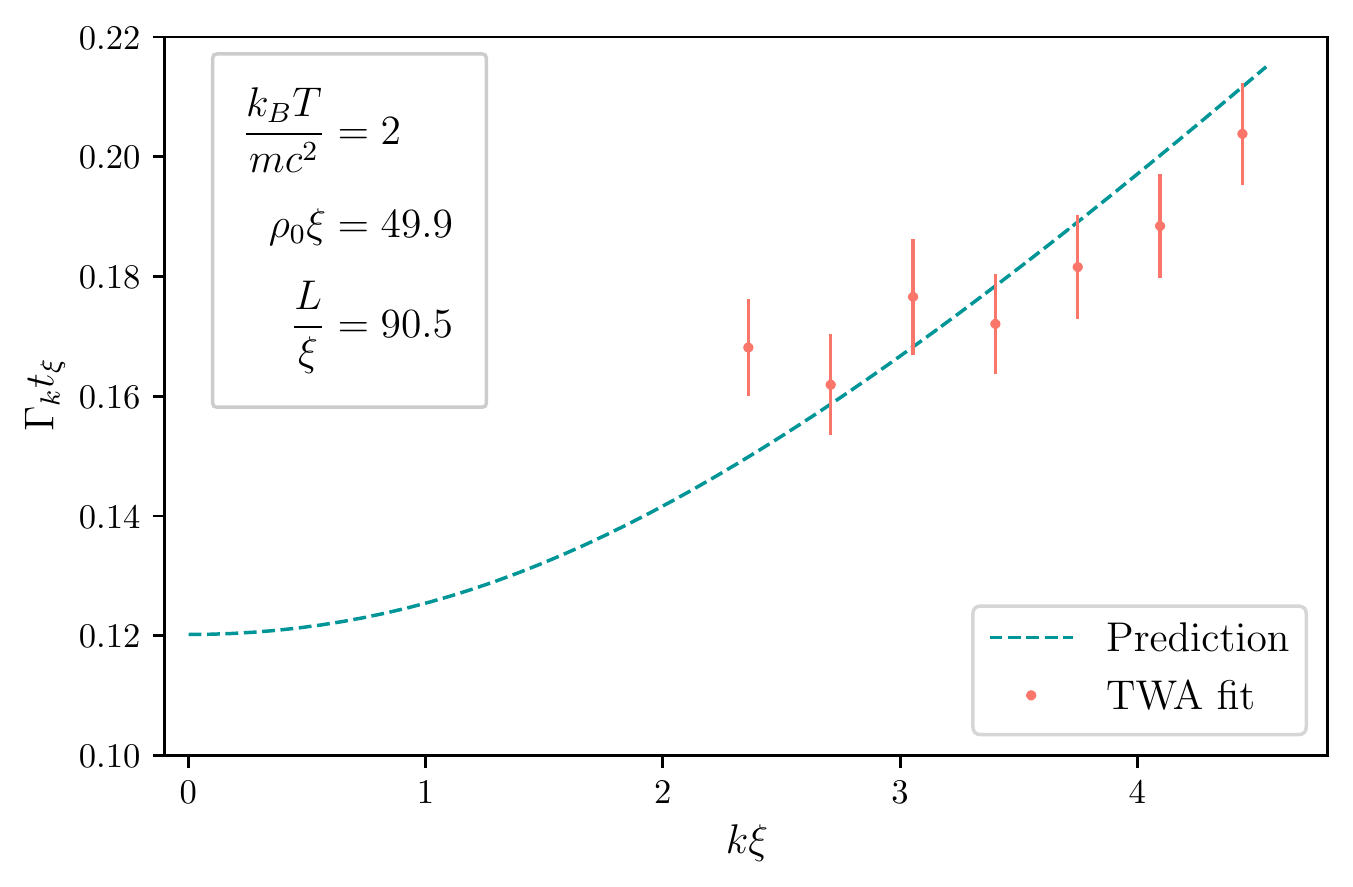}
    \caption{Best fit values for $\hbar \Gamma_k$ extracted from the TWA simulations, as a function of $k \xi$. All other physical parameters are fixed at the values shown, while numerical parameters are the same as in Fig.~\ref{fig:scaling_Gamma_asap_T}.
    Numerically extracted values are shown only for $k\xi \gtrsim 2.5$.  For lower $k\xi$, significant deviations appear due to the longer response time. (A fuller treatment of this regime is given in Appendix~\ref{app:deviations}.)
    \label{fig:scaling_Gamma_k_xi}}
\end{figure}

In Fig.~\ref{fig:scaling_Gamma_asap_T} we demonstrate the linearity of the best-fit values of $\Gamma$ in the overall prefactor in Eq.~(\ref{def:decay_rate}), at fixed $k\xi$.  The numerical results are in good agreement with the prediction.  We illustrate that varying $T$ at fixed $\rho\xi$ (and {\it vice versa}) yields the expected behaviour, and moreover that the fitted decay rate is unaffected by a change in $L$, as expected from prediction~(\ref{def:decay_rate}).  Any $L$-dependence of the observed behavior would then appear in the other fitting parameters like $\gamma$.

In Fig.~\ref{fig:scaling_Gamma_k_xi} we show instead the dependence of these best-fit values for $\Gamma$ on the probe mode $k\xi$, with the prefactor in Eq.~(\ref{def:decay_rate}) fixed.  The numerical observations agree well with the predicted behaviour at modestly high $k\xi \gtrsim 2.5$.
At lower $k\xi$, we expect significant deviations, for as noted in Sec.~\ref{subsec:Response_functions},
the FGR becomes valid only after a critical time $t_{\rm crit}$ (see also~\cite{CohenTannoudji2020,Peres1980}). We show in Appendix~\ref{app:deviations} that $t_{\rm crit}$ diverges like $1/\left(k\xi\right)^{3}$ as $k\xi \to 0$, and for $k\xi \lesssim 2.5$ it is not reached within the sampled time frame. In this regime, the decay proceeds quadratically in time rather than exponentially, with a lifetime that is significantly longer than predicted by~(\ref{def:decay_rate}). This early-time behavior at small $k\xi$ is examined in more detail in Appendix~\ref{app:deviations}.

\subsection{Slowing of exponential growth}
\label{subsec:slowing_of_exponential_growth}

In our second set of simulations, the initial state is simply a thermal state for the quadratic Hamiltonian $\hat{H}_{2}$, with no additional component added by hand.  Instead, during the evolution, parametric resonance is induced by a sinusoidal modulation of the 1D atomic interaction parameter:
\begin{equation}
\label{eq:def:modulation_g}
    g(t) = g \left(1 + a \, {\rm sin}\left(\omega_{p}t\right) \right) \,.
\end{equation}
As $c^{2}(t) \propto g(t)$, this translates into a sinusoidal modulation of the squared phonon frequencies:
\begin{eqnarray}
\label{eq:freq_modulation}
    \omega_{k}^{2}(t) &=& c^{2}k^{2} \, \left[ 1 + a\, {\rm sin}\left(\omega_{p}t\right) + \frac{1}{4} k^{2} \xi^{2} \right] \nonumber \\
    &=& \omega_{k}^{2} \left(1 + A_{k} \, {\rm sin}\left(\omega_{p}t\right) \right) \,,
\end{eqnarray}
where $A_{k} = a/\left(1+k^{2}\xi^{2}/4\right)$.
This is exactly the situation 
modeled in~\cite{Busch-2014}.  The result is an exponential growth of the number of phonons within a resonant frequency window centered at $\omega_{p}/2$.  
In the absence of any damping mechanism, the analysis in Appendix~A of~\cite{Busch-2014} shows that, for the exactly resonant mode at $\omega_{k} = \omega_{p}/2$, the occupation number is parametrically amplified according to 
\begin{align}
\begin{split}
\label{eq:estimate_parametric_growth}
n_{k}(t) & \approx n_{k}^{\rm in} + \left( 2 n^{\mathrm{in}}_{k} +1\right) \sinh^2 \left( \frac{1}{2} G_{k}  t  \right)  \,,
\end{split}
\end{align}
where the growth rate $G_{k} = A_{k} \omega_{k}/2$.
Since the initial state is thermal, the initial occupation number $n_{k}^{\rm in}$ is simply the thermal population of the mode.

\begin{figure}
    \centering
    \includegraphics[width = 0.49\textwidth]{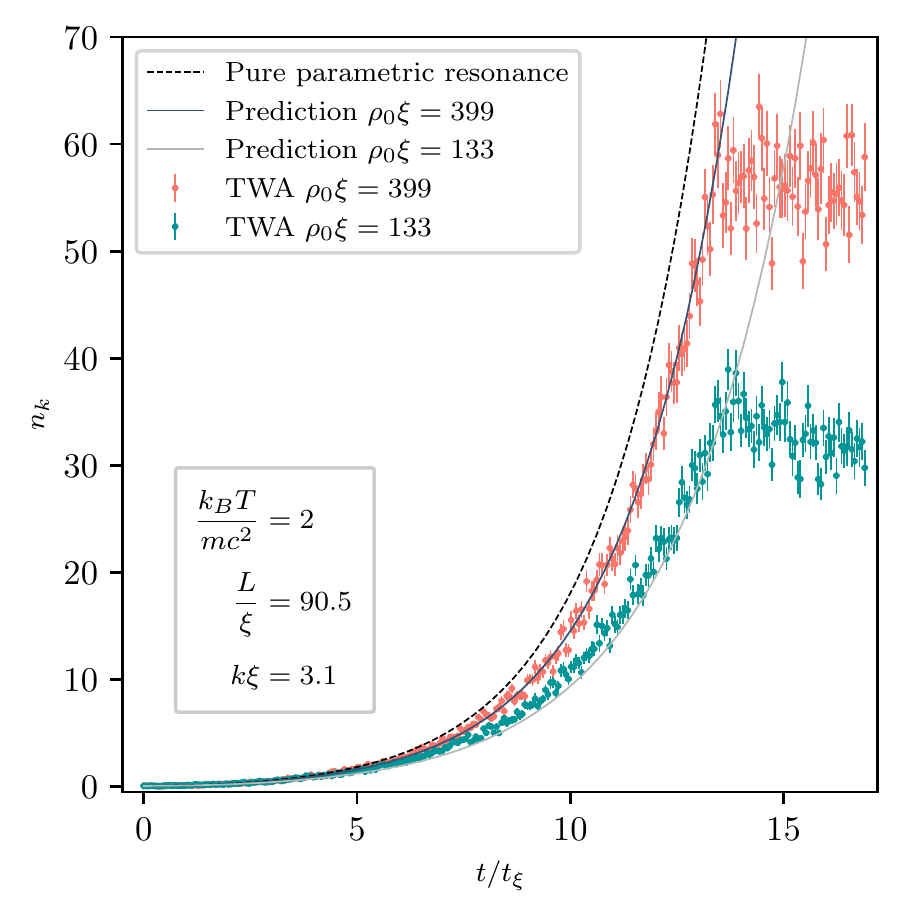}
    \caption{Mean occupation of the resonant mode as a function of time, as extracted from TWA simulations.   We take an initial thermal state at temperature $k_{B} T/m c^2 = 2$, and a modulation of amplitude $a = 0.5$ at frequency $\omega_{p} = 2 \omega_{k}$ where $k \xi = 3.1$ (so that the modulation of $\omega_{k}^{2}$ has amplitude $A_{k} = 0.15$) lasting $24$ periods, {\it i.e.}, $13.5 \, t_{\xi} $. The size of the system is $L/\xi = 90.5$ and its atomic density is $\rho_0 \xi =399$ for the red dots and $\rho_0 \xi = 133$ for the green ones. The dashed black line is the estimate Eq.~(\ref{eq:estimate_parametric_growth}), where $n_{\rm in}$ has been set to the thermal value. The solid lines correspond to the corrected estimate Eq.~(\ref{eq:corrected_estimate_parametric_growth}), with $\Gamma_{k}$ as predicted by Eq.~(\ref{def:decay_rate}). Each data point is calculated independently, averaged over $400$ realizations.
    \label{fig:n_k_parametric_amplification_reduction}}
\end{figure}

When including the effects of phonon interactions, the Beliaev-Landau scattering with the thermal population acts simultaneously with the parametric amplification, kicking phonons out of the resonant mode as they are being produced. The damping mechanism thus acts much like the phenomenological damping introduced in~\cite{Busch-2014}, and is therefore expected to reduce the growth rate. This is illustrated in Fig.~\ref{fig:n_k_parametric_amplification_reduction}, where $n_{k}(t)$ of the exactly resonant mode is extracted from the numerical simulations and shown for two different values of $\rho_{0}\xi$.  We clearly see that the exponential growth rate is lower than predicted by Eq.~(\ref{eq:estimate_parametric_growth}), and that it is further reduced as $\rho_{0}\xi$ is reduced, so that $\Gamma_{k}$ of Eq.~(\ref{def:decay_rate}) is accordingly increased.

We may use the difference between the observed growth rate and the ``pure'' growth rate of Eq.~(\ref{eq:estimate_parametric_growth}) 
as a measure of the damping.~\footnote{Similar simulations were performed in~\cite{Robertson_2018} where, although the focus was on nonlinear effects at large $n_{k}$, hints of an $n_{k}$-independent damping rate were seen at sufficiently early times (see in particular Fig.~6 and footnote~10).  Given the parameters used ($k\xi \sim 1$, $k_{B}T/mc^{2} \sim 1/2$, $\rho_{0}\xi \sim 400$), this intrinsic decay rate has $\hbar\Gamma_{k}/mc^2$ equal to a few times $10^{-3}$, in agreement with prediction~(\ref{def:decay_rate}).}
Assuming that the damping acts straightforwardly as a reduction of the growth rate, we make a slight generalization of Eq.~(\ref{eq:estimate_parametric_growth}) and propose the following ansatz for the occupation number of the resonant mode:
\begin{equation}
\label{eq:corrected_estimate_parametric_growth}
    n_{k} (t) = n^{\mathrm{in}}_{k} + \left( 2 n^{\mathrm{in}}_{k} +1\right) \sinh^2 \left[ \frac{1}{2}  \left(G_{k} - \Gamma_{k}\right)  t  \right] \, .
\end{equation}
Figure~\ref{fig:n_k_parametric_amplification_reduction} shows that this ansatz, taken together with the assumption that $\Gamma_{k}$ is as predicted by Eq.~(\ref{def:decay_rate}), accounts very well for the reduced growth observed in TWA simulations with respect to that of Eq.~(\ref{eq:estimate_parametric_growth}). We stress that this reduction rapidly leads to a sizeable change in the number of produced phonons. 
For example, considering the red data points in Fig.~\ref{fig:n_k_parametric_amplification_reduction} (corresponding to $k\xi = 3.1$, $k_{B}T/mc^{2}=2$, and $\rho_0 \xi = 399$), the relative damping is $\Gamma_{k}/G_{k} = 5 \%$, yet the reduction with respect to the non-damped case is very clear, and we are thus able to extract quite precise values for the damping rate.

\begin{figure}
    \centering
    \begin{minipage}{0.49\textwidth}
        \centering
        \includegraphics[width=\textwidth]{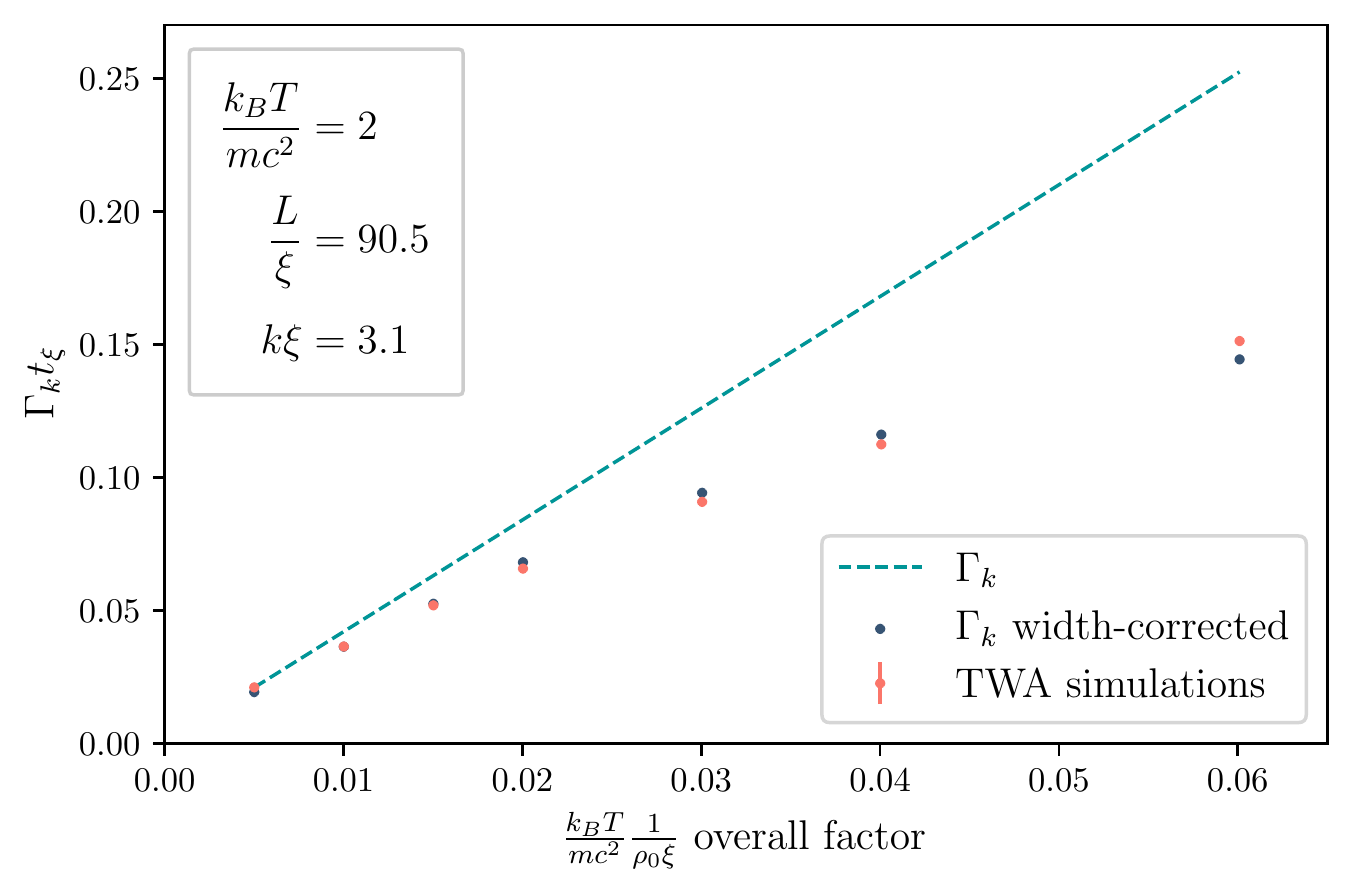}
    \end{minipage} \hfill
    \begin{minipage}{0.49\textwidth}
        \centering
         \includegraphics[width=\textwidth]{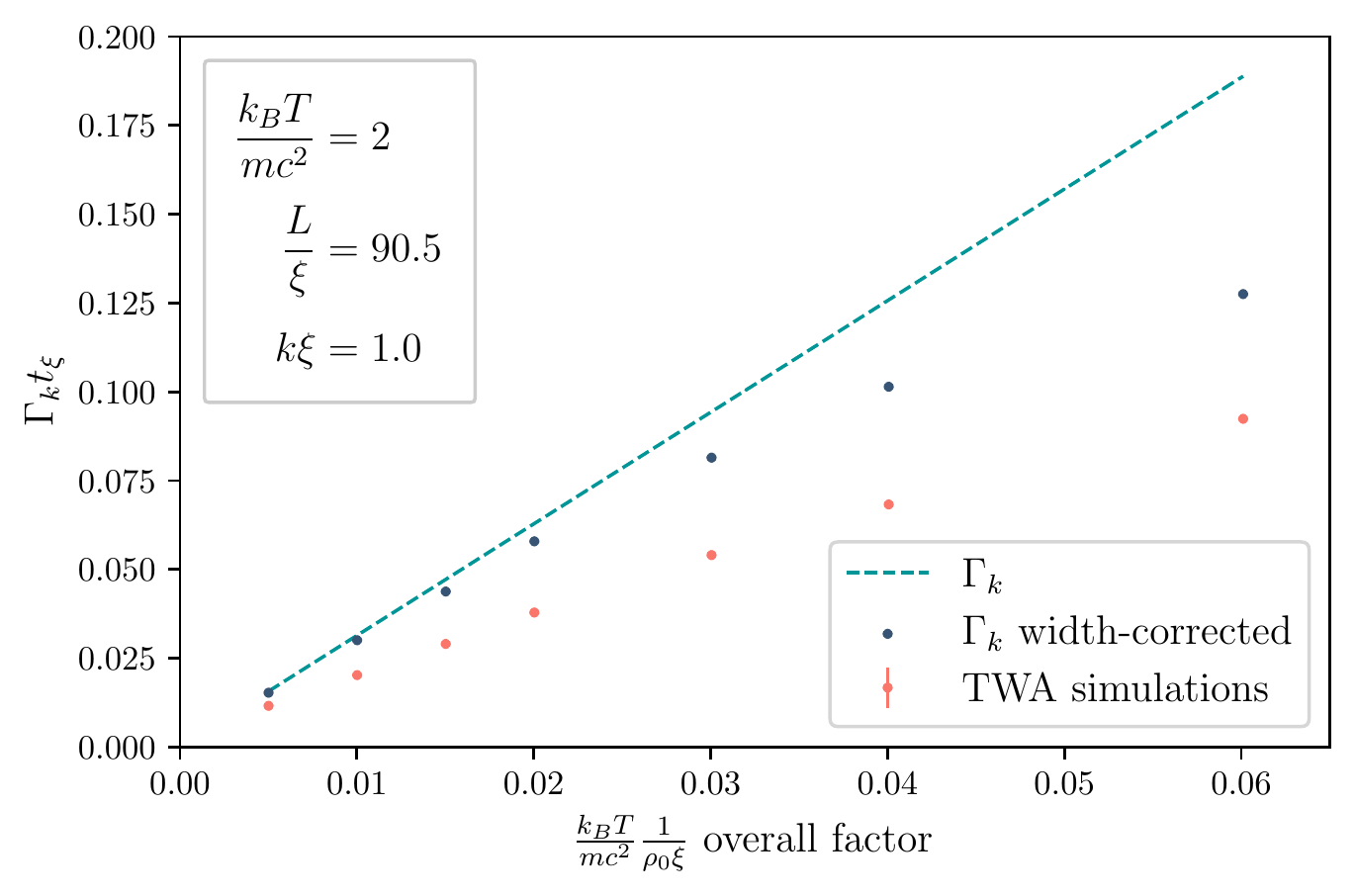}
    \end{minipage}
    \caption{Best fit values for $\hbar \Gamma_{k}$ extracted from the TWA simulations as a function of $k_{B} T/ \rho_{0}\xi$.
    (We divide by $mc^{2}$ to adimensionalize.) The temperature is kept fixed to $k_{B} T/ mc^2 = 2$ and only the density is varied $\rho_0 \xi \in \left[ 33 , 399 \right]$. The parameters that are fixed in each run are listed in the legend.
    All points are extracted from simulations performed using a continuous modulation of amplitude $a = 0.5$ at frequency $\omega_{p} = 2 \omega_{k}$ where $k \xi = 3.1$ (top) and $k \xi = 1.0$ (bottom); the corresponding amplitudes for the modulation of $\omega_{k}^{2}$ are $A_{k} = 0.15$ (top) and $A_{k} = 0.39$ (bottom). We use $n_r = 400$ realisations, a spatial grid with spacing $\Delta x/\xi = 0.35$.
    The fit is performed using the template Eq.~(\ref{eq:corrected_estimate_parametric_growth}) over a time window $t / t_{\xi} \in \left[ 0, 13.5 \right]$ (top) and $t / t_{\xi} \in \left[ 0, 22.5 \right]$ (bottom) with $224$ (top) and $280$ (bottom) time steps. The green dashed line represents the FGR prediction Eq.~(\ref{def:decay_rate}) while the blue dots are the corrected predictions taking into account the effect of a finite width of the resonant peak, see Appendix~\ref{app:deviations}. The peak is assumed to be of Lorentzian shape and its parameters are extracted by a procedure described in Fig.~\ref{fig:Rq_spectrum_different_asap}.
    }
    \label{fig:gamma_k_parametric_amplification_reduction}
\end{figure}

We now perform a more systematic examination of the numerically observed reduction of the growth rate.
Proceeding as for the first set of simulations, we fit $n_{k}(t)$ for the exactly resonant mode to Eq.~\eqref{eq:corrected_estimate_parametric_growth}, where $\Gamma_{k}$ is now treated as a fitting parameter. 
This is done for two different wave vectors, one in the high-$k$ regime ($k\xi = 3.1$) where the first set of simulations worked reasonable well, and one in the low-$k$ regime ($k\xi = 1.0$) where they did not.  The temperature is fixed at $k_{B}T/mc^{2} = 2$, and the 1D density $\rho_{0}\xi$ is varied.  The extracted values of $\Gamma_{k}$ are shown in Fig.~\ref{fig:gamma_k_parametric_amplification_reduction} alongside the prediction of Eq.~(\ref{def:decay_rate}).

\begin{figure}
    \centering
    \begin{minipage}{0.49\textwidth}
        \centering
        \includegraphics[width=\textwidth]{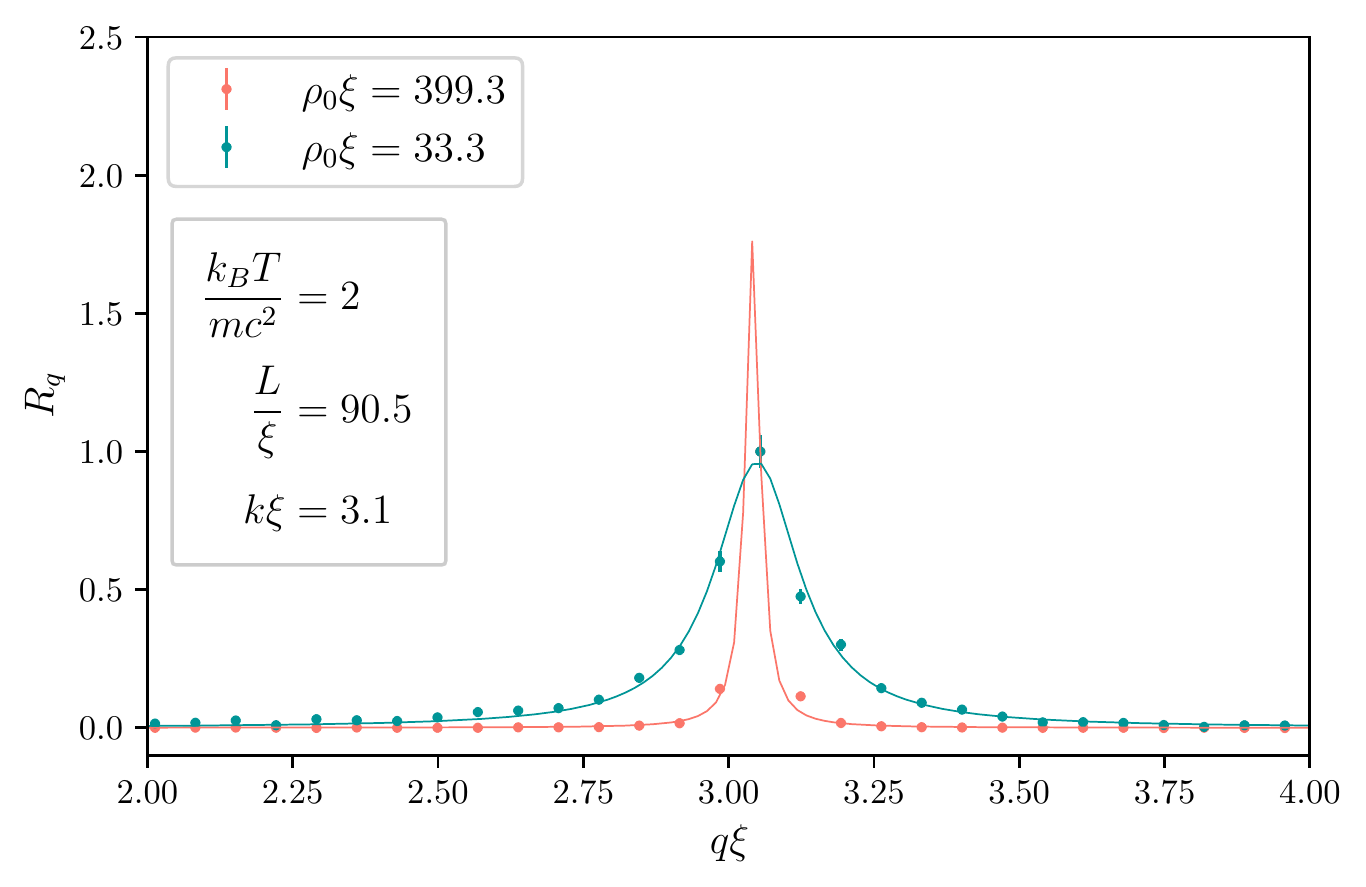}
    \end{minipage} \hfill
    \begin{minipage}{0.49\textwidth}
        \centering
         \includegraphics[width=\textwidth]{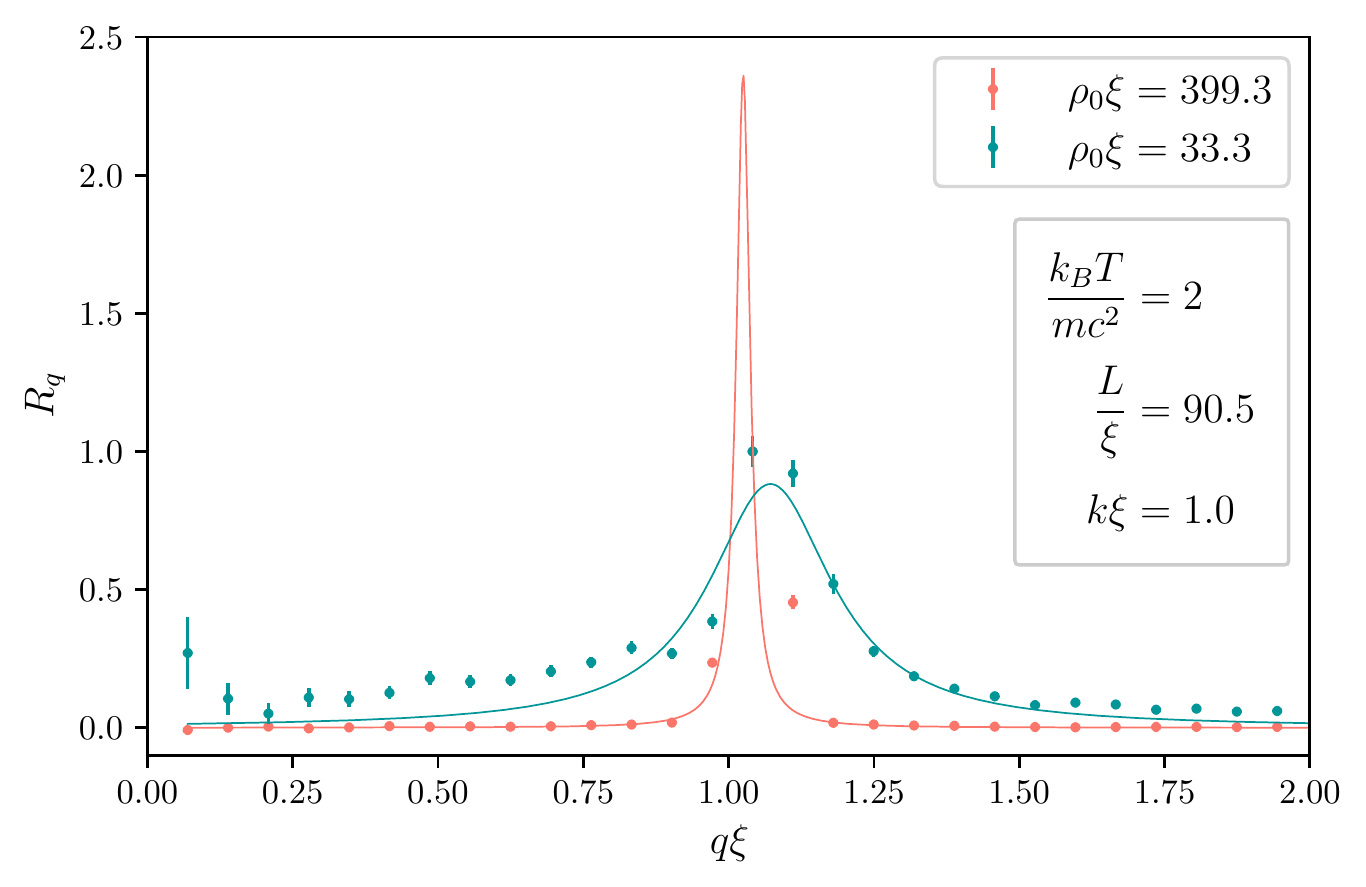}
  
    \end{minipage}
    \caption{Snapshots of the ratio $R_{q} = \frac{\delta n_q}{\delta n_k}$ as a function of $q \xi$ at time $t/t_{\xi}=13.5$ (top) and $t/t_{\xi} = 22.5$ (bottom). The spectrum is plotted for  $\rho_0 \xi = 399$ (red) and $\rho_0 \xi = 33$ (green).
    They are obtained by a continuous modulation of the type Eq.~(\ref{eq:def:modulation_g}) with $a =0.5$ and at the appropriate frequency so that $k \xi = 3.1$ (top) or $k \xi = 3.1$ (bottom) are exactly at the resonance. A Lorentzian is fitted to the distribution taking into account the first five neighbours on each side of the central resonant mode letting the overall amplitude, width and center be free fitting parameters. The best-fit value of the width is then used to correct the prediction of the decay rate, blue dots in Fig.~\ref{fig:gamma_k_parametric_amplification_reduction}. The averages are calculated from an ensemble of $400$ independent realisations, while the error bars represent the standard deviation. There are $256$ points on the grid.
    \label{fig:Rq_spectrum_different_asap}
    }
\end{figure}

A few remarks are in order concerning these results.  We begin by focusing on the high-$k$ mode ($k\xi = 3.1$), the first and third data points of which correspond to the simulations shown in Fig.~\ref{fig:n_k_parametric_amplification_reduction}.
While the extracted decay rate tends towards prediction~(\ref{def:decay_rate}) at large $\rho_0 \xi$, there is a clear trend for it to fall further away from this prediction as $\rho_0 \xi$ is decreased. We attempt to explain this behavior by appealing to the finite width of the resonant peak.
Examining the evolution of the number spectrum shows that, to a good approximation, the shape of the resonant peak saturates such that $n_{k} \sim R_{k} e^{G t}$ at sufficiently late time, for some profile $R_{k}$ and some growth rate $G$.  This is in contrast to the results of the ``pure'' parametric resonance with no phonon interactions, where the growth rate is $k$-dependent and largest at exact resonance~\cite{Busch-2014}, so that the width of the peak approaches zero as $t \to \infty$.  In the present case, in addition to the saturation of the profile, we observe that larger interaction strengths are associated with wider peaks, see Fig.~\ref{fig:Rq_spectrum_different_asap}.  It thus seems likely that $n_{k}$ is also being fed by phonons in neighboring modes through the last term on the right-hand side of Eq.~(\ref{eq:nk_eom_response_functions}). We investigate this effect by adopting a Lorentzian ansatz for the profile of $\delta n_{k}$. Figure~\ref{fig:Rq_spectrum_different_asap} shows how our ansatz compares with the numerically observed number spectrum. Under this assumption, we solve Eq.~(\ref{eq:nk_eom_response_functions}) self-consistently to extract the net growth rate.  The details of this calculation are given in Appendix~\ref{app:deviations}, but the corrected predictions for the decay rate are shown by the blue dots in Fig.~\ref{fig:gamma_k_parametric_amplification_reduction}. For $k \xi = 3.1$ this corrected prediction is found to be in very good agreement with the extracted rate.

For the low-$k$ mode ($k \xi = 1.0$) there remains a clear discrepancy even when accounting for the finite width of the peak. Part of the explanation lies in the non-Lorentzianity of the profile visible in Fig.~\ref{fig:Rq_spectrum_different_asap}, making the correction less valid.  However, note that the extracted decay rate does not tend well to prediction~(\ref{def:decay_rate}) at large $\rho_{0}\xi$, as it seems to approach a line with a different slope.  We believe this to be a consequence of the large critical time in the low-$k$ regime: in effect, the occupation number is growing a little too fast for the system to have time to react, and the amount of damping is thus lower than expected.

The parametric resonance simulations described here yield results that corroborate and complement those found using straightforward phonon injection.
In the high-$k$ regime where the critical response time is sufficiently short, the observed deviations from prediction~(\ref{def:decay_rate}) are well described via corrections due to the finite width of the peak, which only appear in the case of parametric resonance.  
On the other hand, at low $k$ where the critical time is long, the phonon injection method yields a non-exponential behavior that has not been shown here (see instead Appendix~\ref{app:deviations}), whereas the parametric resonance approach still gives a damping rate that has the same qualitative behavior as in the high-$k$ regime (see the near-linear behavior of the red dots in the lower panel of Fig.~\ref{fig:gamma_k_parametric_amplification_reduction}).
Moreover, the method of parametric resonance for exciting phonons has some interesting advantages over that of phonon injection. 
Most notably, the damping mechanism manifests in a rather more dramatic way, as is evident from the difference in the final phonon numbers shown in Fig.~\ref{fig:n_k_parametric_amplification_reduction}.  It is also of considerable experimental relevance: while the injection of phonons is conceptually simple, it is not very practical, whereas parametric resonance is a way of exciting phonons that has already been implemented in experiments \cite{Jaskula2012}.
On the conceptual side, the method of parametric resonance bypasses the finite-size effect.
In Sec.~\ref{subsec:initial_injection}, it arises due to the discretization of a time-dependent continuous profile, the ``binned'' mode of interest containing both the exponentially decaying mode and the nearby modes within $\pi/L$ that are growing in time.  Here, the profile is fixed (the only time-dependence being an overall exponential factor $e^{G t}$), and the discretization is therefore irrelevant.

\section{Concluding remarks}
\label{sec:Conclusion}

We have identified a mechanism whereby interactions with a thermal bath of phonons, through Landau and Beliaev damping mechanisms, yields exponential decay of a singularly occupied phonon mode. 
It is effectively described by an application of the FGR with the resonant channel being the trivial one, in the limit $q \to 0$. 
The result was numerically verified using TWA simulations, following both the decay of a mode excited ``by hand'' and the reduced growth of a parametrically resonant mode.  The prediction works particularly well in the higher-$k$ regime, where the critical response time is relatively short and the discreteness of the thermal population is invisible to the mode in question.  
We also observe a reduction of the damping rate due to the finite width of the peak.

While this limit yields a non-vanishing rate only in 1D (since in higher dimensions the IR divergence of the thermal population is tamed by the volume element in $k$-space), it seems generally applicable to 1D systems with an approximately linear excitation spectrum in the limit of small $k$.   We thus believe that it could appear as an additional contribution to the decay of excitations in such systems, such as the 1D dipolar gas considered in~\cite{Kurkjian-Ristivojevic-2020}.

Comparing with higher dimensionality $D$~\cite{pastukhov_damping_2015},  there are some clear similarities: prediction~(\ref{def:decay_rate}) exhibits an increasing dependence on $k \xi$, is proportional to $\gamma_{\rm LL}^{D/2}$, and  is linear in $T$ (as found in higher dimensions when $k_{B} T/ mc^2 \gtrsim 1$).

The experimental relevance of this decay mechanism was exemplified by numerically studying the parametric growth of phonons in the gas, as experimentally tested in~\cite{Jaskula2012}. We have demonstrated that the numerically observed reduction of the growth rate can be quantitatively explained assuming that the phonons generated by the parametric process simultaneously decay by thermal Landau-Beliaev processes.   
 	
These thermal scatterings of phonons are expected to occur generally in 1D quasicondensate experiments. They can produce important deviations to the dynamics of phonons and should therefore be taken into account in designing and analysing analogue gravity experiments on this platform.

Finally, let us comment on the absence of entanglement between the induced peaks in the experimental observations of~\cite{Jaskula2012}.  In their phenomenological treatment of dissipation, the authors of~\cite{Busch-2014} showed that a dissipative rate $\Gamma/\omega_k \sim 4.2 \% $~\footnote{Since we focus directly on the phonon number $n_{k}$ rather than the amplitudes $\hat{\varphi}_{k}$, there is a difference of a factor 2 between the dissipative rate defined in~\cite{Busch-2014} and the one defined here.} would be sufficient to explain this negative result. Using the relevant parameters ($k\xi \sim 1$, $k_{B}T/\hbar\omega_{k} \sim 1$, $\rho_{0}\xi \sim 60$), Eq.~(\ref{def:decay_rate}) gives $\Gamma/\omega_k \sim 5 \% $.  The decay mechanism identified here thus provides a possible microphysical basis to this scenario. The precise dynamics of the entanglement will be the subject of a future work.

\section*{Acknowledgments}
We thank Christos Charmousis, Alessandro Fabbri, Denis Boiron, Chris Westbrook, and especially Florent Michel for interesting discussions and useful suggestions.
We are particularly grateful to our friend and mentor Renaud Parentani, who was responsible for the genesis of this work and whose guidance and input in its initial stages proved invaluable.  We would like to dedicate the paper to his memory.
This work was supported by the French National Research Agency via Grant No. ANR-20-CE47-0001 associated with the project COSQUA (Cosmology and Quantum Simulation).

\bibliography{Phonon_decay_in_thermal_1D_quasicondensate}

\newpage

\onecolumngrid
\begin{appendices}

\section{Fuller derivation of evolution of $\delta n_{k}$ \label{app:phonon_decay}}

In this appendix we derive Eq.~(\ref{eq:nk_eom_response_functions}), a more precise equation of motion for the phonon spectrum, and we show how the FGR emerges despite there being no exactly elastic scattering channel. We then analyze more fully the deviations with respect to the FGR result.

\subsection{Equations of motion for phonon operators}

We start by computing the Heisenberg equation of motion for $\hat{\varphi}_{k}$. The one for $\hat{\varphi}^{\dagger}_{k}$ is easily deduced by taking the adjoint. We consider only the dynamics under the quadratic~(\ref{eq:Hamiltonian_quadratic}) and cubic~(\ref{eq:Hamiltonian_fluctuations}) Hamiltonians, neglecting higher orders. We have :
\begin{align}
\begin{split}
\label{eq:eom_varphi_k}
    \partial_{t}\hat{\varphi}_{k} = -i \left\{ \omega_{k} \hat{\varphi}_{k} + \frac{1}{\sqrt{N_{\rm at}}} \sum_{q \neq 0  , -k }  2 V_{3}(k,q) \hat{\varphi}_{q}^{\dagger} \hat{\varphi}_{k+q}
    + \frac{1}{\sqrt{N_{\rm at}}} \sum_{q \neq 0 , k } V_{3}(k-q,q) \hat{\varphi}_{k-q} \hat{\varphi}_{q} \right\} \,.
\end{split}
\end{align}
Considering then the full number operator $\hat{n}_{k} = \hat{\varphi}_{k}^{\dagger} \hat{\varphi}_{k}$, we have
\begin{align}
\begin{split}
\label{eq:nk_evolution}
    \partial_{t}\hat{n}_{k} &= \hat{\varphi}_{k}^{\dagger} \cdot \partial_{t}\hat{\varphi}_{k} + \partial_{t} \hat{\varphi}_{k}^{\dagger} \cdot \hat{\varphi}_{k} \, , \\
    &= - \frac{i}{\sqrt{N_{\rm at}}} \sum_{q \neq 0  , -k } 2 V_{3}(k,q) \hat{\varphi}_{k}^{\dagger} \hat{\varphi}_{q}^{\dagger} \hat{\varphi}_{k+q} - \frac{i}{\sqrt{N_{\rm at}}}   \sum_{q \neq 0  , k } V_{3}(k-q,q) \hat{\varphi}_{k}^{\dagger} \hat{\varphi}_{k-q} \hat{\varphi}_{q}   + {\rm h.c.} \, .
\end{split}
\end{align}
On the right-hand side of the equation of motion for $\hat{n}_{k}$ appears the momentum preserving 3-phonon operator $\hat{\varphi}_{k}^{\dagger} \hat{\varphi}_{q}^{\dagger} \hat{\varphi}_{k+q}$, the same that appears in $\hat{V}_{3}$. We are thus compelled to consider the dynamics of this operator as well. 
It is useful to define $\hat{\varphi}_{p}^{\dagger} \hat{\varphi}_{q}^{\dagger} \hat{\varphi}_{p+q} = \hat{c}_{3}(p,q) \exp \left[  -i \left( \omega_{p+q} - \omega_{p} - \omega_{q} \right)  t \right]$, where the oscillations are made explicit. These operators obey
\begin{align}
\begin{split}
\label{eq:c3pq_evolution}
& e^{ - i \left( \omega_{p+q} - \omega_{p} - \omega_{q} \right)  t  } \partial_{t}\hat{c}_{3}(p,q) =   \partial_{t}\hat{\varphi}_{p}^{\dagger} \cdot \hat{\varphi}_{q}^{\dagger} \hat{\varphi}_{p+q} + \hat{\varphi}_{p}^{\dagger} \cdot \partial_{t} \hat{\varphi}_{q}^{\dagger} \cdot \hat{\varphi}_{p+q}  + \hat{\varphi}_{p}^{\dagger} \hat{\varphi}_{q}^{\dagger} \cdot \partial_{t}\hat{\varphi}_{p+q} + i \left( \omega_{p+q} - \omega_{p} - \omega_{q} \right) \hat{\varphi}_{p}^{\dagger} \hat{\varphi}_{q}^{\dagger} \hat{\varphi}_{p+q} \\
& =  \frac{  2  i}{\sqrt{N_{\rm at}}} V_{3}(p,q) \hat{\varphi}_{p+q}^{\dagger} \hat{\varphi}_{p+q} \\
   & + \frac{2 i}{\sqrt{N_{\rm at}}} \sum_{\lambda \neq 0, -p}  V_{3}(p,\lambda) \hat{\varphi}_{p+\lambda}^{\dagger} \hat{\varphi}_{q}^{\dagger} \hat{\varphi}_{\lambda} \hat{\varphi}_{p+q} +  \sum_{\lambda \neq 0,-q}  V_{3}(q,\lambda) \hat{\varphi}_{p}^{\dagger} \hat{\varphi}_{q+\lambda}^{\dagger} \hat{\varphi}_{\lambda} \hat{\varphi}_{p+q} -  \sum_{\lambda \neq 0,-(p+q) } V_{3}(p+q,\lambda) \hat{\varphi}_{p}^{\dagger} \hat{\varphi}_{q}^{\dagger} \hat{\varphi}_{\lambda}^{\dagger} \hat{\varphi}_{p+q+\lambda} \\
   & + \frac{i}{\sqrt{N_{\rm at}}}  \sum_{\lambda \neq 0,p}   V_{3}(p-\lambda,\lambda) \hat{\varphi}_{\lambda}^{\dagger} \hat{\varphi}_{p-\lambda}^{\dagger} \hat{\varphi}_{q}^{\dagger} \hat{\varphi}_{p+q}
    + \sum_{\lambda \neq 0,q} V_{3}(q-\lambda,\lambda) \hat{\varphi}_{p}^{\dagger} \hat{\varphi}_{\lambda}^{\dagger} \hat{\varphi}_{q-\lambda}^{\dagger} \hat{\varphi}_{p+q} - \sum_{\lambda \neq 0,p+q}  V_{3}(p+q-\lambda,\lambda) \hat{\varphi}_{p}^{\dagger} \hat{\varphi}_{q}^{\dagger} \hat{\varphi}_{p+q-\lambda} \hat{\varphi}_{\lambda}  \, .
\end{split}
\end{align}
The extra term on the first line involving the amplitudes for the $p+q$ mode comes from rearranging the one term which is not initially in normal order.

\subsection{Equations of motion for average values}

We want to take average values in both equations of motion above and derive the evolution of $n_k = \left\langle \hat{n}_{k} \right\rangle$. However, in order to get a closed system we have to make some approximations. We assume that the state is initially homogeneous; it will remain so as our Hamiltonian is momentum-conserving. We also assume that the initial state is Gaussian, and that the only deviation from Gaussianity to evolve  is a non-vanishing value of $\hat{c}_{3}(p,q)$, {\it i.e.}, every connected correlation function of order higher than three is negligible. Therefore, when taking average values on the right hand-side of Eq.~(\ref{eq:c3pq_evolution}), the 4-point functions reduce to products of 2-point functions which have to respect the homogeneity of the state:
\begin{equation}
    \left \langle  \hat{\varphi}_{p}^{\dagger} \hat{\varphi}_{q}^{\dagger} \hat{\varphi}_{p+q-\lambda} \hat{\varphi}_{\lambda} \right \rangle =  n_p n_q \delta_{q, \lambda} + n_p n_q \delta_{p, \lambda} + c_q^{\star} c_{\lambda} \delta_{p, -q} \, ,
\end{equation}
where $c_p =  \left \langle  \hat{\varphi}_{p} \hat{\varphi}_{-p} \right \rangle $. The quantities $n_{\pm p}$ and $c_p$ -- respectively, the population and 2-mode correlation of the modes $\pm p$ -- are the only non-vanishing 2-point functions in a homogeneous state.  We shall further assume, for simplicity, that the $c_p$ are all vanishing.~\footnote{This is not actually the case for the 2-mode squeezed state generated by the parametric resonance considered in the second set of simulations. Yet, we achieve a quantitative agreement with the TWA simulations for large enough value of $k \xi$. We relegate the analysis of the influence of the correlation on the decay to a future work dedicated to the dynamics of entanglement in this context.} The equation of motion for $n_p$ and 
$c_{3}(p,q) =  \left \langle \hat{c}_{3}(p,q)  \right \rangle$ then reads :
\begin{align}
\begin{split}
\label{eq:reduced_syst_eom_nk_cpq}
\partial_{t} n_k & =  \frac{1}{\sqrt{N_{\rm at}}} \sum_{q \neq 0 , - k }  4 V_{3}(k,q) \Im \left[ c_{3}(k,q) e^{  - i  \delta \omega_L (q;k) t }  \right] - \frac{1}{\sqrt{N_{\rm at}}} \sum_{q \neq 0 , k  } 2 V_{3}(k-q,q) \Im \left[  c_{3}(k-q,q) e^{  i  \delta \omega_B (q;k) t }  \right]    \, , \\ 
\partial_{t} c_{3}(p,q) & =  2 i \frac{V_{3}(p,q)}{\sqrt{N_{\rm at}}}  \left[ n_{p+q} \left(n_{p}+n_{q}+1\right) - n_{p}n_{q} \right] e^{ i \left( \omega_{p+q} - \omega_{p} - \omega_{q} \right) t  } \, .
\end{split}
\end{align}
We can reduce this to a single equation of motion for $n_{k}$ by writing $c_{3}(p,q)$ explicitly in terms of $n_{k}$:
\begin{equation}
    c_{3}(p,q)(t) = 2 i \, \frac{V_{3}\left(p,q\right)}{\sqrt{N_{\rm at}}}  \int_{0}^{t} dt^{\prime} \, N_{p,q}\left(t^{\prime}\right) e^{i\left(\omega_{p+q}-\omega_{p}-\omega_{q}\right)t^{\prime}} \, ,
    \label{eq:c3_integral}
\end{equation}
where $N_{p,q} = n_{p+q} \left(n_{p}+n_{q}+1\right) - n_{p}n_{q}$. Equation~(\ref{eq:c3_integral}) can now be substituted directly into the equation of motion for $n_{k}$:
\begin{align}
\begin{split}
\partial_{t} n_{k} = 8 \sum_{q \neq 0 , - k} \frac{\left|V_{3}\left(k,q\right)\right|^{2} }{N_{\rm at}} \int_{0}^{t} dt^{\prime} \, N_{k,q}\left(t^{\prime}\right) {\rm cos}\left[\left(\omega_{k+q}-\omega_{k}-\omega_{q}\right)\left(t-t^{\prime}\right)\right] \\
    - 4 \sum_{q \neq 0 , k}  \frac{\left|V_{3}\left(k-q,q\right)\right|^{2} }{N_{\rm at}} \int_{0}^{t} dt^{\prime} \, N_{k-q,q}\left(t^{\prime}\right) {\rm cos}\left[\left(\omega_{k}-\omega_{k-q}-\omega_{q}\right)\left(t-t^{\prime}\right)\right] \,.
\label{eq:nk_eom_full_non_markovian}
\end{split}
\end{align}
Note that the terms depending on the various populations can be rewritten as the difference between the direct process mentioned in the text below Eq.~(\ref{eq:Hamiltonian_fluctuations}) and the corresponding reverse process. For instance in the first sum, corresponding to the Landau scattering, we have
\begin{equation}
   n_{p+q}\left(n_{p}+n_{q}+1\right) - n_{p} n_{q}  = \left(n_{p}+1\right) \left(n_{q}+1\right) n_{p+q} - \left(n_{p+q}+1\right) n_{q} n_{p} \, .
\end{equation}
On the right-hand side the $+1$ terms are associated only with the decay products and allow for spontaneous processes, while the $n$s encode the stimulated part. This generalizes the matrix elements given in the main text, which only include one direction where an excitation at $p$ is removed by the interaction.

Equations~(\ref{eq:reduced_syst_eom_nk_cpq}) and~(\ref{eq:nk_eom_full_non_markovian}) are the key equations governing the system, given our simplifying approximations.  They are entirely equivalent if $c_{3}(p,q)$ is set to zero at $t=0$, though Eqs.~(\ref{eq:nk_eom_full_non_markovian}) can be straightforwardly modified in a more general case.  Since Eqs.~(\ref{eq:reduced_syst_eom_nk_cpq}) are Markovian, they are much more suitable for numerical integration.  On the other hand, Eq.~(\ref{eq:nk_eom_full_non_markovian}) is a nonlinear, non-Markovian equation for the full phonon spectrum $n_{k}$. However, if analyzed using appropriate approximations, we will show that it encodes both the exponential decay of the phononic population, its first deviations and the corrections to the growth of population in a parametric resonance process. 

\subsection{Dynamics of a probe on top of a quasicondensate}
\label{subsubsec:eom_for_probe_new}

Up to this point, we have assumed homogeneity and quasi-Gaussianity of the state but worked with the full phonon spectrum $n_q$. As explained in the main text, the presence of a near-thermal population in the IR modes is instrumental in the decay process that we study. Therefore, as done in Eq.~(\ref{eq:nk_decomposition}), we split $n_q$ into a thermal background $n^{\rm th}_{q} = 1/\left[{\rm exp}\left(\hbar\omega_{q}/k_{B}T\right)-1\right]$ plus a perturbation $\delta n_q$. Physically this setup allows us to analyze the response of a quasicondensate at temperature $T$ to the addition of phonons around a certain mode $k$. This can either be done ``by hand'', as we have done in the first set of simulations presented in the text, or by a parametric amplification as in the second set, see Sec.~\ref{sec:numerical_confirmation}. As illustrated by Fig.~\ref{fig:phonon_spectrum_different_probes} of the main text, we expect these $\delta n$ probe phonons to redistribute across the system and want to compute the ensuing decay rate.  

Notice that the background thermal population $n^{\rm th}_q$ is not strictly stationary due to the addition of the interaction term $\hat{V}_3$, but as a first approximation we will assume it to be a solution of the equation of motion. Since we expect the relevant interactions to be between the peak and the thermal population, and not of the peak on itself, we insert $n_{k} = n^{\rm th}_{k} + \delta n_{k}$ in Eq.~(\ref{eq:nk_eom_full_non_markovian}) and linearise in $\delta n$. We have 
\begin{align}
\begin{split}
    N_{p,q} & = N_{p,q}^{\rm th} +\delta n_{p+q} \left(n^{\rm th}_{p}+n^{\rm th}_{q}+1\right) + \delta n_{p} \left(n_{p+q}^{\rm th}- n_{q}^{\rm th} \right) + \delta n_{q} \left(n_{p+q}^{\rm th}- n_{p}^{\rm th} \right) + O \left( \delta n^2 \right) \, , \\
    & = N_{p,q}^{\rm th} + \delta  N_{p,q} + O \left( \delta n^2 \right) \, , \\
\end{split}
\end{align}
with $N_{p,q}^{\rm th}$ the specific combination of populations evaluated with thermal populations and  
\begin{align}
    \delta N_{k,q} & = \delta n_{k} \left(n_{k+q}^{\rm th}- n_{q}^{\rm th} \right) + \delta n_{k+q} \left(n^{\rm th}_{k}+n^{\rm th}_{q}+1\right) +  \delta n_{q} \left(n_{k+q}^{\rm th}- n_{k}^{\rm th} \right) \, , \\
    \delta N_{k-q,q} & = \delta n_{k} \left(n^{\rm th}_{k-q}+n^{\rm th}_{q}+1\right) + \delta n_{k-q} \left(n_{k}^{\rm th}- n_{q}^{\rm th} \right) + \delta n_{q} \left(n_{k}^{\rm th}- n_{k-q}^{\rm th} \right) \, .
\end{align}
Inserting back in the equations of motion we get
\begin{align}
\begin{split}
\partial_{t} \delta n_{k} & = 8 \sum_{q \neq 0 , - k} \frac{\left|V_{3}\left(k,q\right)\right|^{2} }{N_{\rm at}} \int_{0}^{t} dt^{\prime} \,  \delta  N_{k,q} {\rm cos}\left[\left(\omega_{k+q}-\omega_{k}-\omega_{q}\right)\left(t-t^{\prime}\right)\right] \\
    & - 4 \sum_{ q \neq 0 , k }  \frac{\left|V_{3}\left(k-q,q\right)\right|^{2} }{N_{\rm at}} \int_{0}^{t} dt^{\prime} \, \delta  N_{k-q,q} {\rm cos}\left[\left(\omega_{k}-\omega_{k-q}-\omega_{q}\right)\left(t-t^{\prime}\right)\right] \, , \\
    & = - \int_{0}^{t} dt^{\prime} \, D_{k}\left(t-t^{\prime}\right) \, \delta n_{k}\left(t^{\prime}\right) + \int_{0}^{t} dt^{\prime} \, \sum_{q \neq - k } M_{k,k+q}\left(t-t^{\prime}\right) \, \delta n_{k+q}\left(t^{\prime}\right) \,.
    \label{eq:nk_eom_linearized}
\end{split}
\end{align}
In the last line we have split the RHS into two terms, defining a ``diagonal'' response function $D_{k}(t-t^{\prime})$ that acts only on $\delta n_{k}(t^{\prime})$, and a ``matrix'' response function that includes contributions from other modes  $\delta n_{k+q}(t^{\prime})$. Notice that when defining the response functions we have included the term $q=0$, by extending the summand to its finite limit as $q$ goes to 0, in both the diagonal and the matrix function; one can check that they exactly cancel out. This inclusion allows to separate clearly the exponential decay from its deviations, as laid out below. The last equality in Eq.~(\ref{eq:nk_eom_linearized}) gives Eq.~(\ref{eq:nk_eom_response_functions}) of the main text.

\subsection{Diagonal response function and exponential decay}

Considering first the diagonal response function, we have explicitly:
\begin{multline}
\label{def:diagonal_response_function}
D_{k}\left(\tau\right) = 8 \sum_{q  \neq - k} \left|V_{3}\left(k,q\right)\right|^{2} \left( n_{q}^{\rm th} - n_{k+q}^{\rm th} \right) \, {\rm cos}\left[\left(\omega_{k+q}-\omega_{k}-\omega_{q}\right) \tau \right] \\
+ 8 \sum_{ q \leq k/2  } \left|V_{3}\left(k-q,q\right)\right|^{2} \left( n_{q}^{\rm th} + n_{k-q}^{\rm th} + 1 \right) \, {\rm cos}\left[\left(\omega_{k}-\omega_{k-q}-\omega_{q}\right) \tau \right] \, ,
\end{multline}
where we have used the $q \to k - q$ symmetry in the second sum to fold it on $q \leq k/2$ adding a factor $2$. We want to show that $D_{k}\left(\tau\right)$ reduces to a Dirac delta, and for this purpose it is convenient to first consider its integral. We define
\begin{equation}
    I_{k}(\tau) = \int_{0}^{\tau} d\tau^{\prime} \, D_{k}\left(\tau^{\prime}\right) \qquad \iff \qquad D_{k}(\tau) = I_{k}^{\prime}(\tau) \, , \quad I_{k}(0) = 0 \,.
\end{equation}
Since $D_{k}(\tau)$ is an even function ($D_{k}(-\tau) = D_{k}(\tau)$), we find that $I_{k}(\tau)$ is odd:
\begin{equation}
    I_{k}(-\tau) = \int_{0}^{-\tau} d\tau^{\prime}\,D_{k}\left(\tau^{\prime}\right) = \int_{0}^{\tau} d\left(-\tau^{\prime}\right)\,D_{k}\left(-\tau^{\prime}\right) = -\int_{0}^{\tau} d\tau^{\prime}\,D_{k}\left(\tau^{\prime}\right) = -I_{k}(\tau) \,.
\end{equation}
Explicitly, we note that $\int_{0}^{\tau} {\rm cos}\left(\Omega \tau^{\prime}\right) \, d\tau^{\prime} = \tau \, {\rm sinc}\left(\Omega \tau\right)$.  Therefore,
\begin{multline}
\label{eq:Ik_sinc_discrete}
    I_{k}\left(\tau\right) = 8 \sum_{q \neq - k} \left|V_{3}(k,q)\right|^{2} \left( n_{q}^{\rm th} - n_{k+q}^{\rm th} \right) \, \tau \, {\rm sinc}\left[\left(\omega_{k+q}-\omega_{k}-\omega_{q}\right) \tau \right] \\
    + 8 \sum_{ q \leq k/2  } \left|V_{3}(k-q,q)\right|^{2} \left(n_{q}^{\rm th} + n_{k-q}^{\rm th} + 1\right) \, \tau \, {\rm sinc}\left[\left(\omega_{k}-\omega_{k-q}-\omega_{q}\right) \tau \right] \,.
\end{multline}
It is also useful to take the continuous (large $L$) limit, replacing $\frac{1}{\Delta q} \sum_{q} \Delta q \to \frac{L}{2\pi} \int dq$
\begin{multline}
\label{eq:Ik_sinc}
    I_{k}(\tau)= \frac{4 L}{\pi} \int_{-\infty}^{+\infty} dq \, \left|V_{3}(k,q)\right|^{2} \left( n_{q}^{\rm th} - n_{k+q}^{\rm th} \right) \, \tau \, {\rm sinc}\left[\left(\omega_{k+q}-\omega_{k}-\omega_{q}\right) \tau \right] \\
    + \frac{4 L}{\pi} \int_{-\infty}^{k/2} dq \, \left|V_{3}(k-q,q)\right|^{2} \left(n_{q}^{\rm th} + n_{k-q}^{\rm th} + 1\right) \, \tau \, {\rm sinc}\left[\left(\omega_{k}-\omega_{k-q}-\omega_{q}\right) \tau \right] \,.
\end{multline}
Considering the large $\tau$ limit of Eq.~(\ref{eq:Ik_sinc}), the sinc functions in the integrand become highly peaked around $q=0$, and to a good approximation can be replaced by Dirac deltas proportional to $\delta \left( q \right)$.  However, the limit must be taken with care since the general term has a discontinuity as $q$ goes to $0$. We split the integrals for negative and positive momenta and in the limit of large $\tau$ we have $\int_0^{\epsilon} f(x) \tau \, {\rm sinc}(\tau \, \delta \omega_{L/B} (q))  \mathrm{d}q  \xrightarrow{} f(0^+) \int_0^{\infty} \tau \, {\rm sinc}(\tau \, \delta\omega_{L/B}^{\prime}\left(0^{+}\right) \, q)  \mathrm{d}q = \frac{\pi}{2} \, f(0^+) / \left|\delta\omega_{L/B}^{\prime}\left(0^{+}\right)\right| $, and similarly on the other side.  
Explicitly, for $\tau \to \infty$
\begin{align}
\begin{split}
\label{eq:sinc_to_delta}
& \int_{-\epsilon}^{\epsilon}  \left| V_{3}(k,q) \right|^{2} \left( n^{\rm th}_q - n^{\rm th}_{k+q} \right) \, \tau \, {\rm sinc} \left[ \delta \omega_L (q) \tau \right] \, d q = \frac{\pi}{2} \left( \lim_{q \xrightarrow{} 0^-} + \lim_{q \xrightarrow{} 0^+} \right) \left| V_{3}(k,q) \right|^{2} \frac{n^{\rm th}_q}{ \left| \delta \omega_L^{\prime} \left( q \right) \right| }  \, , \\
& \int_{-\epsilon}^{\epsilon} \left| V_{3}(k-q,q) \right|^{2}  \left(n^{\rm th}_{k-q}+n^{\rm th}_q+1\right)  \, \tau \, {\rm sinc}\left[ \delta \omega_B (q) \tau \right] \, d q  =  \frac{ \pi }{2} \left( \lim_{q \xrightarrow{} 0^-} + \lim_{q \xrightarrow{} 0^+} \right) \left| V_{3}(k-q,q) \right|^{2} \frac{n^{\rm th}_q}{ \left| \delta \omega_B^{\prime} \left( q \right) \right| } \, .
\end{split}
\end{align}
In both integrals, in the limit $q$ to $0$, $n^{\rm th}_{k \pm q}$ and $1$ are negligible compared to $n^{\rm th}_q$ that diverges as $1/q$. In addition, we have $\left|\delta\omega_{L/B}^{\prime}\left(0^{\pm}\right)\right| = \left| v_{\rm gr}(k) \mp c \right|$. Combining the above two limits we get the formula (\ref{def:decay_rate}) for the decay rate of phonon of momentum $k$. Therefore
\begin{equation}
   I_{k}(\tau) = \int_{0}^{\tau} D_{k}\left(\tau^{\prime}\right) \, d\tau^{\prime} \to  \pm \Gamma_{k} \qquad {\rm as} \qquad \tau \to \pm \infty \, ,
   \label{eq:Ik_asymptotic}
\end{equation}
where the opposite limit is obtained by anti-symmetry. $I_{k}$ is asymptotically constant in both directions and only varies in the vicinity of $\tau= 0$. As a consequence $D_k$ is peaked around $\tau =0$. We now assume that $D_k$ is sufficiently peaked compared to the variation of $\delta n_k$ so that the integral in Eq.~(\ref{eq:nk_eom_linearized}) only picks out its instantaneous value $\delta n_k(t)$
\begin{align}
\begin{split}
\int_{0}^{t} dt^{\prime} \, D_{k}\left(t-t^{\prime}\right) \, \delta n_{k}\left(t^{\prime}\right) & \approx \delta n_{k}\left(t \right) \int_{0}^{t} dt^{\prime} \, D_{k}\left(t-t^{\prime}\right) \, , \\
& = \delta n_{k}\left(t \right) I_{k}(t) \, .
\end{split}
\end{align}
The equation of motion becomes
\begin{align}
\begin{split}
\partial_{t} \delta n_{k} & = - \delta n_{k}\left(t \right) I_{k}(t) + \int_{0}^{t} dt^{\prime} \, \sum_{q \neq - k } M_{k,k+q}\left(t-t^{\prime}\right) \, \delta n_{k+q}\left(t^{\prime}\right)  \,.
\label{eq:nk_eom_linearized_diagonal_markovian}
\end{split}
\end{align}

Let us now consider the ideal situation where at initial time every mode but a single mode $k$ is exactly thermal: $n_q(0) = n^{\rm th}_q + \delta n \, \delta_{k,q}$ where $\delta n $ is the number of phonons added in mode $k$ on top of the quasicondensate. This corresponds to the situation analyzed in Sec.~\ref{subsec:initial_injection}. Then we may neglect the matrix response function (since $\delta n_q \ll \delta n_k$):
\begin{align}
\begin{split}
\partial_{t} \delta n_{k} & = - \delta n_{k}\left(t \right) I_{k}(t)  \,.
\label{eq:nk_eom_linearized_only_diagonal}
\end{split}
\end{align}
Finally we take a large time limit to have $I_{k}(t) \to  \pm \Gamma_{k}$ so that in this limit $n_k$ obeys 
\begin{equation}
    \partial_{t} \delta n_{k} = - \delta n_{k}\left(t \right) \Gamma_{k} \, ,
    \label{eq:nk_eom_exponential}
\end{equation}
{\it i.e.}, the population of the mode $k$ decays exponentially at the rate predicted by Eq.~(\ref{def:decay_rate}).

\subsection{Matrix response function: the slowing effect of a finite width}

It remains to give the form of $M_{k,k+q}$ and describe its effects. It reads:
\begin{align}
\begin{split}
\label{def:matrix_response_function}
M_{k,k+q}\left(t-t^{\prime}\right) & = 8 \left|V_{3}\left(k,q\right)\right|^{2} \left( n_{q}^{\rm th} + n_{k}^{\rm th} + 1 \right) \, {\rm cos}\left[\left(\omega_{k+q}-\omega_{k}-\omega_{q}\right)\left(t-t^{\prime}\right)\right] \\ 
& + 8 \left|V_{3}\left(k+q,-q\right)\right|^{2} \left( n_{q}^{\rm th} - n_{k}^{\rm th} \right) \, {\rm cos}\left[\left(\omega_{k}-\omega_{k+q}-\omega_{q}\right)\left(t-t^{\prime}\right)\right] \\
& + 8 \left|V_{3}\left(k,k+q\right)\right|^{2} \left( n_{2k+q}^{\rm th} - n_{k}^{\rm th} \right) \, {\rm cos}\left[\left(\omega_{2k+q}-\omega_{k}-\omega_{k+q}\right)\left(t-t^{\prime}\right)\right] \mathds{1}_{q \neq -2k}
\end{split}
\end{align}
Firstly, a couple of technical remarks. Notice that there is an indicator function in the last term, stating that this piece should not be evaluated at $q = -2k$.  Also, the $q=0$ term has been included to compensate for the inclusion of the opposite term in $D_{k}\left(t-t^{\prime}\right)$. Therefore, even in the case where only a single mode is significantly occupied, the matrix response function can never be completely neglected and \eqref{eq:nk_eom_exponential} has to be amended.  This first limitation is dealt with in Appendix~\ref{app:deviations} below.

$M_{k,k+q}$ represents the indirect interaction of phonons in the modes $k$ and $k+q$ through the thermal population. Equation~\eqref{def:matrix_response_function} shows that there are three such processes. The first term corresponds to the conversion between phonons of wavevectors $k$ and $q$ and that of wavevector $k+q$, which we write symbolically as $\left(k \,,\, q\right) \leftrightarrow k+q$. Similarly, the second term encodes processes of the form $\left(k+q \,,\, -q\right) \leftrightarrow k$, and the third term $\left(k+q \,,\, k\right) \leftrightarrow 2k+q$. Even though the perturbation is initially localised in the mode $k$, the decay process will generate a non-zero $\delta n_q$ in the vicinity of both $q=0$ and $q=k$. We shall assume that the perturbation spectrum is relatively narrow around these two points. Equation~\eqref{eq:nk_eom_linearized_diagonal_markovian} then implies that we need only consider the values of $M_{k,k+q}$ for $k + q \approx k$ and $k + q\approx 0$. 

Consider first $k + q \approx 0$.  In the second and third terms of~(\ref{def:matrix_response_function}), the factor of $\left| V_3 \right|^2$ and the combination of thermal populations independently vanish in the limit $k+q \to 0$, making these terms negligible. In the first term, only the factor of $\left| V_3 \right|^2$ tends to zero, but the rapid oscillations of the cosine function at frequencies close to $2\omega_{k}$ will greatly suppress its contribution to the integral of \eqref{eq:nk_eom_linearized_diagonal_markovian}. Therefore, the back-reaction from the decay products at very low momenta is expected to be negligible. 

On the other hand, for $k + q \approx k$ only the contribution of the third term in~(\ref{def:matrix_response_function}) is expected to be negligible, for the frequency of the cosine function is large (roughly $\omega_{2k} - 2\omega_{k}$) and suppresses its contribution. We are then left with contributions coming from the first and second terms.  Their frequency differences are small, preventing any averaging out due to rapid oscillations, and can be written to first order in $q$. Their coefficients have non-vanishing limit since the vanishing of $\left| V_3 \right|^2$ is compensated by the divergence of the thermal population, and can be approximated by their low-$q$ limits. Finally, we get~\footnote{Notice that we dropped the terms $n^{\rm th}_{k}$ and $+1$ compared to $n^{\rm th}_{q}$ which is divergent as $q \to 0$. However, if we work in the discrete setting and consider a small $k \sim j 2 \pi / L$, then one could argue that $q$ can never be less than $2 \pi /L$ so that $n^{\rm th}_{q}$ is always finite and of the order of $n^{\rm th}_{k}$. Still, in the limit $k \approx 0$, these terms come with opposite signs and practically cancel out unlike the ones in $n^{\rm th}_{q}$.}
\begin{multline}
    M_{k,k+q}\left(t-t^{\prime}\right) \approx \frac{8}{N_{\rm at}} 
    \, {\rm lim}_{q\to 0^{+}} \left[ \left|V_{3}(k,q)\right|^{2} n_{q}^{\rm th} \right] {\rm cos}\left[q \left(v_{\rm gr}(k)-c\right) \left(t-t^{\prime}\right) \right] \\
    + \frac{8}{N_{\rm at}} 
    \, {\rm lim}_{q\to 0^{-}} \left[ \left|V_{3}(k,q)\right|^{2} n_{q}^{\rm th} \right] {\rm cos}\left[q \left(v_{\rm gr}(k)+c\right) \left(t-t^{\prime}\right) \right] \,.
\label{eq:simplified_non_diagonal_response}
\end{multline}

Equation~(\ref{eq:simplified_non_diagonal_response}) will be used to compute the correction to the FGR prediction of the decay rate due to the finite width of the peak, {\it i.e.}, to the back-reaction of neighbouring modes.

\section{Deviations from FGR result} 
\label{app:deviations}

We turn in this appendix to the deviations with respect to the simple exponential decay predicted by the FGR.  Each of these deviations is described by equation of motion~(\ref{eq:nk_eom_linearized}).

\subsection{Non-exponential decay of IR phonons}
\label{subsubsec:non_exp_decay}

\subsubsection{Numerical observations}

\begin{figure}
    \centering
    \includegraphics[width=0.45\textwidth]{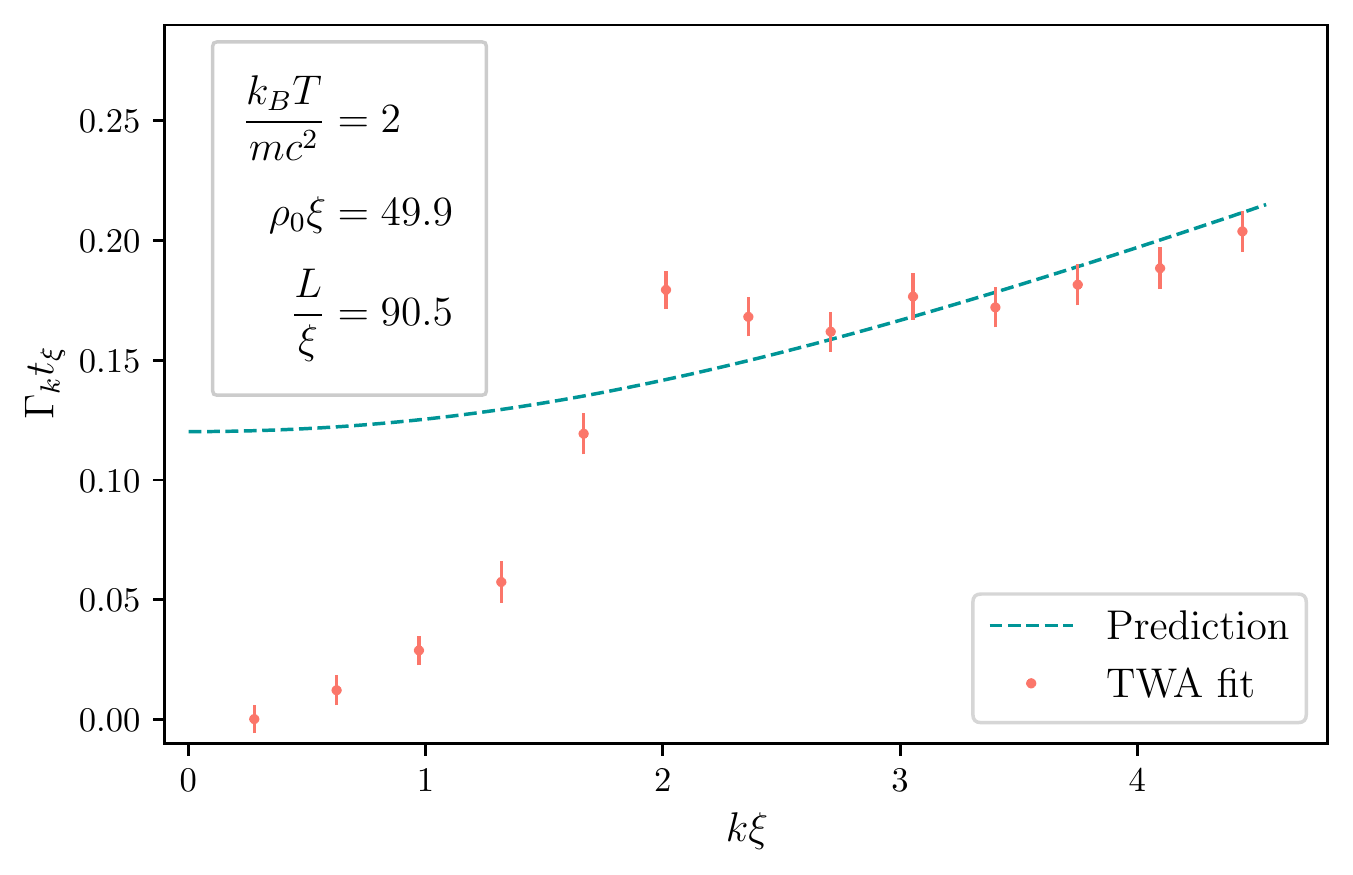}
    \caption{Best fit values for $\hbar \Gamma_k$ extracted from the TWA simulations as a function of $k \xi$. The window shown here is larger than that of Fig.~\ref{fig:scaling_Gamma_k_xi} while all physical parameters are fixed at the same values.
    \label{fig:scaling_Gamma_k_xi_small_and_large}}
\end{figure}

It was shown in the main text that, in the first set of simulations where an initial probe is simply injected into a single phonon mode, the numerically observed decay of modes with a large enough momentum ($k\xi \geq 2.5$) is very well described by an exponential decay at a rate given by Eq.~(\ref{def:decay_rate}). In Fig.~\ref{fig:scaling_Gamma_k_xi_small_and_large}, a more complete set of results is shown, in which the template $\delta n_{k}(t) = A {\rm exp}\left(-\Gamma t + \gamma t^{2}/2\right)$ is fitted to the behavior of a larger set of initial momenta.  This figure demonstrates that there are significant deviations at lower $k\xi$, with the values of $\Gamma_{k}$ extracted from the simulations going to zero as $k\xi \to 0$ instead of the predicted finite limit. These deviations are due to the fact that the FGR result is only valid within a certain time window~\cite{CohenTannoudji2020,Peres1980}.

We review the last approximation made in the passage from Eq.~(\ref{eq:nk_eom_linearized_only_diagonal}) to Eq.~(\ref{eq:nk_eom_exponential}), which is that $t$ be large enough for $I_k(t)$ to be equal to its asymptotic value, or equivalently (as shown in Eqs.~(\ref{eq:sinc_to_delta})-(\ref{eq:Ik_asymptotic})), being able to replace $t \, {\rm sinc}\left(\delta\omega_{L/B}t\right)/\pi$ by a $\delta$ function.
This requires that $\delta\omega_{L/B} \left( k, q \right) t$, which vanishes for the elastic scattering channel $q = 0$, should nevertheless reach a large enough value in its vicinity so that the most significant part of the sinc is squeezed into the region of constant effective interaction strength. 
We may introduce a critical response time, $t_{\rm crit}$, which marks the time after which this condition is satisfied.  This critical time depends on $k$, so that, for a fixed time window $t/ t_{\xi} \in \left[ 0, 10 \right] $ over which the fit is performed, those $k$ for which $t_{\rm crit}/t_{\xi} \ll 10$ will be in the FGR regime, while those $k$ for which $t_{\rm crit}/t_{\xi} \gg 10$ will be in an early-time regime long before the FGR can be applied.

\begin{figure}
\centering
\includegraphics[width=0.45\textwidth]{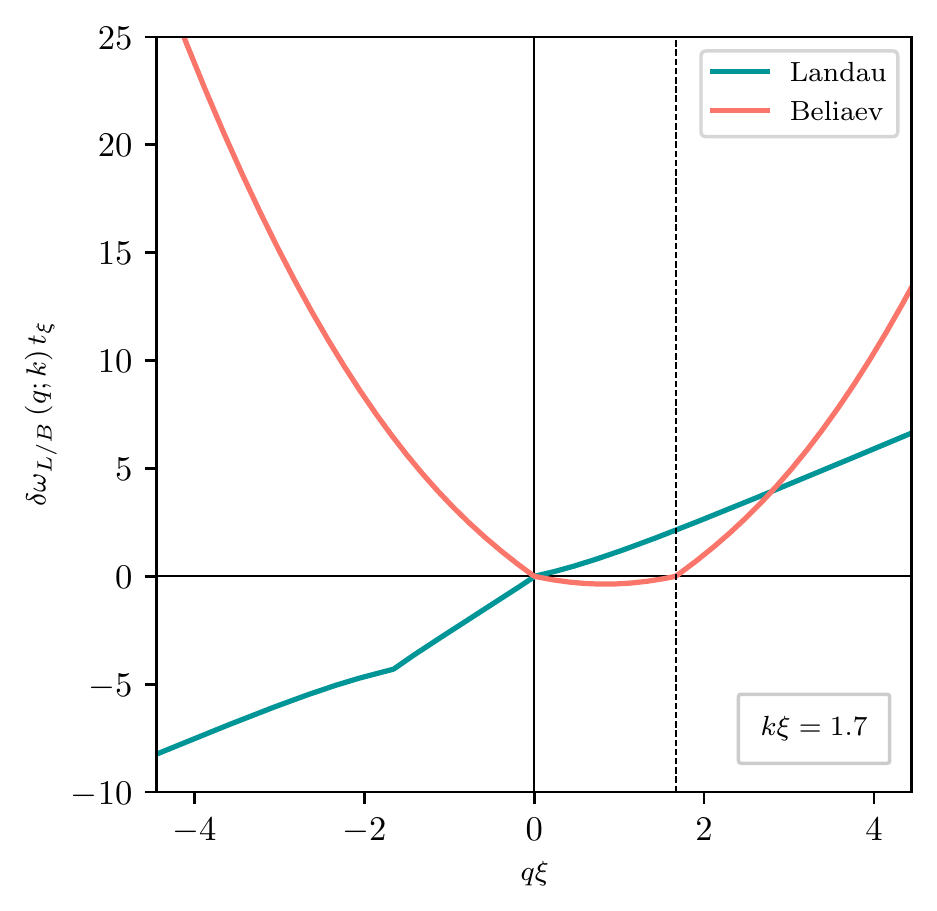}
\includegraphics[width=0.45\textwidth]{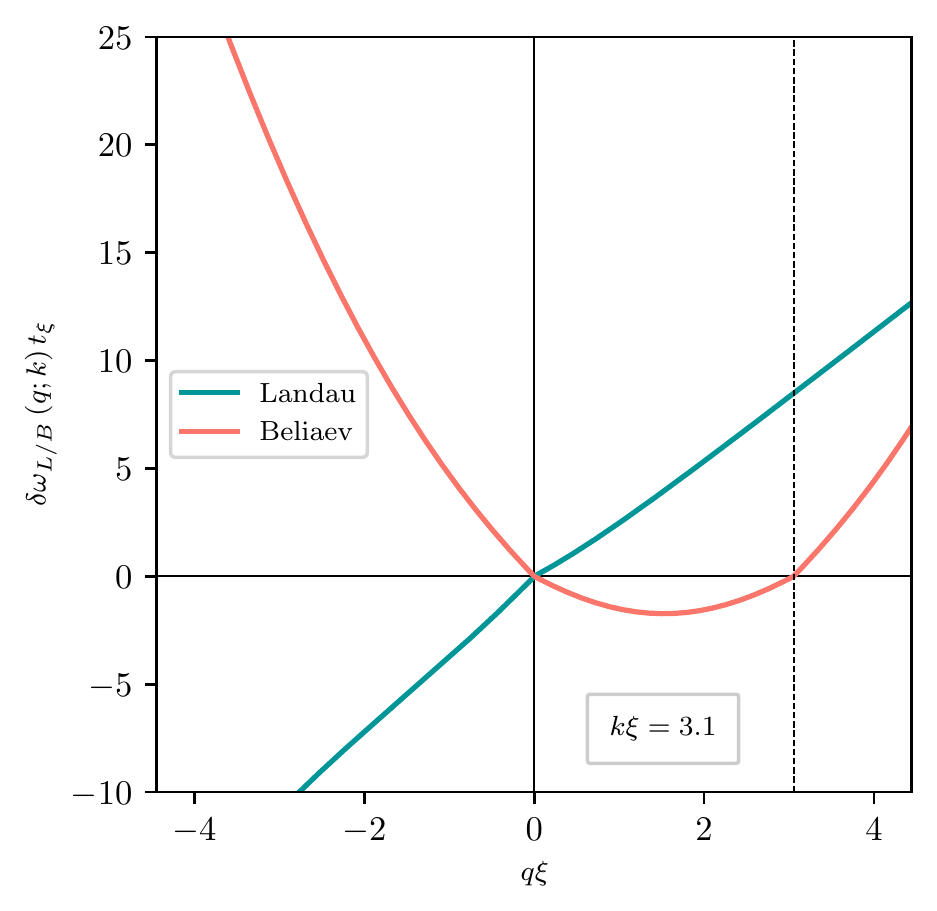}
\caption{Plots of the Rabi frequencies associated to Landau and Beliaev damping  processes $\delta \omega_{L,B} \left( q ; k \right)$ as a function of $q$ for $k \xi = 1.7$ (left.) and $k \xi = 3.1$ (right.). The value of $k$ is shown by the vertical dashed line.}
\label{fig:rabi_frequencies} 
\end{figure}

Examples of $\delta \omega_{L/B} (q,k) t_{\xi}$ are plotted in Fig.~\ref{fig:rabi_frequencies}, for $k\xi = 1.7$ and $3.1$. The key observation is that $\left|\delta\omega_{B}\right|$ is the most significantly constrained, especially in the window $q \in \left[0,k\right]$ where the decay processes are expected to be most efficient (as the effective interaction strength is largest there).  It is this restriction on $\left|\delta\omega_{B}\right|$ that is most clearly responsible for the failure of the approximation. $\left| \delta \omega_{B} \right|$ reaches a maximum $\delta\omega_{B}^{\rm (max)} = \omega_{k} - 2 \omega_{k/2}$ at $q=k/2$.  At lowest order in $k\xi$, we have $\delta\omega_{B}^{\rm (max)} t_{\xi} \approx \frac{3}{32} \left(k\xi\right)^{3}$. 
The critical response time can be (somewhat arbitrarily) defined via $\delta\omega_{B}^{\rm (max)} t_{\rm crit} = 2\pi$, but the key point is that it increases quickly at low momentum: $t_{\rm crit}/t_{\xi} \propto \left(k\xi\right)^{-3}$.
Since the fitting window of Fig.~\ref{fig:scaling_Gamma_k_xi_small_and_large} extends only up to $t/t_{\xi} = 10$, we require at least $10 \, \delta\omega_{B}^{\rm (max)} t_{\xi} \approx \left(k\xi\right)^{3} \geq 2\pi$, which imposes $k\xi \geq 2$.   This estimate is in close correspondence with the onset of deviations seen for $k\xi \sim 2$ in Fig.~\ref{fig:scaling_Gamma_k_xi_small_and_large}. Moreover, as the total duration of the simulation is varied, the estimated threshold for $k\xi$ will vary only slowly, as $\left(t/t_{\xi}\right)^{-1/3}$. 

At lower $k\xi$, we are well outside the validity regime of the FGR, in an early-time regime where a $t^{2}$ behavior is expected~\cite{Peres1980} (see below). Furthermore, for very low-lying modes $k \sim 2 \pi / L$, the range $q \in \left[0 \,,\, k \right]$ will be  poorly sampled.  We thus expect a strong suppression of the Beliaev component of the decay as $k\xi \to 0$ due to the small number of available modes to decay to, in accordance with similar remarks made in \cite{Fedichev_1998, Pitaevskii_1997}.  While such low-lying $k$ mode can still decay via a Landau process we expect that in this regime the system becomes sensitive to the discreteness of excitations and this process would thus require a separate analysis (see, {\it e.g.}, Ref.~\cite{bayocbocFrequencyBeating2022}).

\subsubsection{Approximate analytical description of behavior}

\begin{figure}
     \centering
    
     \begin{minipage}{0.45\textwidth}
        \centering
        \includegraphics[width=\textwidth]{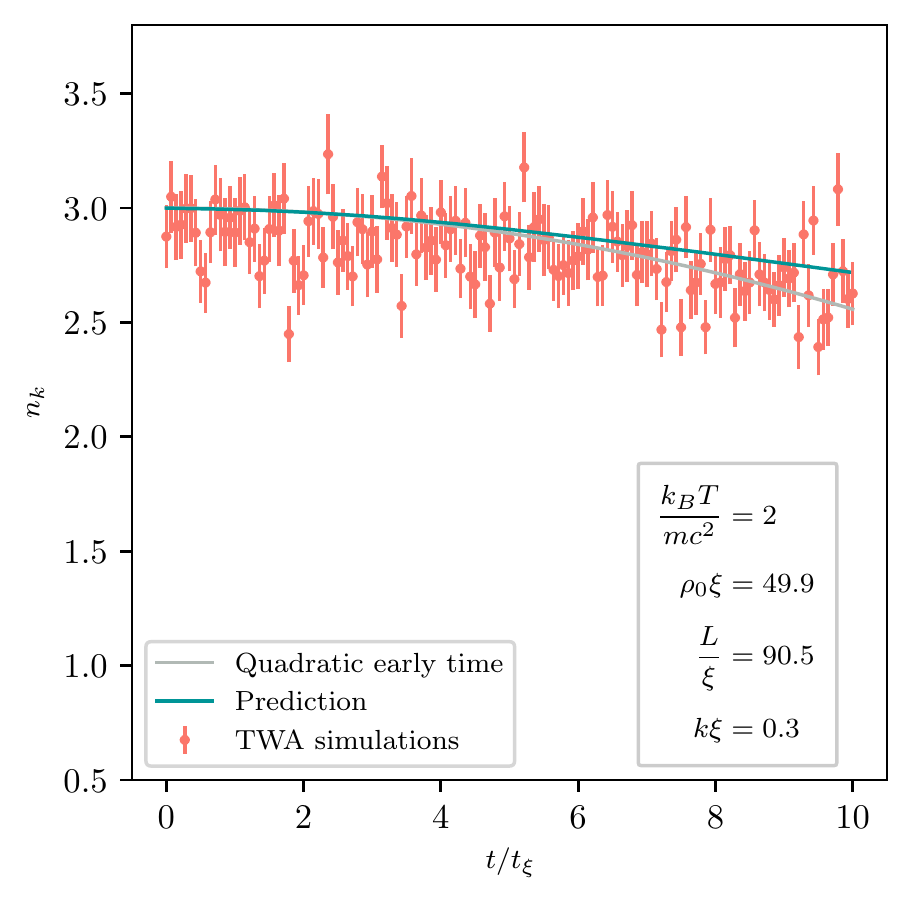}
    \end{minipage} \hfill
          \begin{minipage}{0.45\textwidth}
        \centering
         \includegraphics[width=\textwidth]{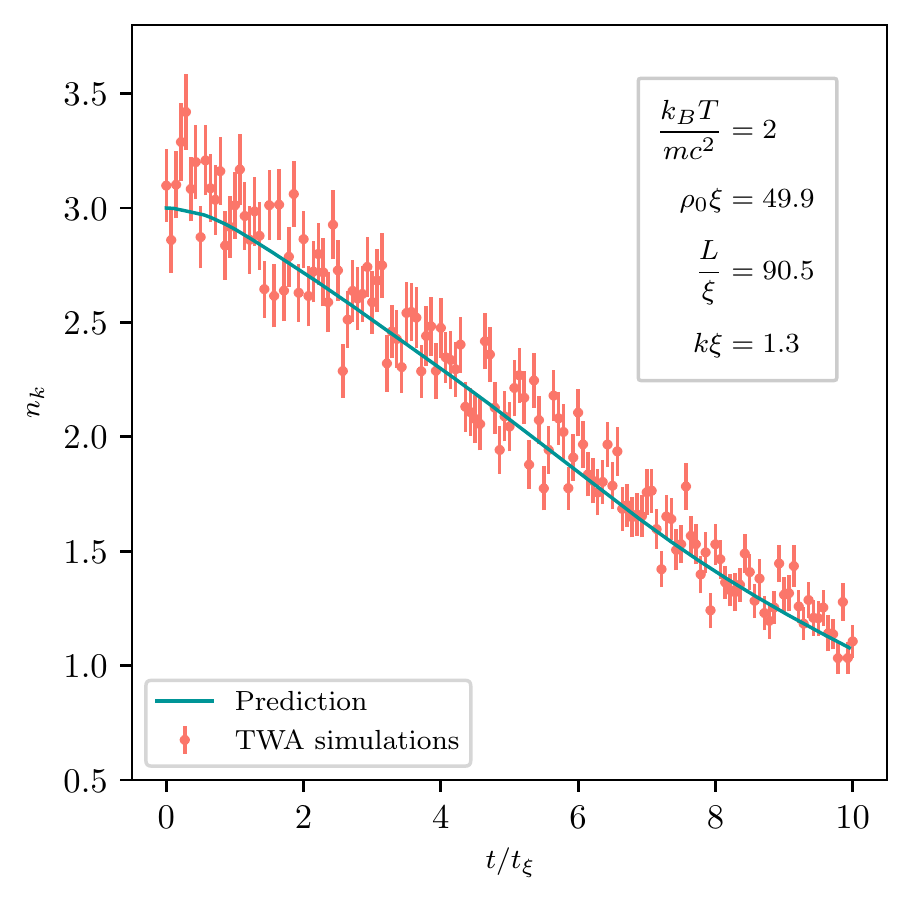}
  
    \end{minipage}

     \begin{minipage}{0.45\textwidth}
        \centering
        \includegraphics[width=\textwidth]{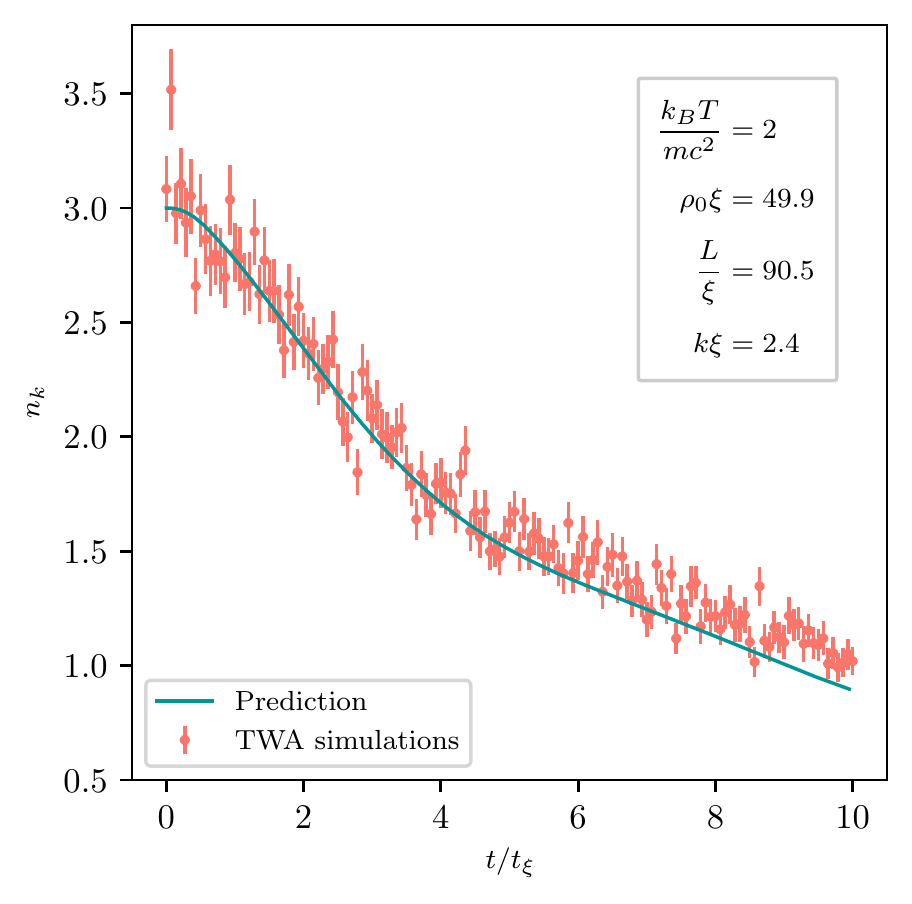}
    \end{minipage} \hfill
          \begin{minipage}{0.45\textwidth}
        \centering
         \includegraphics[width=\textwidth]{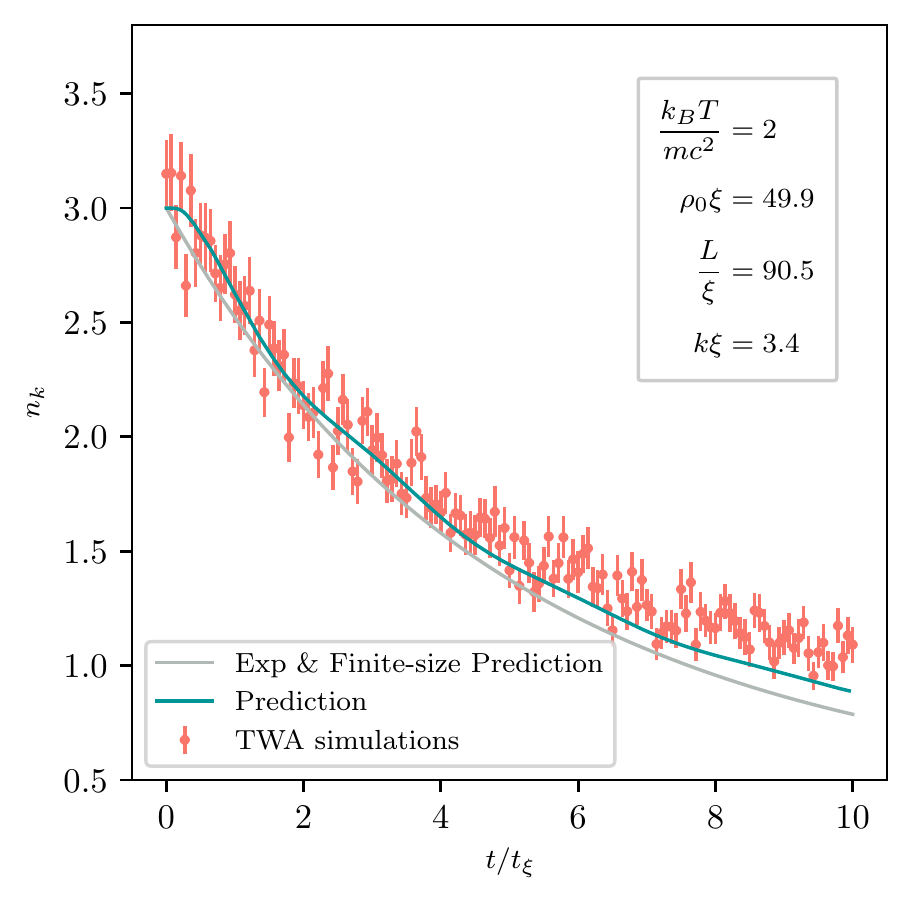}
    \end{minipage}

    \caption{Number of phonons $n_k$ in the mode $k \xi = 0.3$ (top left), $k \xi = 1.3$ (top right), $k \xi = 2.4$ (bottom left), $k \xi = 3.4$ (bottom left) as a function of time $t /t_{\xi}$ for $k_{B} T/ m c^2=2$, $\rho_0 \xi = 49.9$, $L/\xi = 90.5$ and $n_r = 400$ realisations. Each plots is comprised of $n_{t}=140$ points. The red dots correspond to the result of the TWA simulations. The green line is obtained by taking the difference of the prediction of Eq.~(\ref{eq:nkEvolutionDiscreteSinc}) for the background thermal population and the full population. The gray line in the top left panel corresponds to the prediction for an early-time quadratic decay given by Eq.~(\ref{eq:low_k_early_times}), while in the bottom right panel it corresponds to the prediction of an exponential decay at the rate given by Eq.~(\ref{def:decay_rate}) of the text with corrections given by Eq.~(\ref{eq:nk_finite_size_correction}).}
    \label{fig:different_n_k} 
\end{figure}

We expect that these low-$k$ modes can still be well described by Eq.~(\ref{eq:nk_eom_full_non_markovian}).  However, we shall adopt an early-time approximation, assuming that $N_{p,q}(t)$ varies sufficiently slowly for it to be taken out of the time-integral as an overall prefactor. We get
\begin{align}
\begin{split}
\label{eq:nkEvolutionDiscreteSinc}
& \partial_{t}n_{k} = 8  \sum_{q \neq 0, -k} \left\{ \frac{\left| V_{3}(k,q) \right|^{2}}{N_{\rm at}}  \left[ n_{k+q}\left(n_{k}+n_{q}+1\right) - n_{k} n_{q} \right] \, t \, {\rm sinc} \left[ \delta\omega_{L}(q;k) t\right] \right\} \\
    &  -  8  \sum_{ 0 < q \leq k/2 } \left\{ \frac{\left| V_{3}(k-q,q) \right|^{2}}{N_{\rm at}}  \left[ n_{k} \left(n_{k-q} + n_{q} + 1\right) - n_{k-q} n_{q} \right] \, t \, {\rm sinc}\left[ \delta\omega_{B}(q;k) t\right] \right\} \, ,
\end{split}
\end{align}
where we have used the symmetry $k \to k-q$ in the second term to sum over only half the momenta, compensating with the inclusion of a factor $2$. This equation is then numerically solved twice: once using only the thermal population as an initial state, and a second time using the thermal population plus the $\delta n$ phonons added in the mode $k$. We then take the difference to obtain the green curves in Fig.~\ref{fig:different_n_k}. This procedure correct for small variations of the background thermal population. A good agreement is found with the TWA simulations even for large values of $k \xi$. 

Let us try to get an analytical estimate for the behavior of the low-lying modes. Their critical time being very large, it is appropriate to consider the early-time limit of Eq.~(\ref{eq:nk_eom_linearized_only_diagonal}) where $\delta n_k(t) \approx \delta n_k(0)$ and $ {\rm sinc}\left(\delta \omega \, \tau \right) \sim 1$, therefore $I_{k} \left( \tau \right) = \alpha_k \tau $ for a certain constant $\alpha_k$. This gives
\begin{equation}
\label{eq:low_k_early_times}
\delta n_k (t) = \delta n_k(0) \left(1 - \alpha_k t^2 /2 \right) \, , 
\end{equation}
where we now have to calculate $\alpha_k$. 

We cannot simultaneously set all the sinc functions in Eq.~(\ref{eq:Ik_sinc}) to $1$, as the $+1$ term in the second sum would then lead to a divergence. Indeed, for large $q$ the frequency difference diverges and the sinc decay accordingly quickly. We typically consider the evolution of the system over time scales of the order of $t_{\xi}$. Considering $k \xi \sim 0.3$, we want to sort every mode $q$ in two categories. Either $\delta \omega_{L/B} (q,k) t_{\xi} \ll 1$, then this mode can be considered to experience an early time behaviour and the related sinc can be set to $1$, or $\delta \omega_{L/B} (q,k) t_{\xi} \gg 1$ then  the sinc should be set to $0$. We simply exclude the latter modes from the sum. 

On the one hand, as noted above for the Beliaev-type channels it is clear from Fig.~\ref{fig:rabi_frequencies} that the modes $q \in \left[0 , k \right]$  oscillate at frequencies smaller than the others and should be the one kept in the sum. On the other hand there is no clear separation for the Landau-type scatterings. However, the terms associated to $\delta \omega_L$ in Eq.~(\ref{eq:Ik_sinc}) are already suppressed at large $q$ by an exponentially decaying thermal population $n_{q}$ in factor. The inclusion of them in the sum should then be irrelevant for the resulting numerical prediction and we include all the modes in the first sum. We then get:
\begin{align}
\begin{split}
\label{def:prediction_alpha_t_square}
\alpha_k & = \frac{ 8  }{N_{\rm at}} \left\{ \sum_{  q \neq 0, -k  }  \left| V_{3}(k,q) \right|^{2} \left( n_{q}^{\rm th} - n^{\rm th}_{k+q}  \right)    + \sum_{ 0 < q \leq k/2  } \left| V_{3}(k-q,q) \right|^{2} \left(n_{k-q}^{\rm th} + n_{q}^{\rm th} + 1 \right) \right\} \, . 
\end{split}
\end{align}

\begin{figure}
    \centering
    \includegraphics[width=0.45\textwidth]{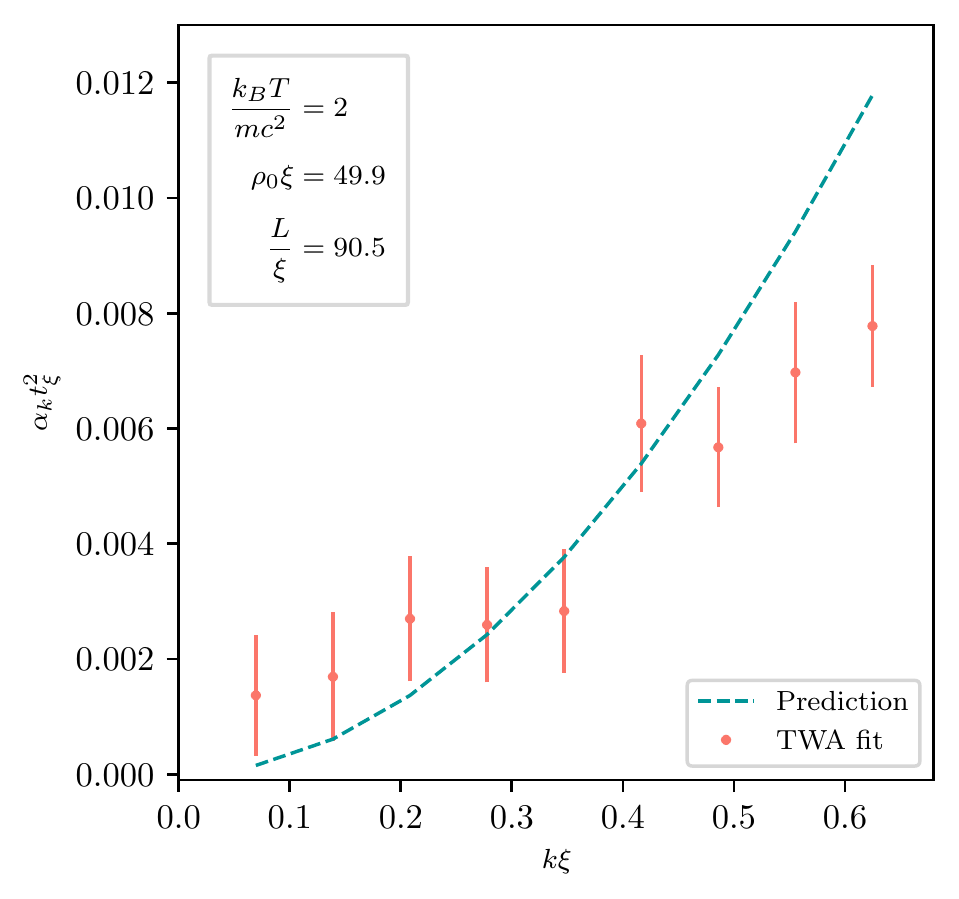}
    \caption{Red dots are the best fit values for $\alpha_k t_{\xi}^2$  extracted from the TWA simulations using a template of the form of Eq.~(\ref{eq:low_k_early_times}). The green dashed line is the prediction for $\alpha_k t_{\xi}^2$ of equation Eq.~(\ref{def:prediction_alpha_t_square}).
    The simulation parameters are $k_{B} T/ m c^2=2$, $\rho_0 \xi = 49.9$ and $L/\xi = 90.5$ with $n_r = 400$. The fits are performed over a time-window $t / t_{\xi} \in \left[ 0, 5\right]$, which is half the time-window used in for the fits of $\Gamma_{k}$ in Figs.~\ref{fig:scaling_Gamma_asap_T} and \ref{fig:scaling_Gamma_k_xi}, but with the same number of points $n_{t}=140$.
    \label{fig:scaling_alpha_k} }
\end{figure}

In Fig.~\ref{fig:scaling_alpha_k} we compare this prediction with the best-fit value of $\alpha$ for the template $A\left( 1 - \alpha t^2 / 2\right)$ applied to the TWA simulations with $A$ and $\alpha$ as fitting parameters. We considered the smallest values of $k$ of Fig.~\ref{fig:scaling_Gamma_k_xi} in the text and used the same time-window $t / t_{\xi} \in \left[ 0, 10\right]$. The agreement is good for very small $k$ and deviations appear around $k \xi = 0.5$. This can be understood using Fig.~\ref{fig:different_n_k}. For $k \xi = 0.3$ the decay is satisfyingly described by the quadratic prediction Eq.~(\ref{def:prediction_alpha_t_square}), while for $k \xi = 3.4$ it is well described by the exponential prediction of Eq.~(\ref{def:decay_rate}) plus the finite size-correction examined in the next section. On the other hand for the intermediate values $k \xi = 1.4$ and $k \xi = 2.4$, corresponding to the second and third panels, the decay from $t/ t_{\xi} = 0$ to $t/ t_{\xi} = 10$ is neither quadratic nor exponential all the way. $k \xi = 2.4$ is precisely the value around which our prediction of exponential decay seems to break down in Fig.~\ref{fig:scaling_Gamma_k_xi_small_and_large}.

\subsection{Finite-size effect}
\label{subsubsec:finite_size_effect}

\subsubsection{Description of effect}

\begin{figure}
    \centering
    \includegraphics[width=0.495\textwidth]{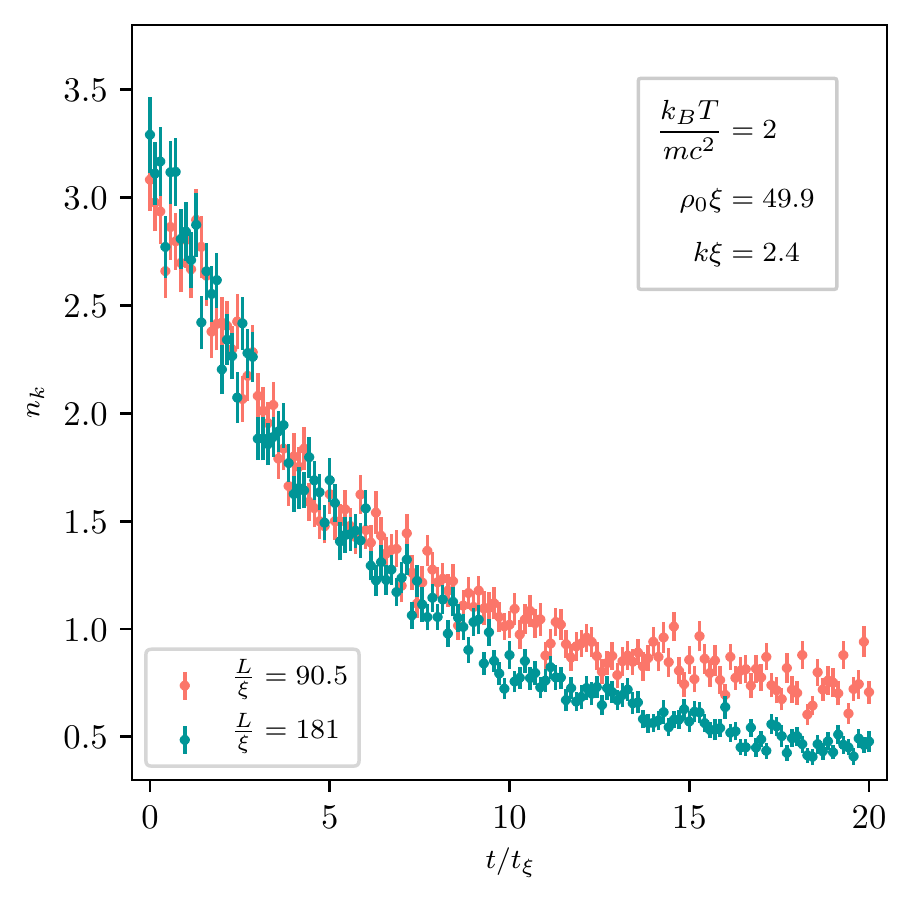}
    \includegraphics[width=0.495\textwidth]{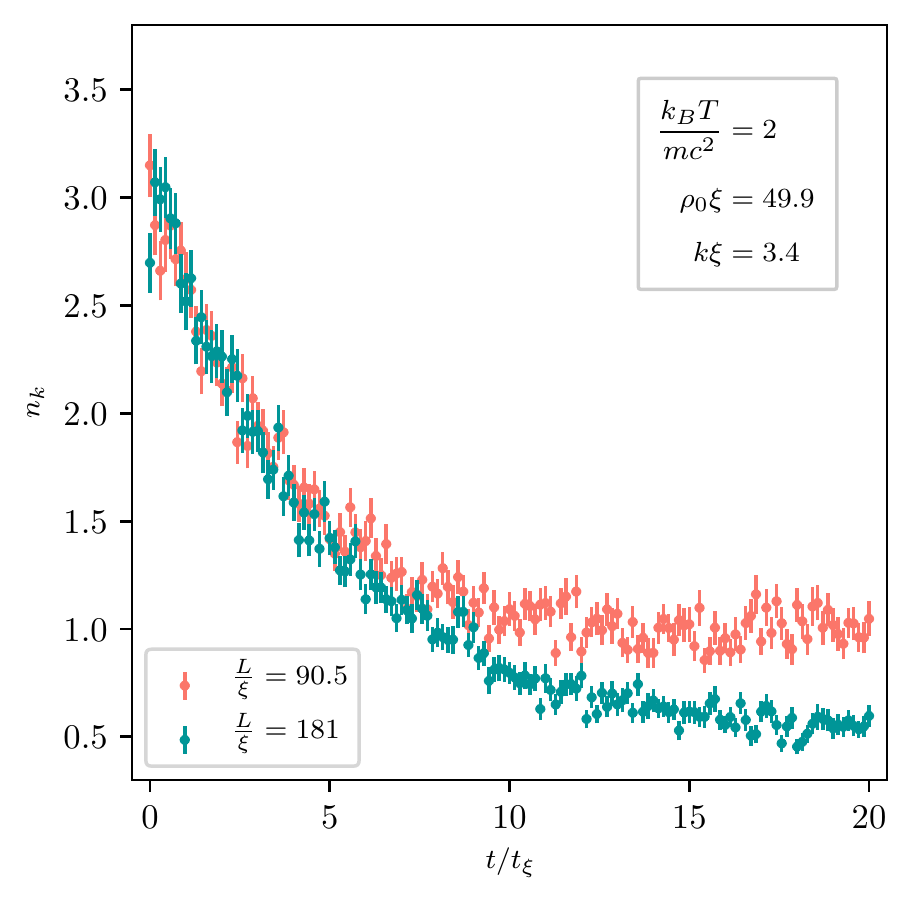}
    \caption{Number of phonons $n_k$ in the mode $k \xi = 2.4$ (left.), $k \xi = 3.1$ (right.) as a function of time $t /t_{\xi}$ for $k_{B} T/ m c^2=2$, $\rho_0 \xi = 49.9$ and $n_r = 400$ realisations. The red dots correspond to $L/\xi = 90.5$ and the green ones to $L/\xi = 181$.
    \label{fig:decay_nk_different_L} }
\end{figure}

The inequivalent curves in Fig.~\ref{fig:decay_nk_different_L} plotted for $L / \xi = 90.5$ and $L / \xi = 181$ demonstrate that the dynamics of $n_k$ in the TWA simulations is not completely insensitive to the size of the system $L$. Therefore, the $L$-independent exponential decay rate of Eq.~(\ref{def:decay_rate}) cannot fit exactly the result of the simulations. 

The key to understanding this effect is the $q=0$ contribution to the matrix part of the response function.  For a singularly occupied mode, this is the only term in the matrix part that plays any significant role.  Recall that it must be included simultaneously in both the diagonal and matrix parts, so that there is no net change in the total response function.  If these $q=0$ terms were not included, the effect would be related to the {\it absence} of the $q=0$ term in the {\it diagonal} response function.

What effect does this term have?  Considering the case where a certain number of phonons are injected at $t=0$, and working in the regime where the diagonal response function reduces to a Dirac delta, we have
\begin{eqnarray}
    \partial_{t}\left(\delta n_{k}\right) &=& -\Gamma_{k} \, \delta n_{k} + \int_{0}^{t} dt^{\prime} \, M_{k,k}\left(t-t^{\prime}\right) \, \delta n_{k}\left(t^{\prime}\right) \nonumber \\
    & = & -\Gamma_{k} \, \delta n_{k} + \gamma_{k} \int_{0}^{t} dt^{\prime} \, \delta n_{k}\left(t^{\prime}\right) \,,
    \label{eq:finite_size_deceleration}
\end{eqnarray}
where we have made it manifest that $\gamma_{k} \equiv M_{k,k}$ does not depend on $t-t^{\prime}$:
\begin{equation}
\gamma_k = 8 \left( \lim_{q \xrightarrow{} 0^{+}} + \lim_{q \xrightarrow{} 0^{-}} \right) \frac{\left| V_{3}(k,q) \right|^{2}}{ N_{\rm at}}   n^{\rm th}_q  \, .
    \label{eq:total_correction_decay_rate}
\end{equation}

At fixed density, $N_{\rm at}$ is proportional to $L$, so $\gamma_{k} \propto 1/L$.
Performing computations explicitly we get
Eq.~\eqref{def:prediction_small_gamma} of the main text.

\begin{figure}
    \centering
    \includegraphics[width=0.495\textwidth]{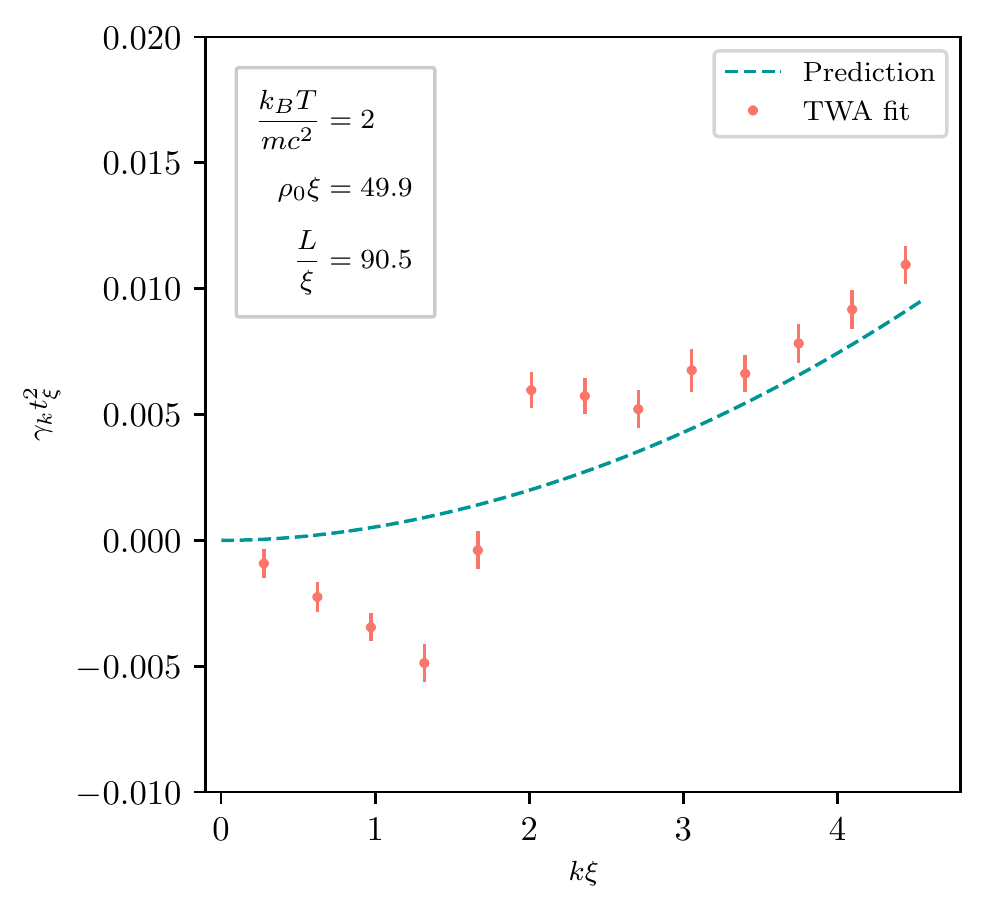}
    \hfill
    \includegraphics[width=0.495\textwidth]{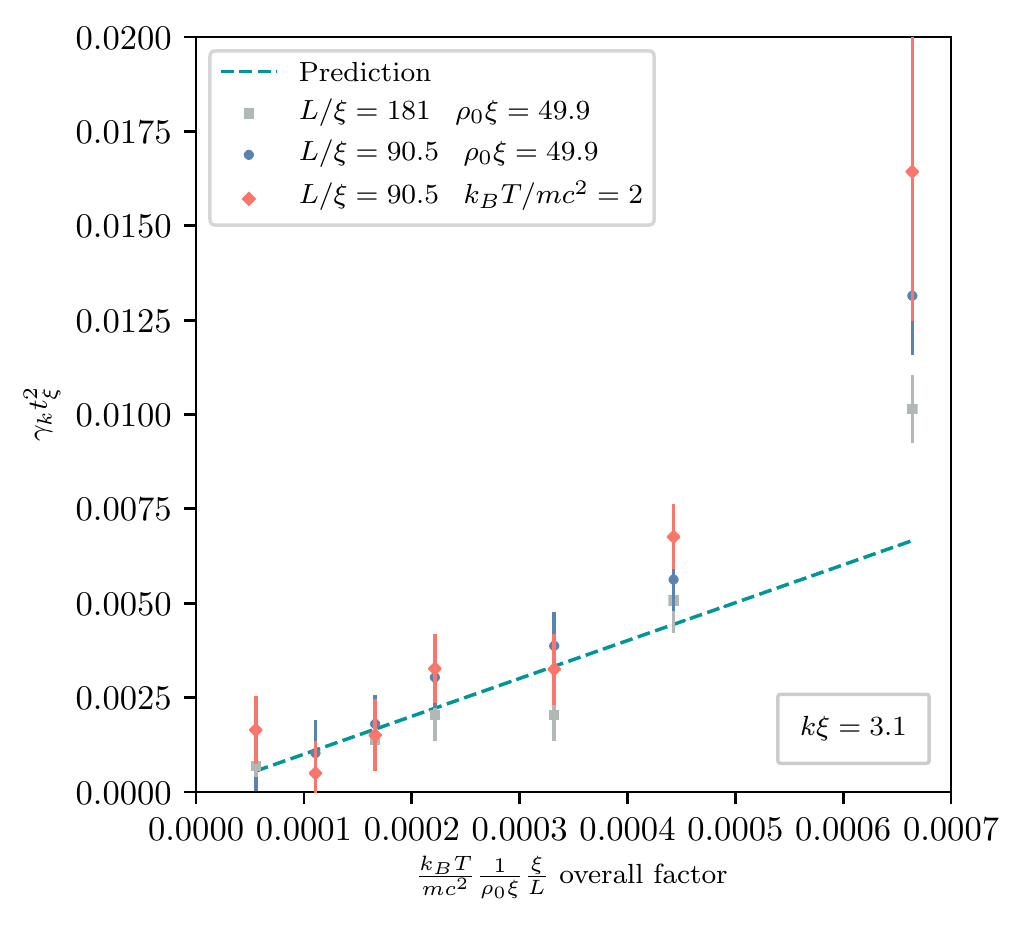}
    
    \caption{(Left) Plot of $\gamma_k  t_{\xi}^2$ as a function of $k \xi$. The parameters and data used are the same as  Fig.~\ref{fig:scaling_Gamma_k_xi}.
   (Right) Plot of $\gamma_k  t_{\xi}^2$ as a function of $k_{B} T/ m c^2$, $ \rho_0 \xi$ and $L/ \xi$. The parameters and data used are the same as  Fig.~\ref{fig:scaling_Gamma_asap_T}. 
   In both panels the red dots are the best fit values extracted from the TWA simulations using Eq.~(\ref{eq:nk_finite_size_correction}) as a template. The green dashed line is the prediction of equation Eq.~(\ref{def:prediction_small_gamma}).
    }
    \label{fig:scaling_small_gamma}

\end{figure}

It is clear from Eq.~(\ref{eq:finite_size_deceleration}) that there is some push-back on the decay.  The simplest solution is when $\delta n_{k}$ does not vary much over the duration of interest, so that we can pull it out of the integral:
\begin{equation}
    \partial_{t}\left(\delta n_{k}\right) \approx \left( -\Gamma_{k} + \gamma_{k} t\right) \delta n_{k} \,,
\end{equation}
with solution given by Eq.~\eqref{eq:nk_finite_size_correction} of the main text.
In this approximation, the $\gamma_{k}$ term simply provides a constant deceleration to the decay rate.~\footnote{Here a time dependence of the rate appears through the $L$-dependence i.e. the finite size effect. The reader might wonder why we never mentioned possible deviations to the exponential decay solely due to the finiteness of $t$ and hence the sinc being not exactly a Dirac delta in Eq.~(\ref{eq:Ik_sinc}). The finite $t$ deviation can be obtained by expanding the effective interaction strength around $q = 0$. Indeed, a Dirac delta would pick up only the value at $q=0$, first term in the expansion, while a sinc is still sensitive to the deviations from the value at $q=0$ which are encoded by the higher order terms. We can then proceed by integration by part in Eq.~(\ref{eq:Ik_sinc}) to find that each term in $q^n$ corresponds to a term decaying as $1/t^{n}$. A careful analysis shows that by combining both integrals the next to leading order term in the expansion around $q=0$ is in $q^2$ and not linear in $q$ as one might have expected. The first deviation is then in $1/t^2$. The reason is that since $n^{\rm th}_q + n^{\rm th}_{k-q} + 1 \approx 1/q + 1/2 + n^{\rm th}_{k}$ and $n^{\rm th}_{q} - n^{\rm th}_{k+q} \approx 1/q - 1/2 - n^{\rm th}_{k}$, the contribution coming from both integrals cancel exactly. Hence, the convergence of the sinc to a Dirac delta is effectively faster than one might expect and not a limitation to our result. Notice that since this term is not continuous we have to perform different expansions on both sides.}
In numerical simulations, we have used Eq.~(\ref{eq:nk_finite_size_correction}) as a template to extract the fitted values of $\Gamma_k$ shown in the figures of the main text.  The values of $\gamma_k$ extracted from the same fits in the time window $t / t_{\xi} \in \left[ 0, 10\right]$ are plotted in Fig.~\ref{fig:scaling_small_gamma}, along with the prediction of Eq.~(\ref{def:prediction_small_gamma}). It demonstrates a reasonable agreement in the same regime of validity as the one of exponential decay ($k \xi \leq 2.5$) shown in Fig.~\ref{fig:scaling_Gamma_k_xi}. At smaller $k\xi$, the system exhibits the $t^{2}$ behavior described above, and this becomes reflected in the fitting parameters.  In particular, $\gamma_{k}$ moves over into the $t^{2}$ coefficient $-\alpha_{k}$, and we see indeed that its sign changes.

That said, there is a difference between how $\gamma_k$ and $\alpha_k$ enter the expression of $n_k(t)$, respectively, linearly and exponentially.  In particular, while the early $t^{2}$ behavior is described by the single fitting parameter $\alpha_{k}$, the inclusion of $\gamma_{k}$ as a correction to the exponential decay means that it is one of two fitting parameters.   The time window of fit is large enough so that the difference between these two templates (applied on the same data) is visible, as illustrated by a direct comparison of Fig.~\ref{fig:scaling_alpha_k} with Fig.~\ref{fig:scaling_small_gamma}.
 
\subsubsection{Relation to finite size and resolution in $k$-space}

\begin{figure}
    \centering
    \includegraphics[width=0.495\textwidth]{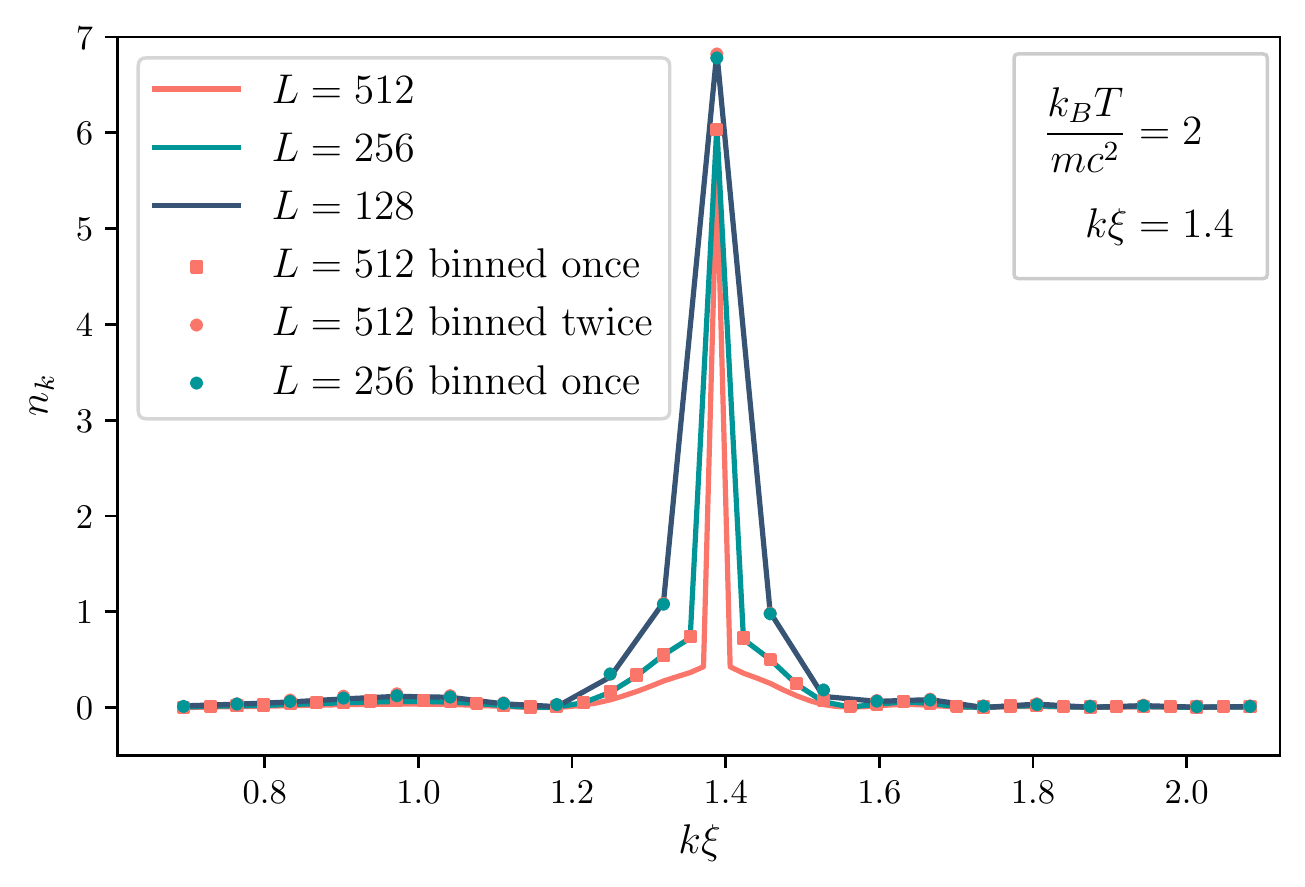}
    \includegraphics[width=0.495\textwidth]{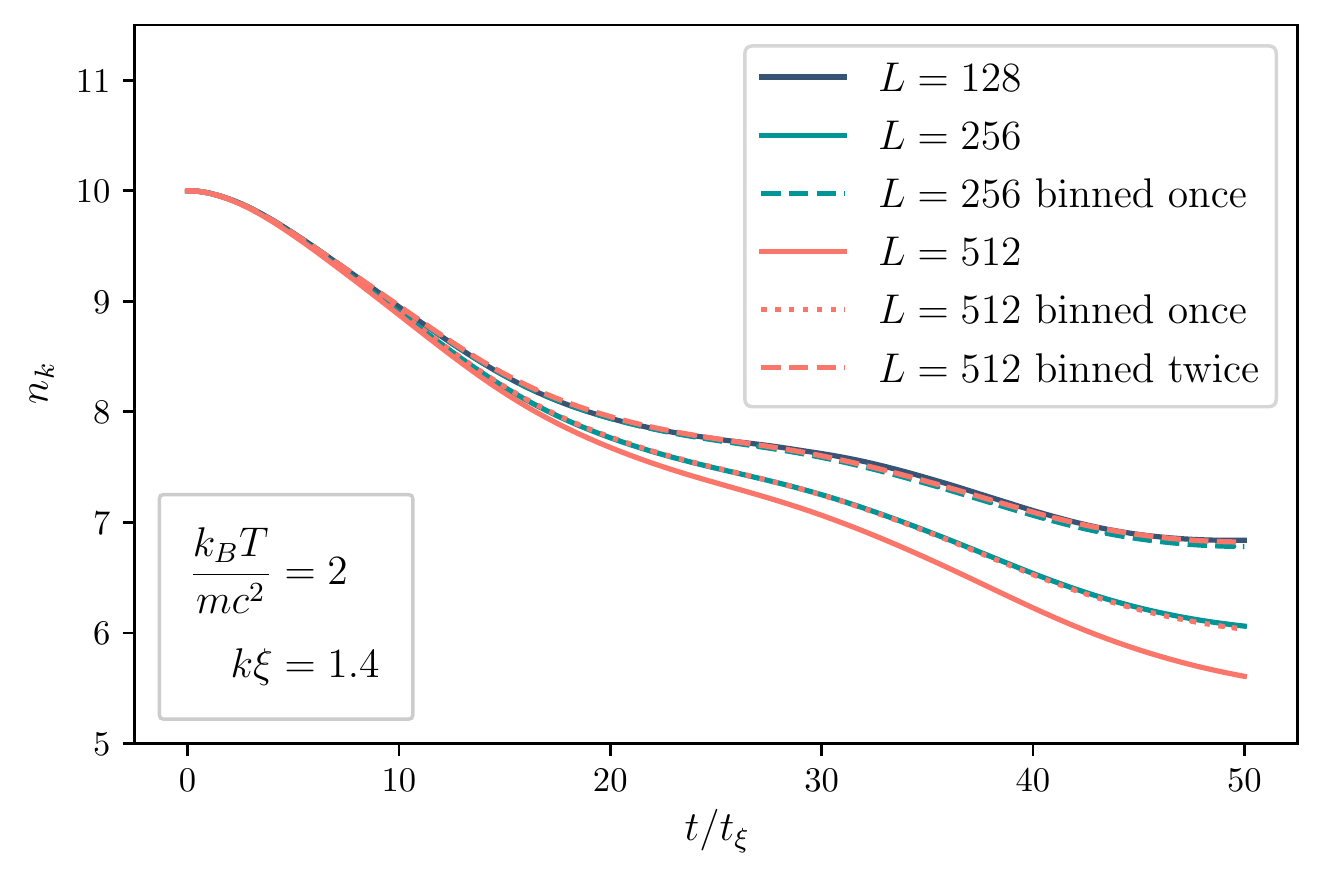}
    \includegraphics[width=0.495\textwidth]{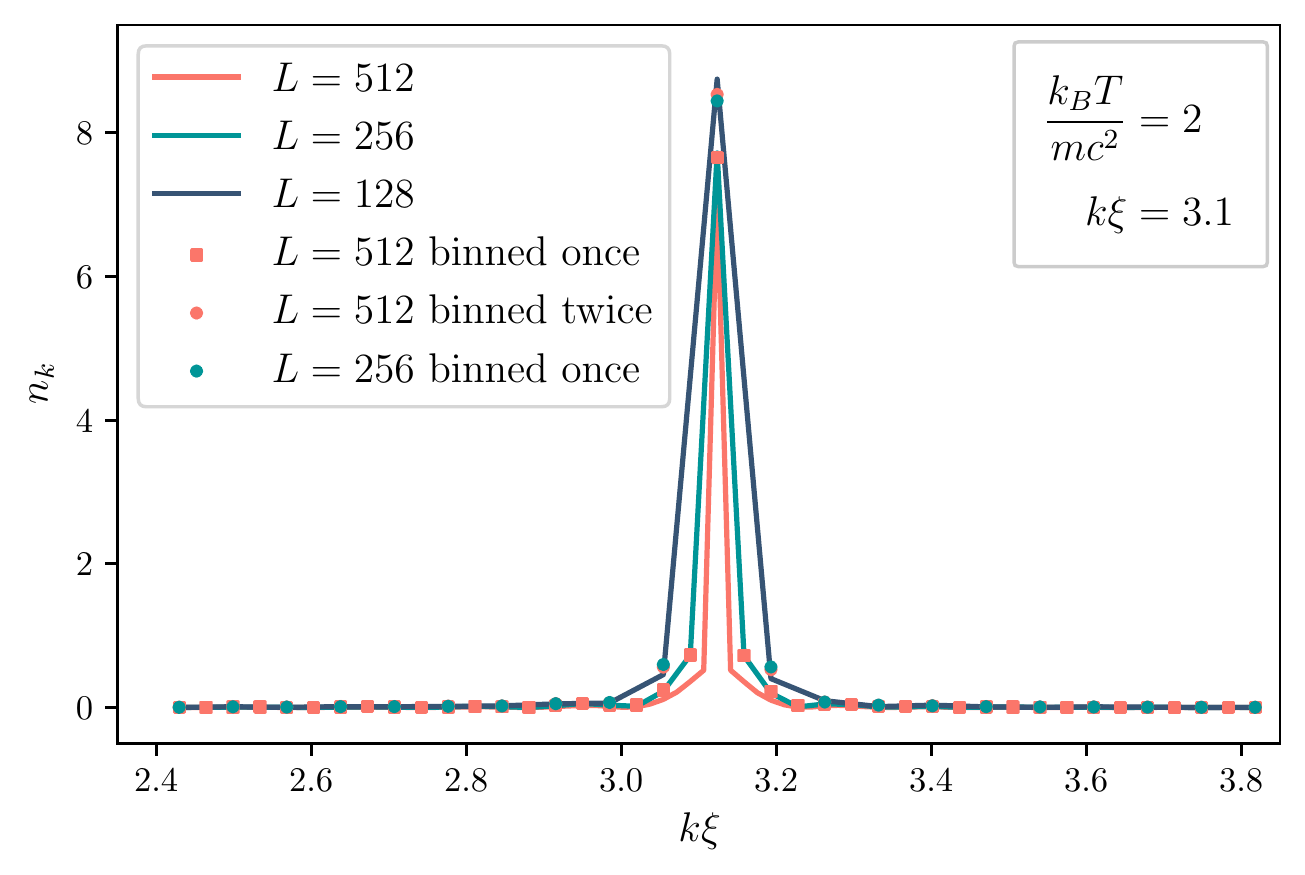}
    \includegraphics[width=0.495\textwidth]{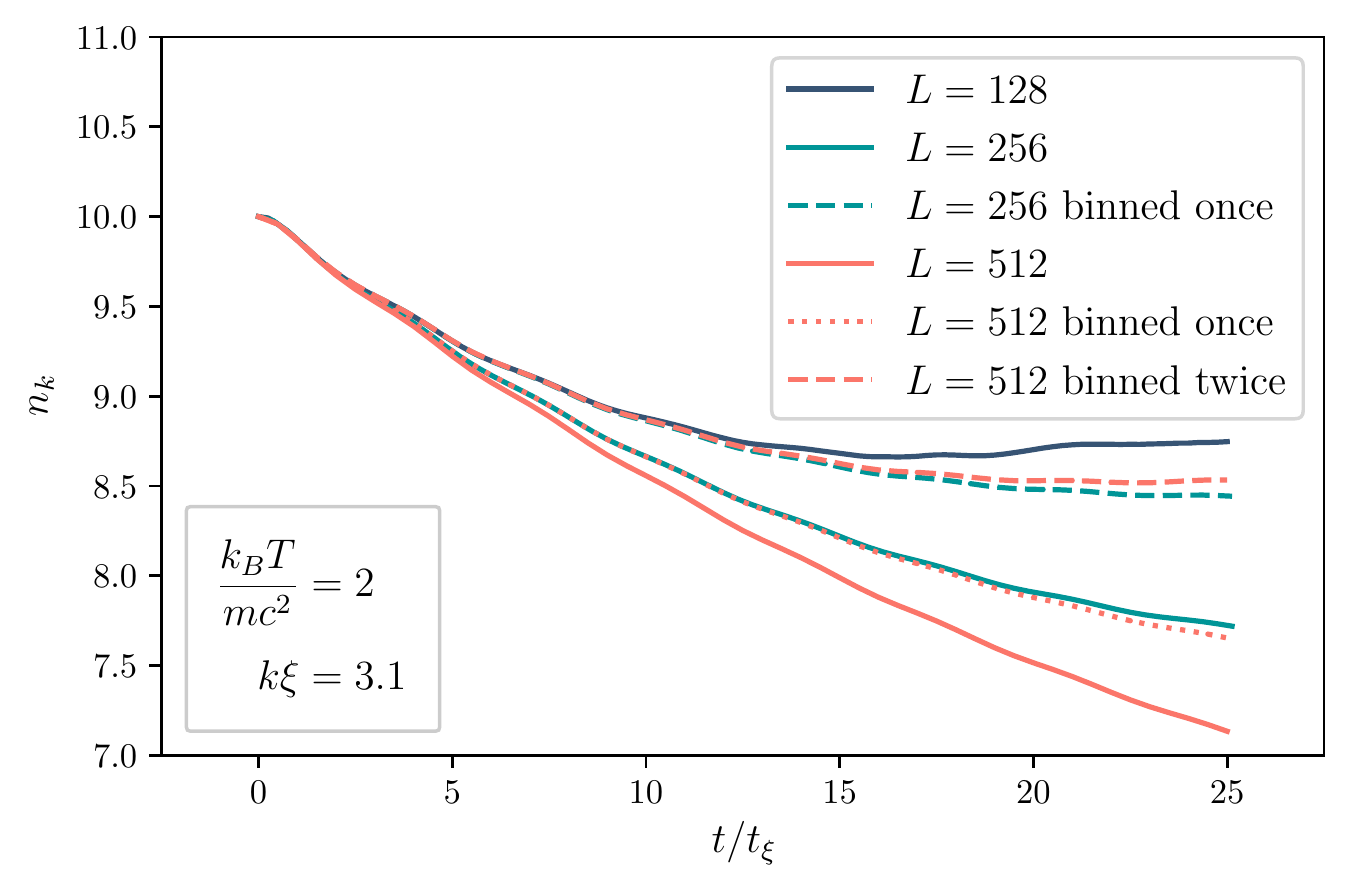}
    \caption{Number of non-thermal phonons $\delta n_k$ as of function of $k \xi$ at $t /t_{\xi} = 50$ (top left.), and $t /t_{\xi} = 25$ (bottom left.), when $\delta n = 10$ phonons were initially added in the mode $k \xi = 1.4$ (top left.), or $k \xi = 3.1$ (bottom left.). This number is shown in full line for different values of density of the grid $\Delta k = 2 \pi / L$ where $L= 128$ (blue), $L= 256$ (green) and $L= 512$ (red).
    The number of modes in the simulation for $L=512$ is then halved by merging nearby modes into a single bin via Eq.~(\ref{eq:binning_once}) and shown in red squares. Equation~(\ref{eq:binning_once}) is then applied on the set of modes for $L=256$ and repeated via Eq.~(\ref{eq:binning_twice}) on the new set of modes to have the same set of modes as $L=128$ shown respectively the red and green dots. They are almost everywhere superimposed so that they can hardly be distinguished.
    The evolution as a function of $t/t_{\xi}$ of the population in the probe mode is shown in full lines on the right figures for $k \xi = 1.4$ (top right.) and $k \xi = 3.1$ (bottom right.). The result of the binning procedure from $L=512$ to the set of modes of $L=256$ is shown in red dotted line. The result of the second binning and the binning of $L=256$ to the set of modes of $L=128$ is shown in red and green dashed lines.
    The initial thermal state used corresponds to $k_{B} T/ m c^2=2$. The evolution is performed using the same numerical strategy as described for the green curves of Fig.~\ref{fig:different_n_k} for  $\rho_0 \xi = 49.9$.
    \label{fig:decay_nk_different_L_binning} }
\end{figure}

Is there an intuitive way to link this apparent deceleration to the finite size of the system?
As our calculations are based in Fourier space, the size of the system enters through the discreteness of the available modes. To some extent, the discretized description should simply provide a finite-resolution view of the continuous-mode (infinite-$L$) case.

We have investigated this by numerically solving Eqs.~(\ref{eq:reduced_syst_eom_nk_cpq}) directly.   We inject a number of phonons in a single mode, and allow the system to evolve in time, for three different values of $L$.  We then compare the evolution by appropriately binning the phonon number spectrum $n_{k}$ when the $k$-space resolution is higher, so that each simulation is ultimately represented using the same set of discrete modes. 
This procedure can also be seen as restricting attention to a section of length $D$ of the system so that the relevant momenta are $2 \pi \, n/ D$ rather than $2 \pi\, n/ L$, for $n \in \mathbb{Z}$. The precise relationship between the two sets of modes is equivalent to a choice of window  in position space, over which the Fourier transform of the field is taken. We take a more heuristic approach here.

 For example, to map the data for a simulation of a system of length $L$ onto the set of modes applicable to a system or section of length $L/2$, the phonons in every second mode must be reallocated to neighbouring modes.  We do this by dividing them symmetrically into their two nearest neighbours, half the phonons going into the mode above, half into the mode below.  So:
\begin{equation}
\label{eq:binning_once}
    n_{k}^{\rm binned} = \frac{1}{2} n_{k-\Delta k} + n_{k} + \frac{1}{2} n_{k+\Delta k} \,.
\end{equation}
Similarly, to map onto a set of modes applicable to a system of length $L/4$, we adopt the following binning procedure:
\begin{equation}
\label{eq:binning_twice}
    n_{k}^{\rm binned} = \frac{1}{2} n_{k-2\Delta k} + n_{k-\Delta k} + n_{k} + n_{k+\Delta k} + \frac{1}{2} n_{k+2\Delta k} \,. 
\end{equation}
Results are shown in Fig.~\ref{fig:decay_nk_different_L_binning}, for two different values of $k$.  Interestingly, the binning procedure applied to larger-$L$ data gives a very good approximation to the data for smaller-$L$.  This suggests that there is no new (relevant) physics due to the discretization in $k$-space, in the sense that we can solve for a continuous $k$-space ({\it i.e.}, in the limit of infinite $L$) and then simply apply a suitable binning procedure to see how the discretized system behaves.  Similarly, the (relevant) physics on a section of length $L$ of a system of infinite size is the same as that of a finite-size system of the same length.

This binning procedure provides an intuitive explanation for the apparent deceleration of the decay encoded in Eq.~(\ref{eq:finite_size_deceleration}).  For, while the singularly occupied mode decays exponentially, the lost phonons are kicked into nearby modes, whose occupation numbers grow in time.  This is what is captured by the second term of Eq.~(\ref{eq:finite_size_deceleration}): it represents the growth of those modes in the continuous spectrum that are very near $k$, but which, due to the finite resolution in $k$-space, are included in the same bin.
By local conservation of the number of phonons described in the text, if we bin all modes within the width of the peak into a single mode, we effectively suppress the decay, {\it i.e.}, if we consider a sufficiently short section we will not witness the decay of the phononic excitations. The time-scale comparison of Eq.~\eqref{eq:relevant_finite_size} shows that the relevant length scale is the coherence length of the quasicondensate $r_0$. For shorter distances we do not expect to be able to resolve the decay of the phonons and the broadening of the peak in momentum space.

Of course, we do not expect the good correspondence shown in Fig.~\ref{fig:decay_nk_different_L_binning} between simulations of different $L$ to last indefinitely.  Eventually, the relevant components of the system -- namely, the thermal spectrum and the probe -- will be able to tell that they are on a finite torus, rather than a finite section of an infinite-size system, and we can then expect the different simulations to diverge significantly.  The critical time can be conceptualized as the re-crossing of the relevant components, which would not re-cross if the system were truly infinite in length.  There are two such times, corresponding to the re-crossing of the probe with the positive and negative wave vector components of the thermal spectrum, which propagate and speeds $c$ and $-c$, respectively.  Then the re-crossing times are
\begin{equation}
    t_{+}^{\rm rec} = \frac{L}{v_{\rm gr}(k)-c} \,, \qquad t_{-}^{\rm rec} = \frac{L}{v_{\rm gr}(k)+c} \,.
\end{equation}
Since these are simply proportional to $L$, it makes sense that the simulation with smallest $L$ should be the first to show deviations, as we see in the lower-right panel of Fig.~\ref{fig:decay_nk_different_L_binning}.  (In this example, $t_{-}^{\rm rec}/t_{\xi} = 30.1$.)  However, the deviations remain small, likely due to the weaker coupling between the probe and the negative part of the thermal spectrum.  We expect more significant differences after time $t_{+}^{\rm rec}$; indeed, we have observed some recurrence of $\delta n_{k}$ on this time scale.

\subsection{Finite width of the resonant peak \label{app:finite_width}}

Here we will discuss the corrections to the decay rate induced by a finite peak width.  In particular, we will show how the approximate corrections shown in Fig.~\ref{fig:gamma_k_parametric_amplification_reduction} were calculated.

The reduction of the decay rate is induced by the off-diagonal part of the matrix response function.  Phonons next to the mode of interest also decay due to interaction with the thermal population, and some of these are thus transferred into the mode of interest.  Since they can only contribute positively to the occupation number of the main mode, they decelerate the decay of $n_k$, thereby effectively reducing the decay rate $\Gamma_{k}$.

In our first set of simulations, we impose a very narrow initial peak, and this effect does not have a chance to build up significantly.  By contrast, in the second set of simulations where the peak is induced via parametric resonance, it naturally has a finite width which is observed to depend on the interaction strength.  We thus see signs of the overall decay rate being smaller than the predicted value at larger interaction strengths, as seen in Fig.~\ref{fig:gamma_k_parametric_amplification_reduction}.

\begin{figure}
    \centering
    \begin{minipage}{0.49\textwidth}
        \centering
        \includegraphics[width=\textwidth]{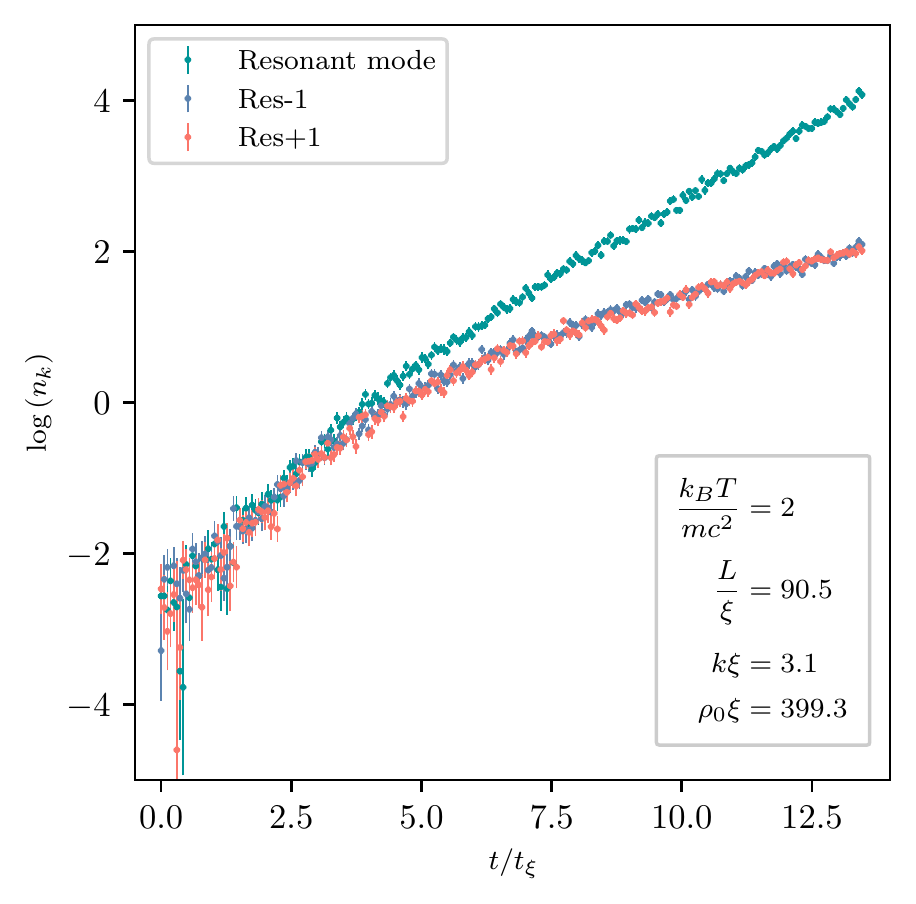}
    \end{minipage} \hfill
    \begin{minipage}{0.49\textwidth}
        \centering
        \includegraphics[width=\textwidth]{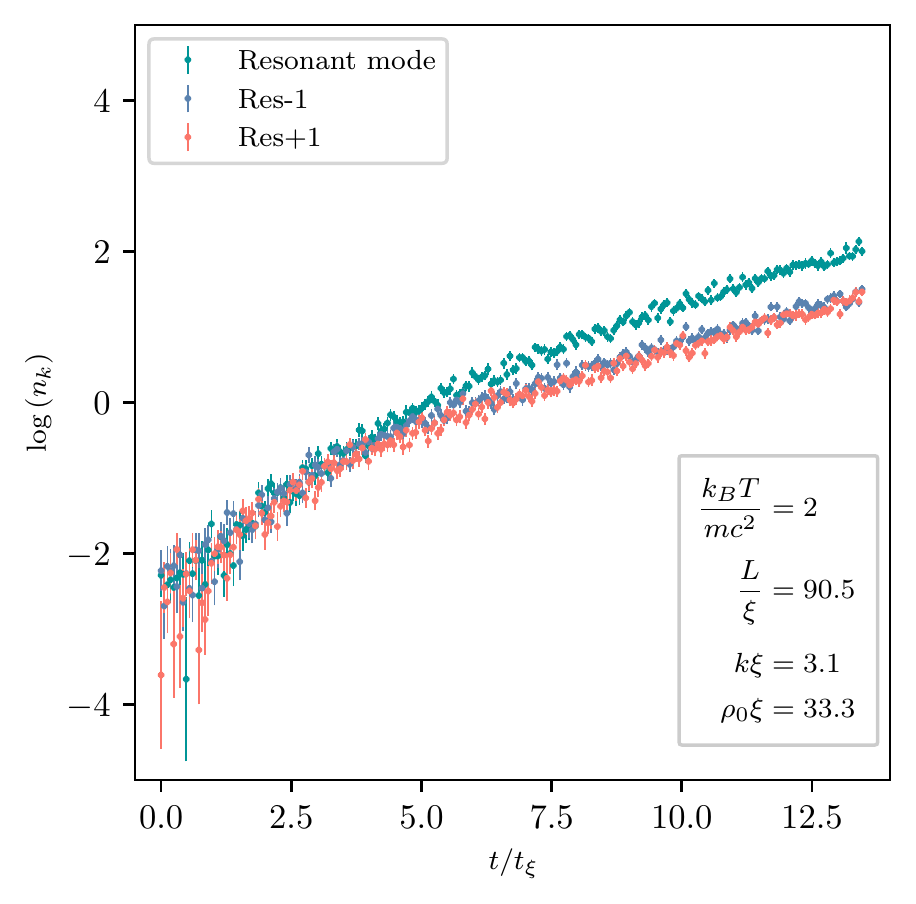}
    \end{minipage}
    \caption{
    Logarithm of the mean occupation of the resonant mode (green) and its nearest neighbours on either sides (red, blue) as a function of time extracted from the TWA simulations. The parameters for the initial state, the amplitude and the duration of the modulation are the same as in Fig.~\ref{fig:n_k_parametric_amplification_reduction}. We see a quicker convergence of the growth rates within the peak on the right plot, which is due to a smaller atomic density resulting in a larger effective interaction strength.
    \label{fig:log_nk_different_asap}
    }
\end{figure}

We wish to predict the expected reduction in $\Gamma_{k}$ given the observed peak width.  To this end we employ Eq.~\eqref{eq:nk_eom_response_functions} with two key assumptions.  First, we modify the equation to include a source term that induces exponential growth, modeling the parametric resonance.  This is achieved by adding $G_{k} \, \delta n_{k}$ to the right-hand side, where $G_{k}$ is the predicted growth rate in the absence of any phonon-phonon interactions:
\begin{equation}
G_{k} = \frac{1}{2} A_{k} \omega_{k} \,,
\end{equation}
which is consistent with Eq.~(\ref{eq:estimate_parametric_growth}) for sufficiently large $t$.  Second, we assume that a steady state is reached where the occupation numbers of all relevant nearby modes grow at the same net rate, $G_{\rm net}$.  Therefore, the shape of the peak is assumed constant, and evolves in time simply according to
\begin{equation}
    \delta n_{k}(t) = f_{k} \, e^{G_{\rm net} t} \,.
\end{equation}
This appears to be consistent with numerical observations see Fig.~\ref{fig:log_nk_different_asap}. 

Incorporating these assumptions in Eq.~\eqref{eq:nk_eom_response_functions} for the occupation number of the resonant mode yields a consistency equation that determines the expected net growth rate, $G_{\rm net}$:
\begin{eqnarray}
    \partial_{t}\left(\delta n_{k}\right) = G_{\rm net} \, \delta n_{k} &=& \left( G_{k} - \Gamma_{k} \right) \delta n_{k} + \int_{0}^{t} dt^{\prime} \, \sum_{q \neq -k} M_{k,k+q}\left(t-t^{\prime}\right) \, \delta n_{k+q}\left(t^{\prime}\right) \nonumber \\
    &\approx& \left(G_{k}-\Gamma_{k}\right) \delta n_{k} + \int_{0}^{t} dt^{\prime} \int dq \, \frac{L}{2\pi} \, M_{k,k+q}\left(t-t^{\prime}\right) \, \frac{\delta n_{k+q}\left(t^{\prime}\right)}{\delta n_{k}\left(t^{\prime}\right)} \, \frac{\delta n_{k}\left(t^{\prime}\right)}{\delta n_{k}(t)} \, \delta n_{k}(t) \nonumber \\
    &=& \left(G_{k}-\Gamma_{k}\right) \delta n_{k} + \delta n_{k}(t) \, \int_{0}^{t} dt^{\prime} \int dq \, \frac{L}{2\pi} \, M_{k,k+q}\left(t-t^{\prime}\right) \, R_{k+q} \, e^{-G_{\rm net}\left(t-t^{\prime}\right)} \,,
\end{eqnarray}
where in the second line we have taken the continuum limit to replace the sum over wave vectors $q$ by an integral.  The exponential suppression in $t-t^{\prime}$ allows us to replace the lower limit of the $t^{\prime}$ integral by $-\infty$, which in turn allows us to write the integral independently of $t$ and to divide through the whole equation by $\delta n_{k}(t)$:
\begin{equation}
    G_{\rm net} = G_{k}-\Gamma_{k} + \int_{0}^{\infty} d\tau \int dq \, \frac{L}{2\pi} \, M_{k,k+q}\left(\tau\right) \, R_{k+q} \, e^{-G_{\rm net} \tau} \,.
\end{equation}
This self-consistency equation can be solved for $G_{\rm net}$, given that we know $G_{k}$, $\Gamma_{k}$, and the profile shape $R_{k+q}$.
If we assume that the peak is narrow enough so that only small $q$ are relevant, we may replace $M_{k,k+q}$ with its small-$q$ limit, given in Eq.~(\ref{eq:simplified_non_diagonal_response}).  This gives
\begin{eqnarray}
    G_{\rm net} &\approx& G_{k}-\Gamma_{k} + \int_{0}^{\infty} d \tau \int dq \left\{ f_{+} \, {\rm cos}\left[q \left(v_{\rm gr}(k)-c\right) \tau \right] + f_{-} \, {\rm cos}\left[q \left(v_{\rm gr}(k)+c\right) \tau \right] \right\} \, R_{k+q} \, e^{-G_{\rm net} \tau} \nonumber \\
    &=& G_{k}-\Gamma_{k} + \int dq \left\{ f_{+} \, \frac{G_{\rm net}}{G_{\rm net}^{2} + q^{2}\left(v_{\rm gr}(k)-c\right)^{2}} + f_{-} \, \frac{G_{\rm net}}{G_{\rm net}^{2} + q^{2}\left(v_{\rm gr}(k)+c\right)^{2}} \right\} \, R_{k+q} \,,
\end{eqnarray}
where we have defined
\begin{equation}
    f_{\pm} = \frac{4}{\pi} \, \frac{L}{N_{\rm at}} {\rm lim}_{q \to 0^{\pm}} \left\{ \left|V_{3}\left(k,q\right)\right|^{2} n_{q}^{\rm th} \right\} \,.
\end{equation}
All that remains is to plug in a suitable profile $R_{k+q}$ for the profile of the peak.  Over a significant range of parameters, we observe numerically that the profile is fairly well described by a Lorentzian:
\begin{equation}
    R_{k+q} = \frac{1}{\left(q/\sigma\right)^{2}+1} \,.
\end{equation}
Adopting this as an ansatz and plugging it into the integral above, we find
\begin{equation}
    G_{\rm net} - \Gamma_{+} g\left(\frac{G_{\rm net}}{\left(v_{\rm gr}(k)-c\right) \sigma}\right) - \Gamma_{-} g\left(\frac{G_{\rm net}}{\left(v_{\rm gr}(k)+c\right) \sigma}\right) = G_{k}-\Gamma_{k} \,,
    \label{eq:self-consistency}
\end{equation}
where $\Gamma_{\pm}$ are the predicted decay rates given in Eq.~(\ref{def:decay_rate}) (recall that $\Gamma_{k} = \Gamma_{+}+\Gamma_{-}$), and where we have defined
\begin{equation}
    g(x) = \frac{1}{1+x} \,.
\end{equation}
The effective decay rate $\Gamma_{\rm eff}$ is then defined as $\Gamma_{\rm eff} = G - G_{\rm net}$, thus capturing the reduction of the growth rate due to phonon-phonon interactions. Note that Eq.~(\ref{eq:self-consistency}) yields the expected behavior in the limits of very narrow and very broad peaks:  
\begin{itemize}
    \item When the peak is very narrow, we take $\sigma \to 0$ so that the argument of $g(x)$ becomes very large, and we can set $g(x)$ to 0.  Then we just get $G_{\rm net} = G_{k}-\Gamma_{k}$, so $\Gamma_{\rm eff} = \Gamma_{k}$; this is exactly our prediction for the decay rate of a narrow peak.
    \item Conversely, when the peak is very broad so that we can send $\sigma \to \infty$, the argument of $g(x)$ becomes very small and we can replace $g(x)$ by 1.  Since $\Gamma_{k} = \Gamma_{+}+\Gamma_{-}$, Eq.~(\ref{eq:self-consistency}) tells us that $G_{\rm net} = G_{k}$, {\it i.e.}, there is no reduction of the growth rate.  This corroborates our expectation that the broadness of the peak tends to suppress the decay rate.
\end{itemize}

To determine the corrected decay rates shown in Fig.~\ref{fig:gamma_k_parametric_amplification_reduction}, we first perform a fit of a Lorentzian profile to the occupation numbers around the resonant peak (including five data points on either side of the resonant mode), in order to extract the width $\sigma$.  This is then used in Eq.~(\ref{eq:self-consistency}) on order to determine the expected net growth rate $G_{\rm net}$, and thereby the effective decay rate $\Gamma_{\rm eff}$.

\section{Monte-Carlo simulations : Truncated Wigner Approximation (TWA)}
\label{app:TWA}

In order to assess the validity of our predictions we compare them to the results of \textit{ab initio} Monte-Carlo simulations of the system. To simulate the evolution of our quasicondensate we use the Truncated Wigner Approximation (TWA) also known as the classical field approximation. This method is based on the description of the state of the system by means of a quasi-probability distribution, the Wigner function. The TWA has been repeatedly used to describe Bose gas \cite{steelDynamicalQuantumNoise1998,Sinatra2002}, specifically one-dimensional quasicondensate \cite{Ruostekoski2013,Ruostekoski2010}, as well as a wide variety of other systems (polaritons, spins etc.) \cite{Carusotto2005,Huber2021}. We will restrict our presentation of the TWA to the necessary minimum and we refer to \cite{steelDynamicalQuantumNoise1998} for further details.

\subsection{TWA in a nutshell}

The Wigner function $W \left( \Psi, \Psi^{\star} \right)$ is a quasi-probability distribution in phase space defined by a bijective transformation of the density matrix $\hat{\rho}$.  Under this transformation the von Neumann equation of motion on the density matrix then translates into a partial differential equation for $W$ \cite{curtrightConciseTreatiseQuantum2014}. For the Hamiltonian~(\ref{eq:Hamiltonian_full}) the equation reads 
\begin{equation}
\label{eq:eom_wigner_function}
    i \hbar \dot{W}\left( \Psi, \Psi^{\star} \right) = - \int_{- \infty}^{+ \infty} \left\{ \frac{\delta}{\delta \Psi} \left[ \frac{1}{2m} \partial^2_{x} \Psi + g \left( \left| \Psi \right|^2 -1 \right) \Psi   \right] - \frac{1}{4} \frac{\delta^3}{\delta^2 \Psi \delta \Psi^{\star}} \Psi \right\} W \left( \Psi, \Psi^{\star} \right)  + c.c. \, ,
\end{equation}
where the derivatives act on the element inside of the brackets multiplied by $W \left( \Psi, \Psi^{\star} \right)$, see Eq.~(23) of \cite{steelDynamicalQuantumNoise1998}. The truncation giving its name to the Truncated Wigner Approximation consists in neglecting terms with three derivatives or more. The resulting equation is then solved by using the method of characteristics. Practically, the Wigner function at time $t$ is found by first sampling the Wigner function at initial time i.e. drawing a set of values $\Psi_i(x,t=0)$ according to the initial Wigner function and evolving these realizations under an equation of motion which is simply the classical counterpart of the Heisenberg equation of motion of $\hat{\Psi}$ under the Hamiltonian~(\ref{eq:Hamiltonian_full})
\begin{equation}
\label{eq:TWA_eom}
    i \hbar \dot{\Psi} = - \frac{1}{2m} \partial^2_{x} \Psi + g  \left| \Psi \right|^2 \Psi   \, .
\end{equation}
The resulting $\Psi_i(x,t)$ represent a sampling of the Wigner function of the state at time $t$. This sample can be used to compute average values of observables $\left \langle A \left(  \varphi_k \right) \right \rangle_{\rm TWA}$ built from $\varphi_k$. It can be shown that expectation values computed treating the Wigner function as a \textit{bona fide} probability distribution are equal to quantum expectation values of the associated operator when the expression is completely symmetrized in $\hat{\varphi}_k$ and $ \hat{\varphi}_k^{\dagger}$. We will not discuss here the conditions of validity of this truncation and of its numerical implementation; some considerations can be found in \cite{Mora_2003,steelDynamicalQuantumNoise1998,Van_Regemortel_2017}.

\subsection{Numerical implementation }

The numerical implementation of the TWA is a straightforward application of the above program. The evolution is performed using a discretized version of Eq.~(\ref{eq:TWA_eom}) and a split-step Fourier algorithm~\cite{AgrawalBook}.
The initial state is taken to be thermal at a temperature $T$, up to the addition of a few ``probe'' phonons in the mode $k$ for the first set of simulations. The exact thermal state of the system is approximated by the one associated to the quadratic Hamiltonian $\hat{H}_2$ {\it i.e.}, the phonon modes are completely uncorrelated to each other and their number spectrum is set equal to the Bose-Einstein distribution at temperature $T$. The realisations of the atomic field are built from $n_r$ realisations of the phonon modes $\varphi_k$. These are themselves built drawing independently $\Re \left[ \varphi_k \right]$ and $\Im \left[ \varphi_k \right]$ according to a centered Gaussian distribution with variance
\begin{equation}
   \sigma_k^2 = \frac{1}{2} \left( n_k + \frac{1}{2} \right) =  \frac{1}{4} \coth \left( \frac{\hbar \omega_k}{2 k_B T} \right)  \, .
\end{equation}
It is then straightforward to check that the TWA averages reproduce the averages of the symmetrized quantum operators in a thermal state:
\begin{align}
\left \langle \varphi_k \right \rangle_{\rm TWA} & = \left \langle \varphi_k^{*} \right \rangle_{\rm TWA} = 0 \, , \\
\left \langle \varphi_k^{*} \varphi_k \right \rangle_{\rm TWA} & = n_k^{\rm th}+\frac{1}{2} = \frac{1}{2} \left \langle  \left\{ \hat{\varphi}_k , \hat{\varphi}_k^{\dagger} \right\} \right \rangle \, ,  
\end{align}
where $\left\{ \hat{A} , \hat{B} \right\} = \hat{A} \hat{B} + \hat{B}\hat{A}$ is the anti-commutator. Since this is not the exact thermal state we let the system evolve for a certain duration to be as close as possible to a stationary state that we use as an initial state.
For the first set of simulations phonons are then injected in the mode $k$ by transforming its amplitude $\varphi_k(0)$ according to
\begin{equation}
    \varphi_k(0) \xrightarrow[]{} \sqrt{ 1 + \frac{\delta n}{n_k^{\rm th}+\frac{1}{2} }  }  \varphi_k(0)  \, , 
\end{equation}
so that $\varphi_k(0)$ still has Gaussian statistics but with a variance corresponding to $n_k = n_k^{\rm th} + \delta n$.
We then evolve under Eq.~(\ref{eq:TWA_eom}) both the realisations with and without the addition of probe phonons using the same code. We repeat this process for the $n_r$ realisations. Finally we compute the average values for $n_q$ with and without the probe at any time and take the difference to get the evolution of the average value $\delta n_q (t)$. 

This method closely mirrors the step of the derivation of the decay rate laid out around Eq.~(\ref{eq:nk_eom_linearized}) where we subtract the evolution of the background thermal population to single out the evolution of the probe. However the TWA encodes also the deviations to the linear case considered in our equations. 
For the second set of simulations we simply evolve the initial quasi-stationary state according to \eqref{eq:TWA_eom} where $g$ is modulated according to \eqref{eq:freq_modulation}, and proceed to the same type of averaging as in the first set. 
We want to stress that in any figure of this work, for both set of simulations, a data point at time $t + \mathrm{d}t$ is \textit{not} obtained simply by evolving the realisations at $t$ time by an extra-step $\mathrm{d}t$ as this would result in strongly correlated data points. We rather start the whole evolution process from new realisations of the initial state and evolve them until $t + \mathrm{d}t$.

\end{appendices}

\end{document}